\documentclass[fleqn,usenatbib]{mnras}

\usepackage{newtxtext,newtxmath}

\usepackage[T1]{fontenc}

\DeclareRobustCommand{\VAN}[3]{#2}
\let\VANthebibliography\thebibliography
\def\thebibliography{\DeclareRobustCommand{\VAN}[3]{##3}\VANthebibliography}

\usepackage{graphicx}	
\usepackage{amsmath}	
\usepackage{comment}
\usepackage{xspace} 
\usepackage{mathtools}
\usepackage{hyperref}
\usepackage{float}
\usepackage{bm}
\usepackage{pdflscape}
\usepackage{pdfpages}
\DeclareFontFamily{U}{mathx}{}
\DeclareFontShape{U}{mathx}{m}{n}{<-> mathx10}{}
\DeclareSymbolFont{mathx}{U}{mathx}{m}{n}
\DeclareMathAccent{\widecheck}{0}{mathx}{"71}

\usepackage{multirow} \usepackage{threeparttable}
\usepackage{booktabs} \usepackage{stfloats}

\usepackage[ruled,vlined]{algorithm2e}
\SetKwRepeat{Do}{do}{while}%
\usepackage{algpseudocode}

\SetKwComment{Comment}{$\triangleleft$ }{}

\newcommand{\AlgoStep}[2]{$\blacktriangleright$ \textbf{Step #1.} #2\;}

\newcommand*{\V}[1]{\boldsymbol{#1}}   
\newcommand*{\M}[1]{\mathbf{#1}}       
\newcommand*{\TransposeLetter}{\hspace*{-.25ex}\top\hspace*{-.25ex}}
\newcommand*{\T}{^{\TransposeLetter}} 
\DeclareFontFamily{U}{mathx}{\hyphenchar\font45}
\DeclareFontShape{U}{mathx}{m}{n}{<-> mathx10}{}
\DeclareSymbolFont{mathx}{U}{mathx}{m}{n}

\DeclarePairedDelimiterX{\Paren}[1]{(}{)}{#1}
\DeclarePairedDelimiterX{\Brace}[1]{\{}{\}}{#1}
\DeclarePairedDelimiterX{\Brack}[1]{[}{]}{#1}
\DeclarePairedDelimiterX{\Abs}[1]{\rvert}{\lvert}{#1}
\DeclarePairedDelimiterX{\Norm}[1]{\lVert}{\rVert}{#1}
\DeclarePairedDelimiterX{\Avg}[1]{\langle}{\rangle}{#1}
\DeclarePairedDelimiterX{\Round}[1]{\lfloor}{\rceil}{#1}
\DeclarePairedDelimiterX{\Floor}[1]{\lfloor}{\rfloor}{#1}
\DeclarePairedDelimiterX{\Ceil}[1]{\lceil}{\rceil}{#1}
\DeclarePairedDelimiterX{\Inner}[2]{\langle}{\rangle}{#1,#2}
\DeclarePairedDelimiterX{\Group}[1]{.}{.}{#1}

\DeclareMathOperator*{\argmin}{arg\,min}

\DeclareMathOperator{\Trace}{tr}

\DeclarePairedDelimiterXPP{\Expect}[1]{\mathbb{E}}(){}{#1}

\newcommand*{\estim}[1]{\widehat{#1}}

\newcommand{\obj}{\V{u}}

\newcommand{\data}{\V{v}}

\newcommand{\dd}{\partial}
\newcommand{\Weight}{\M \Psi}

\newcommand*{\spec}{\mathrm{spec}}
\newcommand*{\spat}{\mathrm{spat}}

\usepackage[squaren,Gray]{SIunits}

\usepackage{numprint}

\makeatletter
\def\widebreve{\mathpalette\wide@breve}
\def\wide@breve#1#2{\sbox\z@{$#1#2$}%
     \mathop{\vbox{\m@th\ialign{##\crcr
\kern0.08em\brevefill#1{0.8\wd\z@}\crcr\noalign{\nointerlineskip}%
                    $\hss#1#2\hss$\crcr}}}\limits}
\def\brevefill#1#2{$\m@th\sbox\tw@{$#1($}%
  \hss\resizebox{#2}{\wd\tw@}{\rotatebox[origin=c]{90}{\upshape(}}\hss$}
\makeatletter

\title[Disk reconstruction leveraging spectral data]{REXPACO ASDI: Joint unmixing and deconvolution of the circumstellar environment by angular and spectral differential imaging}

\author[O. Flasseur et al.]{
Olivier Flasseur$^{1}$\thanks{E-mail: olivier.flasseur@univ-lyon1.fr},  Loïc Denis$^{2}$, {\'E}ric Thiébaut$^{1}$, Maud Langlois$^{1}$
\\
$^{1}$Université de Lyon, Université Lyon1, ENS de Lyon, CNRS, Centre de Recherche Astrophysique de Lyon UMR 5574, Saint-Genis-Laval, France\\
$^{2}$Université de Lyon, UJM-Saint-Etienne, CNRS, Institut d Optique Graduate School, Laboratoire Hubert Curien UMR 5516, Saint-{\'E}tienne, France
}

\pubyear{2024}

\begin{document}
\label{firstpage}
\pagerange{\pageref{firstpage}--\pageref{lastpage}}
\maketitle

\begin{abstract}
Angular and spectral differential imaging is an observational technique of choice to investigate the immediate vicinity of stars. By leveraging the relative angular motion and spectral scaling between on-axis and off-axis sources, post-processing techniques can separate residual star light from light emitted by surrounding objects such as circumstellar disks or point-like objects. This paper introduces a new algorithm that jointly unmixes these components and deconvolves disk images. The proposed algorithm is based on a statistical model of the residual star light, accounting for its spatial and spectral correlations.
These correlations are crucial yet remain inadequately modeled by existing reconstruction algorithms. We employ dedicated shrinkage techniques to estimate the large number of parameters of our correlation model in a data-driven fashion. We show that the resulting separable model of the spatial and spectral covariances captures very accurately the star light, enabling its efficient suppression. We apply our method to datasets from the VLT/SPHERE instrument and compare its performance with standard algorithms (median subtraction, PCA, PACO). We  demonstrate that considering the multiple correlations within the data significantly improves reconstruction quality, resulting in better preservation of both disk morphology and photometry. With its unique joint spectral modeling, the proposed algorithm can reconstruct disks with circular symmetry (e.g., rings, spirals) at intensities one million times fainter than the star, without needing additional reference datasets free from off-axis objects.
\end{abstract}

\begin{keywords}
techniques: high angular resolution -- techniques: image processing -- methods: numerical -- methods: statistical -- methods: data analysis 
\end{keywords}


\section{Introduction}
\label{sec:introduction}

Direct imaging is a recent observational technique allowing to probe the close environment of young stars \citep{traub2010direct, bowler2016imaging}. The targeted tasks are threefold (see e.g., \cite{pueyo2018direct, currie2022direct, follette2023introduction} for reviews): (i) detecting (massive) exoplanets, (ii) characterizing their physical properties by estimating their spectral energy distribution (SED), and (iii) reconstructing the flux distribution image of the circumstellar environment surrounding young nearby stars. In this paper, we  primarily focus on the latter objective and we also address the unmixing of spatially resolved disks from point-like sources.

Circumstellar disks are key components of the intricate processes governing planetary formation. As an illustration, several studies performed in total intensity or in polarimetry \citep{Esposito2020,garufi2020disks,Langlois2020} have revealed the presence of a diversity of structures such as spirals, warps, rings, gaps, shadows and asymmetries, which are considered as potential indicators for the presence of exoplanets \citep{benisty2015asymmetric, muro2020shadowing}. High-quality reconstruction of the circumstellar environment from high-contrast data thus offer a unique vantage point to understand the physical processes governing these objects \citep{keppler2018discovery, haffert2019two, mesa2019vlt}. It also allows to study the intricate interactions between exoplanets and disks, and to provide critical insights into the mechanisms steering the evolution of exoplanetary systems.

In this context, direct imaging faces two observational challenges. First, the objects of
interest (i.e., spatially resolved disks and exoplanets appearing as point-like sources) have a very low contrast\footnote{Throughout this paper, we define the contrast of the objects of interest as the ratio of their peak intensity to the star peak intensity. 
This also corresponds to the classical definition of contrast for single-pixel point objects.} (typically lower than $10^{-4}$ in the infrared). Second, these off-axis objects are located in
the immediate vicinity of the star, thus necessitating high angular
resolution to separate them from the star (disks are generally observed inside an angle of less than
1 arcsecond).
The angular resolution requirement can be achieved using large ground-based telescopes equipped with extreme adaptive optics systems to compensate in real-time for atmospheric turbulence. The contrast is further improved by filtering most of the star light with a coronagraph. However, this is not sufficient to recover interpretable images of the circumstellar environment, as residual star light still dominates (see Fig. \ref{fig:data}). To further reduce the impact of star light, differential imaging is employed. This observational technique involves capturing several images in configurations that introduce diversity (e.g., relative motion in ADI or SDI) between the objects of interest and the star diffraction patterns, known as \textit{speckles}, caused by diffraction effects in the telescope pupil.
There are two primary configurations for differential imaging. In angular differential imaging (ADI; \cite{marois2006angular}), a sequence of images is acquired over a few hours of observation. During the acquisition, the pupil of the telescope keeps a
constant orientation (so-called \emph{pupil-tracking mode}), while the field of view
rotates due to Earth's rotation. This leads to a rotation of the objects of
interest around the optical axis in the images, while quasi-static speckles
created by uncorrected optical aberrations stay mostly fixed between individual exposures. In spectral differential imaging (SDI; \cite{sparks2002imaging,thatte2007very}), images are captured simultaneously in several spectral bands. Due to diffraction, the speckle pattern scales linearly with wavelength, in first approximation. By properly rescaling the images spectrally, the speckle patterns are aligned, while the objects of interest undergo radial motion and homothety due to the scaling transform. ADI and SDI can be advantageously combined to form angular and spectral differential imaging (ASDI) sequences, see e.g. \cite{vigan2010photometric,christiaens2019separating,kiefer2021spectral}. The images recorded in ADI, SDI, or ASDI are then combined in a post-processing step to enhance contrast and obtain interpretable images of the circumstellar environment.

The classical post-processing pipeline typically begins by estimating the stellar component, for instance, by averaging a stack of images with aligned speckles. This stellar component is then subtracted from the data, followed by the alignment and stacking of the residuals to compensate for rotations and scaling of the field of view. Beyond simple averaging, the stellar component can be estimated using various techniques: a median approach \citep{marois2006angular,lagrange2009probable}), a weighted linear combination (LOCI methods; \cite{lafreniere2007new, marois2013tloci, marois2014gpi, wahhaj2015improving}), or principal component analysis (PCA-based methods; \cite{soummer2012detection, amara2012pynpoint}). All of these methods can be applied on spatio-temporo-spectral data from IFS by leveraging differential diversity in various ways, see e.g. \cite{christiaens2019separating,kiefer2021spectral} for PCA. This can be done using ADI alone (i.e., a different model for each spectral channel), SDI alone (i.e., a different model for each temporal frame), ADI+SDI (i.e., two models: the first exploiting ADI diversity and the second exploiting SDI diversity to the ADI residuals), SDI+ADI (i.e., two models applied in reverse order of ADI+SDI models), or ASDI (i.e., a single model that jointly leverages both angular and spectral diversities). This latter strategy combined to the specific case of PCA is known as COmbined Differential Imaging (CODI; \cite{kiefer2021spectral}). However, these methods share a common drawback: part of the signal of interest is included in the estimated stellar component, resulting in its loss when the star component is subtracted from the data. This critical phenomenon, known as \emph{self-subtraction} \citep{milli2012impact, pairet2019iterative}, is particularly problematic close to the star, where the diversity between the disk and the star light is more limited (the apparent displacement of the off-axis objects due to the rotations and scaling transforms being separation-dependent). Consequently, disentangling the component of interest from the star light is even more difficult nearer the star. Self-subtraction can introduce various artifacts, such as partial replicas, suppression of some smooth extended structures, and smearing or non-uniform attenuations of disk features \citep{milli2012impact}.

To mitigate the impact of self-subtraction, several approaches were considered. Some works perform iterative PCA in which the current disk reconstruction is subtracted from the measurements to improve progressively the estimation of the star light \citep{pairet2019iterative, stapper2022iterative}. In the same vein, data imputation strategies \citep{ren2020using, ren2023karhunen} discard measurements affected by the disk, either through a data-driven approach or based on prior knowledge of its shape and location, during the estimation of the star light contribution. 
This type of approaches remains limited by the strategy designed to discard fractions of the field of view impacted by the disk on each image.
Other works consider a parametric model of a disk and iteratively adjust its
parameters \citep{esposito2013modeling, currie2017subaru, milli2017near} by minimizing the resulting residuals, possibly by modeling the effect of the self-subtraction \citep{lawson2020scexao, mazoyer2020diskfm, hom2024uniform}. These approaches are mainly applicable to simple disk structures, such as ellipses, which are typical morphologies of debris disks. Another technique, Reference Differential Imaging (RDI; see \cite{smith1984circumstellar,lafreniere2009hst,lagrange2010giant} for some first examples of applications), employs additional images of one or more reference stars without known exoplanets or disks. These additional data can be captured simultaneously with the observation of the target star using the star-hopping technique \citep{wahhaj2021search}, or they can be drawn from a large library of archival observations \citep{ren2018non, xuan2018characterizing}. RDI can be effectively combined with other observing strategies to simultaneously exploit their diversity. For instance, when integrating RDI with ADI, the nuisance component can be estimated and suppressed using PCA \citep{ruane2019reference, xie2022reference, juillard2024combining} or deep learning techniques \citep{chintarungruangchai2023possible, wolf2024direct, bodrito2024modelco}. RDI reconstructions can also be constrained by additional observations from other imaging modalities, such as polarimetry, where speckles and disk components behave differently as in total intensity images recorded with ADI/ASDI \citep{lawson2022constrained}. In practice, the effectiveness of RDI approaches depends heavily on the similarity between the reference and the actual observations, including factors such as star brightness, spectrum, and observation conditions. This degree of similarity becomes increasingly critical as we search for fainter objects.
Finally, a last category of approaches jointly addresses the problem of estimating star light residuals and reconstructing the flux distribution of the disk and exoplanets. In ADI, three approaches based on an inverse-problems formulation were recently proposed: MAYONNAISE \citep{pairet2020mayonnaise}, MUSTARD \citep{juillard2022analysis, juillard2023inverse}, and REXPACO \citep{flasseur2021rexpaco,flasseur2022multispectral}. These algorithms employ different strategies and regularization penalties of the inversion for separating the components of interest. 
In a first step, MAYONNAISE uses iterative PCA to initialize the inversion process. Building on this preliminary reconstruction, a second step involves estimating and unmixing multiple components by jointly minimizing a data fidelity term. The unmixed components are the star light residuals (restricted to lie within the subspace identified in the first step), the disk (enforced to have a sparse representation in a shearlet basis), and the exoplanets (restricted to be sparse). Non-negativity constraints are also enforced
during the minimization. MUSTARD is a variant of MAYONNAISE that primarily differs in the formulation of the direct model. The reconstructed speckles field is enforced to be identical along the temporal axis to account explicitly for its quasi-static behavior. Unlike MAYONNAISE, MUSTARD does not use iterative PCA for initialization, nor does it enforce sparsity of the disk component in a shearlet basis. Additionally, MUSTARD can incorporate a regularization term based on a predefined mask, which helps resolve ambiguities between the speckle field and portions of the disk that are rotation invariant. Both MAYONNAISE and MUSTARD assume noise to be white, independent, and identically distributed. REXPACO follows quite a different modeling as it does not explicitly estimate the residual star light in each image. Instead, it builds a statistical and local model of all fluctuations other than the component of interest (i.e., noise and star light). REXPACO learns the spatial correlations of these fluctuations at the scale of 2D image patches, following an approach initially introduced for exoplanet detection in the PACO algorithm \citep{flasseur2018exoplanet}, based on PAtch COvariances. The component of interest is deconvolved with an edge-preserving smoothness regularization and a positivity constraint. Further extending REXPACO for ADI post-processing, a recent enhancement replaces its multivariate Gaussian model of the nuisance with a scaled mixture of multivariate Gaussian models \citep{flasseur2022multispectral}. This improved model offers better fidelity to the observations and enhanced  robustness against outlier data (e.g., defective pixels or large stellar leakages), which are identified and neutralized in a data-driven manner. In ADI, REXPACO can be combined with PACO to disentangle user-identified candidate point-like sources from the circumstellar environment.

In this paper, we address the problem of reconstructing circumstellar disks from ASDI sequences through joint multi-spectral post-processing. Compared to ADI, this raises several challenges: (i) modeling the temporal and spectral fluctuations of the residual star light, (ii) jointly exploiting both temporal and spectral information to effectively extract the component of interest, and (iii) ensuring the tractability of estimating high-dimensional models from large datasets. As an illustration of point (iii), typical ASDI datasets produced by the Integral Field Spectrograph (IFS) of the Spectro-Polarimetry High-contrast
Exoplanet Research instrument (SPHERE; \cite{beuzit2019sphere}) at the Very
Large Telescope (VLT) are $N=290\times 290$ pixels, have $L=39$
spectral bands and $T\approx 100$ individual exposures. Several hundred million pixel
measurements must then be combined to produce a multi-spectral reconstruction
of the component of interest. Modeling the full covariance associated with this volume of measurements theoretically involves estimating $N(N+1)/2$ degrees of freedom from the data, which is not feasible without making approximations to the covariance.

\medskip

\begin{table} \centering \caption{Summary of the main notations.}
	\begin{tabular}{@{}c@{}c@{\hspace*{1ex}}l@{}} \toprule \textbf{Not.}\; &
				\textbf{Range}\; &\textbf{Definition} \\ \midrule
				\multicolumn{3}{c}{$\triangleright$ Constants and related indexes}\\
				\midrule
				$K$ & $\mathbb{N}^*$ & number of pixels in a patch\\
				$N$ & $\mathbb{N}^*$ & number of pixels in a dataset\\
				$N'$ & $\mathbb{N}^*$ & number of pixels in a reconstructed image\\
				$T$ & $\mathbb{N}^*$ & number of temporal frames\\
				$L$ & $\mathbb{N}^*$ & number of spectral channels\\
				$L_\text{eff}$ & $\mathbb{N}^*$ & effective  number of spectral channels\\
				$n^{(')}$ & $\llbracket 1,N^{(')} \rrbracket$ & pixel index\\
				$t$ & $\llbracket 1,T \rrbracket$ & temporal index\\
				$\ell$ & $\llbracket 1,L \rrbracket$ & spectral index\\
				$\mathbb{K}$ & -- & set of patch locations\\
				\midrule
				\multicolumn{3}{c}{$\triangleright$ Data quantities}\\
				\midrule
				$\V v$ & $\mathbb{R}^{NTL}$ & ASDI sequence (with speckles aligned)\\
				$\V f$ & $\mathbb{R}^{NTL}$ & nuisance component\\
				$\M E_{n(,t)(,\ell)}$ & $\mathbb{R}^{NTL \times K (L) (T)}$ & patch extractor at pixel $n$ (, time $t$) (, channel $\ell$)\vspace{0.5mm}\\
				$\M V_{n,t}$ & $\mathbb{R}^{K\times L}$ & residual multi-spectral patch at pixel $n$, time $t$\\
				$\obj$ & $\mathbb{R}_+^{N'L}$ & spatio-spectral flux distribution\\
				\midrule
				\multicolumn{3}{c}{$\triangleright$ Operators}\\
				\midrule
				$\M M$ & $\mathbb{R}^{N'L\times NTL}$ & direct image formation model: $\M M =\M S \, \M Z \, \M A \, \M B \, \M R$\\
				$\M F_t$ & $\mathbb{R}^{N'L\times NL}$ & sparse operator at time $t$: $\M F_t = (\M S \, \M Z \, \M A \, \M R)_t$\\
				$\M B$ & $\mathbb{R}^{N'TL\times N'L}$ & convolution by off-axis PSF\\
				$\M R$ & $\mathbb{R}^{N'TL\times N'TL}$ & apparent field rotation\\
				$\M A$ & $\mathbb{R}^{N'TL\times N'TL}$ & coronagraph attenuation\\
				$\M Z$ & $\mathbb{R}^{N'TL\times MTL}$ & field of view cropping\\
				$\M S$ & $\mathbb{R}^{MTL\times NTL}$ & spectral scaling\\
				$\odot$ & $\mathbb{R}^{X \times X}\,, X \in \mathbb{N}^*$ & Hadamard (element-wise) product\\
				$\otimes$ & $\mathbb{R}^{X \times X}\,, X \in \mathbb{N}^*$ & Kronecker product\\
				\midrule
				\multicolumn{3}{c}{$\triangleright$ Estimated quantities}\\
				\midrule
				$\widehat{\V \mu}^{\,\spec}$ & $\mathbb{R}^{NL}$ & multi-spectral mean of $\V f$\\
				$\widetilde{\V \mu}^{\,\spec}$ & $\mathbb{R}^{NL}$ & shrunk multi-spectral mean of $\V f$\\
				$\widehat{\M C}^\spat$ & $\mathbb{R}^{K\times K}$ & local empirical spatial covariance of $\V f$\\
				$\widehat{\M C}^\spec$ & $\mathbb{R}^{L\times L}$ & local empirical spectral covariance of $\V f$\\
				$\widetilde{\M C}^\spat$ & $\mathbb{R}^{K\times K}$ & local shrunk spatial covariance of $\V f$\\
				$\widetilde{\M C}^\spec$ & $\mathbb{R}^{L\times L}$ & local shrunk spectral covariance of $\V f$\\
				$\widetilde{\rho}^{\,\spat}$ & $\left[ 0, 1 \right]$ & spatial shrinkage coefficient\\
				$\widetilde{\rho}^{\,\spec}$ & $\left[ 0, 1 \right]$ & spectral shrinkage coefficient\\
				$\widehat{\V \sigma}$ & $\mathbb{R}_+^T$ & temporal weights ($\widehat{\V \sigma} = \lbrace \widehat{\sigma}_t \rbrace_{t=1:T}$)\\
				$\widetilde{\V \sigma}$ & $\mathbb{R}_+^T$ & shrunk temporal weights ($\widetilde{\V \sigma} = \lbrace \widetilde{\sigma}_t \rbrace_{t=1:T}$)\\
				${\M \Psi}^\spat$ & $\mathbb{R}^{K\times K}$ & matrix of spatial shrinkage coefficients\\
				${\M \Psi}^\spec$ & $\mathbb{R}^{L\times L}$ & matrix of spectral shrinkage coefficients\\
				$\M \Gamma$ & $\mathbb{R}^{KL \times KL}$ & shrunk spatio-spectral precision matrix\\
				$\widehat{\obj}$, $\widetilde{\obj}$ & $\mathbb{R}_+^{N'L}$ & reconstructed spatio-spectral flux distribution\\
				$\widehat{\V \beta}$ & $\mathbb{R}_+^{2}$ & regularization hyper-parameters\\
				\midrule
				\multicolumn{3}{c}{$\triangleright$ Other quantities and metrics}\\
				\midrule
				$\obj_{\text{inv}}$ & $\mathbb{R}^{N'}$ & flux distribution invariant by ASDI\\
				$\obj_{\text{gt}}$ & $\mathbb{R}_+^{N'L}$ & ground truth flux distribution\\
				$\alpha_{\text{gt}}$ & $\mathbb{R}_+$ & maximum ground truth contrast (disk \textit{vs} star)\\
				MSE & $\mathbb{R}$ & mean square error\\
				N-RMSE & $\mathbb{R}_+$ & normalized root mean square error\\	
				SURE & $\mathbb{R}$ & Stein's unbiased risk estimator\\
				\bottomrule
	\end{tabular}
	\label{tab:notation_reminder}
\end{table}

\noindent \emph{Our contributions:} This paper extends the REXPACO algorithm
\citep{flasseur2021rexpaco,flasseur2022multispectral} to ASDI sequences.
This extension, named REXPACO ASDI, involves several specific methodological developments, including:
\begin{itemize}
    \item a spatio-spectral separable model of the covariances of the nuisance,
    \item a spatio-temporal weighting of the measurements based on their relative quality,
    \item a technique to estimate the components of the covariances and weights model,
    \item a regularization strategy of the (noisy) sample covariances,
    \item a strategy to jointly refine the model of the residual star light
    and reconstruct the disk of interest,
    \item a spatio-spectral regularization of the reconstructed multi-spectral images,
    \item a strategy to unmix point-like sources from the disk material.
\end{itemize}
Beyond these methodological developments, the proposed approach is, to the best of our knowledge, the first one to leverage joint processing of multi-spectral data through an inverse problem framework for reconstructing circumstellar disks in high-contrast imaging. We illustrate in this paper the benefits of an accurate exploitation of the spectral diversity to improve reconstruction fidelity. In particular, we show that REXPACO ASDI can faithfully reconstruct disks with near-circulo-symmetric morphologies (e.g., spiral and rings). Such morphological structures are especially challenging to reconstruct without additional data diversity  complementary to A(S)DI (e.g., based on RDI techniques) to build an unbiased model of the nuisance component.

Section \ref{sec:model} develops the statistical model for the residual star
light and different noise contributions. Building on this model, Sect. \ref{sec:recons} presents a reconstruction method that jointly extracts and deconvolves the component of interest: the multi-spectral image of the disk  surrounding the target star. Section \ref{sec:results} showcases reconstruction results on several ASDI sequences obtained with the VLT/SPHERE instrument. Additionally, Sect. \ref{sec:unmixing_ps_disk} describes an iterative method to unmix the contribution of candidate point-like sources from the circumstellar disk. Finally, Sect. \ref{sec:conclusion} draws the conclusions of this work.

\medskip

\noindent Throughout the text, the reader can refer to Table \ref{tab:notation_reminder} summarizing the main notations.

\section{Statistical model of the nuisance}
\label{sec:model}

\begin{figure*}
	\centering
	\includegraphics[width=\textwidth]{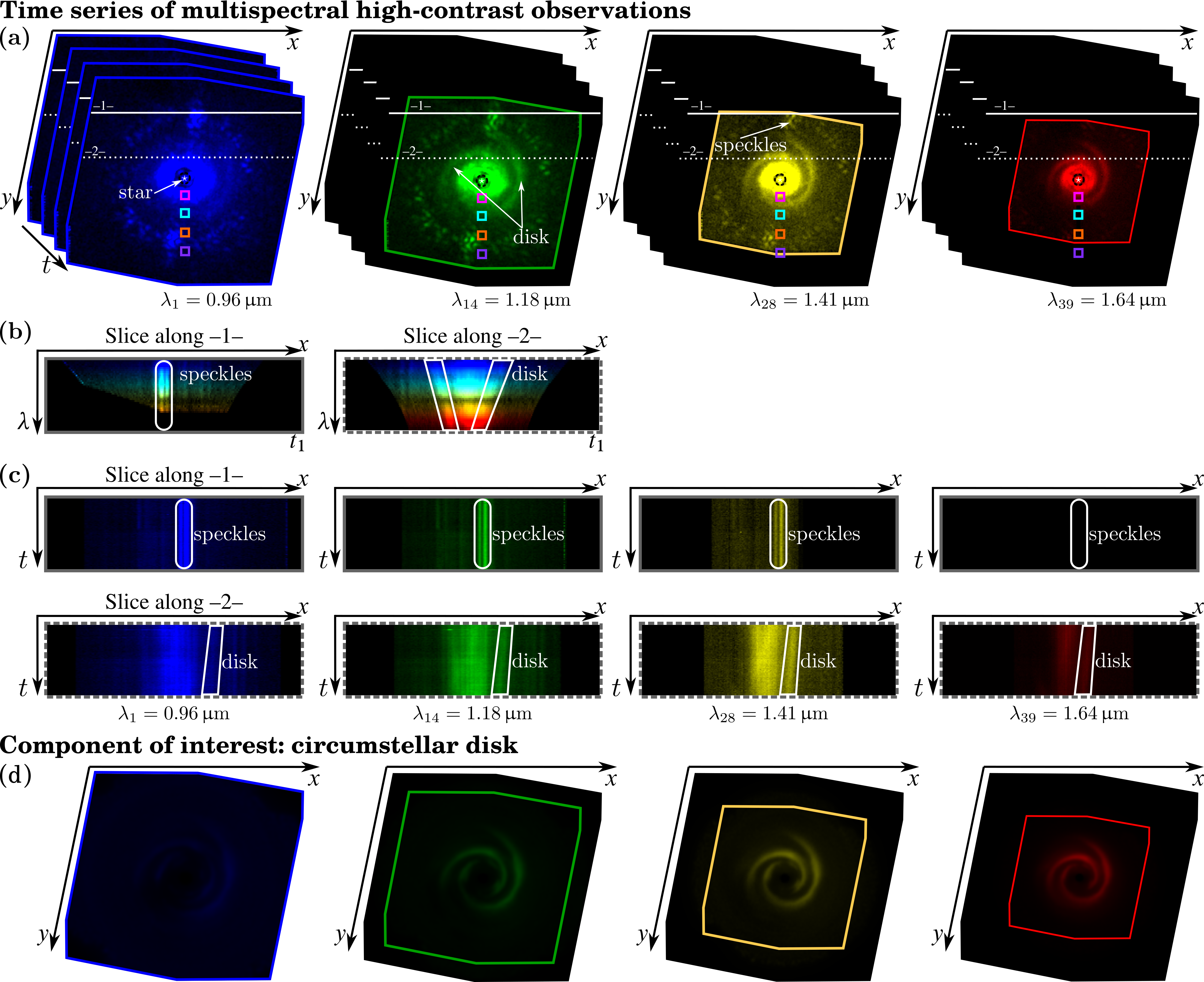}
	\caption{Illustration of a dataset acquired with ASDI: (a) images captured at different wavelengths; (b) spatio-spectral slices along the two lines --1-- and --2-- drawn in (a); (c) spatio-temporal slices along the lines --1-- and --2--. The four square areas define four regions studied in more details in Fig. \ref{fig:covmat}. The component of interest, a spiral-shaped circumstellar disk, is shown in (d) based on REXPACO ASDI reconstruction given in Sect. \ref{subsec:recons_real_disks}. In the first channel, shown in blue, the signal of the disk is faint (contrast about $1.5\times 10^{-6}$) compared to the stellar leakages. Images are displayed using pseudo-colors (ranging from blue to red) chosen to cover the infrared spectrum. Colored polygons delimit the common field of view seen in all spectral channels. Dataset: SAO 206462 (2015-05-15), see Table \ref{tab:dataset_logs} for the observation parameters.}
	\label{fig:data}
\end{figure*}

In contrast to other methods in the literature, we do not
explicitly extract the residual star light component from the data but rather
develop a statistical model to describe both 
the residual star light
(i.e., the speckles) and the various stochastic noise contributions (thermal noise, detector readout noise, photon noise). With pupil tracking mode and after chromatic speckle alignment by rescaling the images according to the
wavelength, residual star light is very similar from one spectral channel to the next
(up to some chromatic factor).
There are, however, some fluctuations due to noise, chromatic phenomena, and
the evolution of the phase aberrations during observation.
These fluctuations
display some spatial and spectral correlations and are highly non-stationary. In particular, they are much stronger close to the star.
We describe in Sects. \ref{subsec:patch_based_statistical_modeling} and \ref{sec:covstruct}
 the rationale of the statistical model embedded in REXPACO ASDI, and we develop in Sect. \ref{sec:nuisancemodelestim}
 a methodology to estimate the resulting large number of parameters directly from the data.

Figure \ref{fig:data} shows a dataset of a star (SAO 206462) surrounded
by a bright disk observed using the ASDI technique. Slices along different dimensions of this 4D dataset are
displayed. The coronagraphic mask is aligned with the star, at the center of
the field of view (center of the images shown in Fig. \ref{fig:data}(a)).
Residual star light dominates the central area and extends over most of the
field of view. It takes the form of granular intensity structures (speckles). During a
pre-processing step, all images were rescaled by a wavelength-specific factor
$\lambda_{\text{ref}}/\lambda$ to compensate for diffraction and spatially
align the speckles. 
The solid line --1-- drawn in the $x$ direction in Fig. \ref{fig:data}(a) crosses a bright speckle. This speckle is visible in the
left part of Fig. \ref{fig:data}(b) and the first row of Fig. \ref{fig:data}(c).
It remains at the same spatial location for all wavelengths $\lambda$ and all
 times $t$. Structures of interest, such as the disk that surrounds the
star SAO 206462, undergo a rotation about the image center throughout time and
a scaling with the wavelength (due to the rescaling applied in the
pre-processing step). These spatial transformations are visible in the slices
along the dotted line --2-- drawn in the images of Fig. \ref{fig:data}(a): the
line crosses the disk (as well as an area with strong residual star light,
close to the image center). The spatio-spectral slice shown at the right of Fig.
\ref{fig:data}(b) displays a scaling of the disk with respect to the
wavelength (shorter wavelengths are dilated due to the speckle-aligning
pre-processing), whereas the rotation motion can be noted in the
spatio-temporal slices shown at the bottom of Fig. \ref{fig:data}(c), in
particular for a bright structure of the disk highlighted within a white box, which is moving
closer to the image center during the sequence. Figure \ref{fig:data}(d) shows only the component of interest: the circumstellar disk. The images were obtained with the reconstruction method introduced in this paper, see Sect. \ref{subsec:recons_real_disks} for a spectrally combined visualization of the reconstructed disk. Comparing Figs. \ref{fig:data}(a) and \ref{fig:data}(d), illustrates that high-contrast observations suffer from a strong nuisance component which has to be numerically suppressed in order to reconstruct the component of interest.

The accuracy of residual star light and noise model has a strong impact on the
reconstruction of the component of interest $\obj$, as further discussed in
 Sect. \ref{sec:results}.
In the following of this section, we first assume that the object $\obj$ has a
negligible impact on
the statistical distribution of the nuisance term, i.e., the statistical
distribution of the aligned data $\text{p}_V(\data)$ in the absence of disk or
exoplanet
is nearly identical to the distribution $\text{p}_V(\data-\M M\,\obj)$ of the nuisance component $\V f = \V v - \M M\,\obj$
obtained when the modeled contribution $\M M\,\obj$ of the component of interest
has
been subtracted from the data $\V v$ (the direct model, $\M M$, is presented in
 Sect. \ref{sec:directmodel}). This
assumption is made in order to initiate the estimation of the model parameters, 
 and we introduce in
Sects. \ref{sec:nuisancemodelestim}--\ref{sec:joint-estimation} several strategies to jointly estimate the statistical
distribution of the nuisance terms and reconstruct the component of interest.
These joint and iterative strategies significantly enhance the fidelity of the reconstruction by explicitly accounting for the bias induced by the disk on the nuisance model. 

\subsection{Patch-based statistical modeling}
\label{subsec:patch_based_statistical_modeling}

Image patches (i.e., neighborhoods of a few tens to a hundred pixels) offer an
interesting trade-off between locality (small enough to capture a local
behavior) and complexity (they include enough pixels to collect geometrical
and textural information). Their use has been very successful in image
restoration, from methods based on image self-similarity \citep{buades2005non},
collaborative filtering \citep{dabov2007image}, sparse coding
\citep{aharon2006k,mairal2009non}, mixture models
\citep{zoran2011learning,yu2011solving}, or Gaussian models
\citep{lebrun2013nonlocal}. Whereas deep neural networks have become the
state-of-the-art approach to learn rich models (either
generative or discriminative) of natural images, patch-based models retain serious advantages
when the number of training samples is limited or in the case of highly
non-stationary images.
As can be seen in Fig. \ref{fig:data}(a), images in an
ASDI dataset are far from stationary: residual star light is the strongest at
the center of the image (at the actual location of the star).
Observations
made during separate nights around different stars also often display
significantly
different structures because of changes in the observing conditions (which
impact the residual aberrations uncorrected by adaptive optics, and hence
the spatial distribution of speckles due to star light) and star
brightness. This limits the possibility to use external observations (e.g., using the RDI technique, see Sect. \ref{sec:introduction}) to learn
a model to process a specific ASDI sequence and motivates the development of a
patch-based approach based solely on the ASDI sequence of interest.

Under our patch-based model, the distribution of an ASDI sequence
$\data\in\mathbb{R}^{NLT}$, formed by the collection of $T$ multi-spectral images with $L$ spectral bands
and $N$ pixels in each band, after chromatic speckles alignment and \emph{without disk or exoplanet}
 is given by:
\begin{align}
    \text{p}_V(\data)\approx\prod_{n\in\mathbb{K}}\text{p}_{V_n}(\M
    E_n\data)\,,
    \label{eq:decomp_patches}
\end{align}
where $\text{p}_V$ is the joint distribution of the whole ASDI dataset, $\M
E_n$ is the linear operator that extracts a $K\times L_\text{eff} \times T$-pixel
spatio-spectro-temporal patch centered at the $n$-th spatial location of the
field of view (i.e., $\V v_n=\M E_n \data$ is a 4D-patch\footnote{Throughout the text, we do not
differentiate the
$x$ and $y$ spatial dimensions to simplify the notations but rather use 2D
spatial indices $n$.}). The set of spatial locations $\mathbb{K}$ is defined to
prevent patch overlapping while tiling the whole field of view (i.e.,
$\text{Card}(\mathbb{K})\times K=N$ and juxtaposed square patches
are used).

The model (\ref{eq:decomp_patches}) assumes a statistical independence between patches,
which is a simplifying hypothesis that eases a data-driven learning of the
distribution $\text{p}_V$. In the sequel, each distribution $\text{p}_{V_n}$
is modeled by a different multivariate Gaussian in order to capture the
correlations between observations within a spatio-spectro-temporal patch.
By adapting the parameters of these Gaussian distributions to the spatial
location $n$, a non-stationary model is obtained, with the capability to
capture the variations between areas close to the star (at the center of the
image) and areas farther away. The statistical model of a patch is thus given
by its assumed distribution:
\begin{align}
	\label{eq:gauss_patch}
        \text{p}_{V_n}(\data_n)=\frac{1}{\sqrt{|2\pi\M
            C_n|}}\exp\left(-\tfrac{1}{2}\bigl\|\data_n-\V \mu_n\bigr\|_{\M
    C_n^{-1}}^2\right)\,,
\end{align}
with $\|\V a\|_{\M B}^2=\V a\T \, \M B \, \V a$ and $|\M
C_n|$ the determinant of matrix $\M C_n$. The Gaussian distribution
$\text{p}_{V_n}$ is defined by the patch expectation $\V \mu_n\in\mathbb{R}^{K
L T}$ and the covariance matrix $\M
C_n\in\mathbb{R}^{K L T\times K L T}$. In order to estimate these two
quantities at each location $n$, additional hypotheses and an estimation
technique are required.

\subsection{Constraining the structure of the average vector and of the covariance matrix}
\label{sec:covstruct}

Estimating and handling different Gaussian parameters for each patch location
is not feasible given the number of parameters involved: the set of all mean
vectors $\{\V \mu_n\}_{n\in\mathbb{K}}$ has as many free parameters as the
total number of measurements in $\data$ (i.e., $NLT$) and the set
of all
covariance matrices $\{\M C_n\}_{n\in\mathbb{K}}$ represents many times the
number of measurements in $\data$ (more precisely, $NLT(KLT+1)/2$
free parameters,
which represents
more than 300,000 times the size of $\data$ for typical values of $K\approx
13\times 13$, $L\approx 39$,
and $T\approx 100$).

There are two options to reduce the number of parameters in the Gaussian
models of Eqs. (\ref{eq:decomp_patches}) and (\ref{eq:gauss_patch}). Approach (i) involves assuming a certain level of stationarity for the means or covariances with respect to
the spatial location $n$. Strategy (ii) is to impose a structure on the mean $\V \mu_n$ and on
the covariance $\M C_n$. 
Beyond obtaining more tractable models, these assumptions are also
indispensable, for a single ASDI dataset $\data$, to constrain the estimator of
the parameters of the Gaussian models.

The strong spatial non-stationarity of ASDI datasets led us to favor option
(ii). We considered several ways to select a structure suitable to ASDI
observations and built on our experience of
point-source detection in ASDI datasets
\citep{flasseur2020paco}. We
found that it is preferable to use a common mean vector $\V \mu_n$ for all
times $t$ rather than a time-specific mean vector common to all wavelengths
(the spectral variations being stronger than the temporal fluctuations):
\begin{equation}
	\text{Mean}\left[ \V v_{n, (k, \ell, :)} \right] = \V \mu_{n, (k, \ell, :)} = \frac{1}{T}\sum\limits_{t'=1}^T \V v_{n, (k, \ell, t')} = \V \mu_{n, (k,\ell)}^\spec\,,
	\label{eq:mumodel_indexes}
\end{equation}
where $\V \mu_n$ represents the mean vector at patch location $n$, and $\V \mu_{n,(k,\ell,t)}$ denotes its specific entry at pixel $k$, spectral channel $\ell$ and time $t$. Equation (\ref{eq:mumodel_indexes}) can be rewritten in the more concise form:
\begin{align}
    \V \mu_n = \text{vec}\!\left(
    \begin{pmatrix}
        |\\
    \V \mu_{n}^{\text{spec }}\\ |
\end{pmatrix}
\overset{\text{\scriptsize$\longleftarrow T\longrightarrow$}}{
\begin{pmatrix}
    1&\cdots&1
\end{pmatrix}}
\right)\,,
\label{eq:mumodel}
\end{align}
where $\V \mu_{n}^\spec$ is a $KL$-pixel
multi-spectral vector that represents the temporal average of the
multi-spectral patches and $\text{vec}(\cdot)$ performs the vectorization of a matrix by stacking
its columns (it transforms a $KL\times T$ matrix into a vector of
dimension $KLT$).

To capture the structures of both the spatial and the spectral covariances, we
model the covariance between two pixels of the patch $\data_n$ by:
\begin{multline}
	\label{eq:covmodel}
	\text{Cov}\!\left[\data_{n,(k_1,\ell_1,t_1)},\,\data_{n,(k_2,\ell_2,t_2)}\right] \\=
  \begin{cases}
   0        & \text{if } t_1\neq t_2\,, \\
   \sigma_{n,t}^2\M C_{n,\,(k_1,k_2)}^\spat\M
    C_{n,\,(\ell_1,\ell_2)}^\spec & \text{if } t_1= t_2=t\,,
  \end{cases}
\end{multline}
where $\sigma_{n,t}^2$ is a scalar that represents the
global level of fluctuation in the multi-spectral slice at time $t$,
$\M C_n^\spat$ is a $K\times K$ covariance matrix encoding the spatial
structure of the fluctuations (a $K$-pixel spatial patch corresponds to a 2D
square window, so this covariance matrix contains information about 2D spatial
structures), and matrix $\M C_n^\spec$ is an $L\times L$ covariance
matrix encoding spectral correlations. To prevent a degeneracy by
multiplicative factors, we
normalize covariance matrices $\M C_n^\spat$ and $\M
C_n^\spec$ such that their trace be equal to $K$ and $L$,
respectively. In the covariance model of Eq. (\ref{eq:covmodel}),
multi-spectral slices at different times $t_1$ and $t_2$ are considered
uncorrelated (and, thus, mutually independent given the joint Gaussian assumption of
Eq. (\ref{eq:gauss_patch})). The time-varying variance parameter
$\sigma_{n,t}^2$ plays the role of a scale parameter in a compound-Gaussian
model \citep{381910}, also known as a Gaussian scale mixture model \citep{wainwright1999scale}.
A large value of parameter $\sigma_{n,t}^2$
almost discards the time frame $t$ from the $n$-th 4D
patch, which limits the impact of possible outliers and thus makes the estimator (more) robust
\citep{flasseur2020robustness}.

The covariance structure given in Eq. (\ref{eq:covmodel}) corresponds to
the following separable covariance matrix:
\begin{align}
    \text{Cov}\!\left[\V v_n\right]=\text{diag}(\V\sigma_n^2)\otimes \M
    C_n^\spec\otimes  \M C_n^\spat\,,
    \label{eq:covsep}
\end{align}
where $\text{diag}(\V\sigma_n^2)$ is a $T\times T$ diagonal matrix whose
$t$-th diagonal entry is $\sigma_{n,t}^2$ and $\otimes$ is Kronecker matrix
product: $\M A\otimes\M B$, with $\M A\in\mathbb{R}^{n\times n}$ and $\M
B\in\mathbb{R}^{m\times m}$, is the $nm\times nm$ matrix with a $n\times n$
block structure such that the $ij$-th block is the $m\times m$ matrix
$A_{ij}\M B$. Note that this is equivalent to modeling each multi-spectral
slice $\V v_{n,t}\in\mathbb{R}^{KL}$ of $\V v_n$ as random vectors following
the compound-Gaussian model
$\mathcal{N}(\V\mu_{n}^\spec,\sigma_{n,t}^2\M
C_n^\spec\otimes  \M C_n^\spat)$, i.e. the scaled and
centered vectors $\frac{1}{\sigma_{n,t}}(\V
v_{n,t}-\V\mu_{n}^\spec)$ are independent and identically
distributed for all $1\leq t\leq T$ according to the centered Gaussian
$\mathcal{N}(\V 0,\M
C_n^\spec\otimes  \M C_n^\spat)$.

With the structure of the mean vector $\V\mu_n$ given in Eqs.
(\ref{eq:mumodel_indexes}) and (\ref{eq:mumodel}), corresponding to a multi-spectral patch constant through
time, there are only $NL$ free parameters to estimate all mean
vectors from $\data\in\mathbb{R}^{NLT}$. The covariance structure
defined in Eqs. (\ref{eq:covmodel}) and (\ref{eq:covsep}) leads to
$T+K(K+1)/2+L(L+1)/2-2$ free
parameters per 4D patch (the -2 comes from the two normalization constraints),
which leads to approximately $NK/2$ free
parameters for the whole ASDI dataset (because $K\gg L$ and $K^2/2 \gg T$), and is typically
one to two orders of
magnitudes smaller than the total number of measurements in $\data$
($K/2$ is typically less than one hundred whereas $LT$ is several thousands).
Jointly
with an adequate estimation method, the structures assumed in Eqs.
(\ref{eq:mumodel_indexes}), (\ref{eq:mumodel}), (\ref{eq:covmodel}) and (\ref{eq:covsep}) can thus be used
to derive a \emph{non-stationary} model of the nuisance terms.

\subsection{Estimation of the model parameters} 
\label{sec:nuisancemodelestim}

The estimation of the parameters of a separable covariance model has been
studied
by several previous works from the signal-processing community, see for example
\cite{lu2005likelihood,genton2007separable,werner2008estimation}. We build on
these works and
introduce
several additional elements specific to high-contrast imaging: (i) whereas most works consider decompositions of
the covariance matrix as a Kronecker product of two factors, we also include
in Eqs. (\ref{eq:covmodel}) and (\ref{eq:covsep})
 the temporal scaling factors $\sigma_{n,t}$ for increased robustness
\citep{flasseur2020robustness, flasseur2022multispectral}; (ii)
given the limited number of samples, we replace maximum likelihood estimates
by shrinkage covariance estimators
\citep{ledoit2004well,chen2010shrinkage,flasseur2024shrinkage} to ensure that all estimated
covariance matrices are definite positive and to reduce estimation errors;
 (iii) to account for the superimposition of a component of interest and
 nuisance terms, we develop a joint estimation strategy in
 Sect. \ref{sec:joint-estimation} based on the estimation technique developed in this
 section.

\medskip

\subsubsection{Maximum likelihood estimators}
\label{subsubsec:mles}

A first possiblity is to determine the parameters of the model of the nuisance statistics
so as to maximize the likelihood of the data knowing the object of interest $\obj$.
According to the considered problem and to the assumed independence of the patches, this
amounts to minimizing the following co-log-likelihood:
\begin{align}
	& \mathscr{L}\Paren*{\Brace*{\V \mu_n^\spec, \Brace[\big]{\sigma_{n,t}^2}_{t \in 1:T}, \M C_n^\spec, \M C_n^\spat}_{n \in \mathbb{K}}, \obj} \notag \\
    & \hspace*{7mm} = \sum_{n \in \mathbb{K}}
      \mathscr{L}_n\!\left(\V \mu_n^\spec, \Brace[\big]{\sigma_{n,t}^2}_{t \in 1:T}, \M C_n^\spec, \M C_n^\spat, \obj \right),
      \label{eq:totalcologlikelihood}
\end{align}
where $\mathscr{L}_n $ is the co-log-likelihood of the patch at location $n$:
\begin{align}
  &\mathscr{L}_n\!\left(\V \mu_n^\spec, \big\lbrace\sigma_{n,t}^2\big\rbrace_{t \in 1:T},
    \M C_n^\spec, \M C_n^\spat, \obj \right) = \notag\\
  &\hspace*{5mm}\frac{1}{2} \sum_{t = 1}^{T} \left(
    \left\Vert \V v_{n,t} - \V\mu_{n}^\spec - [\M M\,\obj]_{n,t}\right\Vert_{\M C_{n,t}^{-1}}^{2}
    + \log\Abs*{\M C_{n,t}}
    \right),
    \label{eq:patchcologlikelihood}
\end{align}
with $\M C_{n,t} = \sigma_{n,t}^{2}\,\M C^\spec_{n} \otimes \M C^\spat_{n}$ the assumed
covariance of the patch data $\V v_{n,t}$ and $\Abs*{\M C_{n,t}}$ its determinant.
The term $\M M\,\obj$ accounts for the contribution of the object of
  interest in the data, the linear model matrix $\M M$ is detailed in
  Sect.~\ref{sec:directmodel}. The maximum likelihood estimators (MLEs) of the parameters
of the nuisance statistic are then given by:
\begin{align}
  \widehat{\V\mu}_n^{\,\spec}
  & = \argmin_{\V\mu_n^\spec} \mathscr{L}_n\!\Paren*{\V \mu_n^\spec,
    \Brace[\big]{\widehat{\sigma}_{n,t}^2}_{t \in 1:T}, \widehat{\M C}_n^\spec,
    \widehat{\M C}_n^\spat, \widehat{\obj}}
  \notag\\
  &= \frac{\sum_{t=1}^{T} \widehat{\sigma}_{n,t}^{-2} \,
    \Paren*{\V v_{n,t} - [\M M\,\widehat{\obj}]_{n,t}}}%
    {\sum_{t=1}^T \widehat{\sigma}_{n,t}^{-2}}\,,
    \label{eq:MLE_mu}
  \\
  \widehat{\sigma}_{n,t}^2
  &= \argmin_{\sigma_{n,t}^2} \mathscr{L}_n\Paren*{\widehat{\V \mu}_n^\spec,
    \big\lbrace\sigma_{n,t'}^2\big\rbrace_{t' \in 1:T}, \widehat{\M C}_n^\spec,
    \widehat{\M C}_n^\spat, \widehat{\obj}}
  \notag\\
   &= \frac{1}{K\,L}\left\|\V v_{n,t} - \widehat{\V\mu}_{n}^\spec - [\M M\,\widehat{\obj}]_{n,t}\right\|_{
    \big( \widehat{\M C}_n^\spec\big)^{-1} \otimes \big(\widehat{\M C}_n^\spat\big)^{-1}}^2\,,
    \label{eq:MLE_sigma}
  \\
  \widehat{\M C}_n^\spec
  &= \argmin_{\M C_n^\spec} \mathscr{L}_n\!\left(\widehat{\V \mu}_n^\spec,
    \Brace[\big]{\widehat{\sigma}_{n,t}^2}_{t \in 1:T}, \M C_n^\spec,
    \widehat{\M C}_n^\spat, \widehat{\obj}\right)
  \notag\\
  &= \frac{1}{T\,K}\sum_{t=1}^T \widehat{\M V}_{n,t}\T
    \left( \widehat{\sigma}_{n,t}^{2} \, \widehat{\M C}_n^\spat \right)^{-1} \, \widehat{\M V}_{n,t}\,,
    \label{eq:MLE_Cspec}
  \\
  \widehat{\M C}_n^\spat
  &= \argmin_{\M C_n^\spat} \mathscr{L}_n\!\left(\widehat{\V \mu}_n^\spec,
    \Brace[\big]{\widehat{\sigma}_{n,t}^2}_{t \in 1:T}, \widehat{\M C}_n^\spec,
    \M C_n^\spat, \widehat{\obj}\right)
    \notag\\
  &= \frac{1}{T\,L}\sum_{t=1}^T \widehat{\M V}_{n,t}
    \left(\widehat{\sigma}_{n,t}^{2} \, \widehat{\M C}_n^\spec\right)^{-1} \, \widehat{\M V}_{n,t}\T\,,
    \label{eq:MLE_Cspat}
\end{align}
with $\widehat{\obj}$ the estimator of the object of interest and where
$\widehat{\M V}_{n,t}$ is a $K\times L$ matrix corresponding to the residual
multi-spectral patch at pixel $n$ and time $t$: at row $k$ and column $\ell$ it is equal
to $\big[\V v_{n,t} - \widehat{\V\mu}_{n}^\spec - [\M M\,\widehat{\obj}]_{n,t}\big]_{k,\ell}$.
The complete derivation of these expressions is given in Appendix \ref{sec:app_proofMLE}.
These equations are generally interdependent which has an incidence on the optimization
strategy, see Sect. \ref{sec:joint-estimation}.

\begin{figure}
	\centering
	\includegraphics[width=0.47\textwidth]{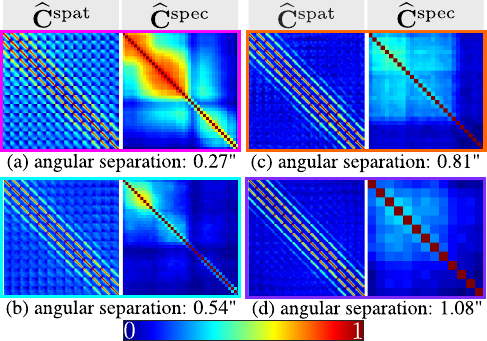}
	\caption{MLEs of the spatial and spectral correlation matrices given by Eqs.~\eqref{eq:MLE_Cspec}--\eqref{eq:MLE_Cspat} computed in the four regions of interest, indicated by small colored squares in Fig. \ref{fig:data}(a), with each matrix corresponding to its respective color-coded region. The angular separation with respect to the star (i.e., image center) increases from the region in (a), close to the star, to the region in (d), which is farther away. Dataset: SAO 206462 (2015-05-15), see Table \ref{tab:dataset_logs} for the observation parameters.}
	\label{fig:covmat}
\end{figure}

The multi-spectral mean $\widehat{\V\mu}_n^{\,\spec}$ in Eq.~\eqref{eq:MLE_mu} is obtained by
weighted averaging, with weights inversely proportional to the patch-wise variance
$\sigma_{n,t}^{2}$: this limits the impact of outliers. The patch-wise variance
$\sigma_{n,t}^{2}$ in Eq.~\eqref{eq:MLE_sigma} corresponds to the average squared deviation to the mean,
computed after spatial and spectral whitening.
The estimator $\widehat{\M C}_n^\spec$  of the spectral covariance given in Eq.~\eqref{eq:MLE_Cspec} is readily the sample covariance of the residuals $\widehat{\M V}_{n,t}\T$ whitened for the spatial covariances by  $\widehat{\sigma}_{n,t}^2 \, \widehat{\M C}_n^\spat$.
 Conversely, the estimator $\widehat{\M C}_n^\spat$  of the spatial covariance given in Eq.~\eqref{eq:MLE_Cspat} corresponds to the sample covariance of the residuals $\widehat{\M V}_{n,t}$ whitened for the spectral covariances by $\widehat{\sigma}_{n,t}^2 \, \widehat{\M C}_n^\spec$.

In practice, the whitening operation by either $\widehat{\sigma}_{n,t}^2 \, \widehat{\M C}_n^\spec$ or $\widehat{\sigma}_{n,t}^2 \, \widehat{\M C}_n^\spat$ is done by first computing the Cholesky's decompositions $\widehat{\sigma}_{n,t}^2 \, \widehat{\M C}_n^\spec = \mathbb{W}_n^\spec \big( \mathbb{W}_n^\spec \big)\T$ and $\widehat{\sigma}_{n,t}^2 \, \widehat{\M C}_n^\spat = \mathbb{W}_n^\spat \big( \mathbb{W}_n^\spat \big)\T$ with $\mathbb{W}_n^\spec$ and $\mathbb{W}_n^\spat$ triangular matrices. We then compute $\left( \mathbb{W}_n^\spec \right)^{-1} \,\widehat{\M V}_{n,t}\T$ and $\left( \mathbb{W}_n^\spat \right)^{-1} \,\widehat{\M V}_{n,t}$, which respectively amounts to spectral and spatial whitening of the residuals $\widehat{\M V}_{n,t}$. We finally take the sample covariances of these whitened residuals.

\medskip

\noindent From the expressions in Eqs. (\ref{eq:MLE_mu})-(\ref{eq:MLE_Cspat}), it is not possible to
derive a closed-form expression of each parameter that does not also depend on other
parameters (i.e., estimators (\ref{eq:MLE_mu})-(\ref{eq:MLE_Cspat}) are interdependent). Yet, these formulae can be
applied alternately until convergence, a method called \emph{flip-flop} in
\cite{lu2005likelihood} where a faster convergence is reported compared to
maximizing the log-likelihood using an iterative optimization algorithm
(Newton's method).

Figure \ref{fig:covmat} illustrates the spatial and spectral covariance
matrices estimated under this model from an ASDI dataset of the VLT/SPHERE-IFS instrument.
MLEs of the spatial and spectral correlation matrices were computed with the flip-flop method for the four regions of interest indicated by small colored squares in Fig. \ref{fig:data}(a).
To compare matrices with very different variances, we
normalized each covariance $\text{Cov[a,b]}$ by
$\sqrt{\text{Cov[a,a]}\text{Cov[b,b]}}$, i.e. we show the correlation
coefficients.
Due to the vectorization of 2D spatial patches, the spatial correlations
display a blocky structure. The spatial correlations within a patch globally
decrease with
the 2D distance between pixels. They are stronger in the area (a) which is the
closest to the star. Spectral correlations are also much stronger close to the
star.
As can be observed in Fig. \ref{fig:data}(a), after the scaling transform
applied in the pre-processing step, regions far from the star are not seen at
the longest wavelengths. The size of multi-spectral patches extracted in these
regions is reduced from $KL$ to $KL_{\text{eff}}$ pixels (with
$L_{\text{eff}}<L$ the effective number of wavelengths seen at location $n$)
and the size of the spectral covariance matrix
$\widehat{\M C}_n^\spec$ is reduced accordingly, from $L\times L$
to $L_{\text{eff}}\times L_{\text{eff}}$.

\subsubsection{Shrinkage estimator of covariances}
\label{subsubsec:reg_est_shrinkage}

Given that the numbers $T$ of exposures and $L$ of spectral channels are limited, the empirical covariance estimates $\widehat{\M C}^\spat_n$ and $\widehat{\M C}^\spec_n$ (indifferently $\widehat{\M C}_n$ in the following) are very noisy (when $T \simeq K$ or $L_{\text{eff}} \simeq K$) and can be even rank-deficient (in particular when $T < K$ or $L_{\text{eff}} < K$). To reduce the estimation error on $\widehat{\M C}_n$ and ensure its definite-positiveness, shrinkage techniques combine the maximum likelihood
estimator with another estimator of smaller variance \citep{ledoit2004well}. Like in our previous
works \citep{flasseur2018exoplanet,flasseur2020paco,flasseur2021rexpaco,flasseur2023combining,flasseur2024deep}, we
consider the convex combination between the low-bias/high-variance sample covariance $\widehat{\M C}_n$ and a high-bias/low-variance matrix $\widehat{\M F}_n$:
 \begin{align}
        \widetilde{\M
            C}_n= \gamma ( (1-\widetilde{\rho}_n)\widehat{\M
            C}_n+\widetilde{\rho}_n\widehat{\M F}_n)\,,
        \label{eq:shrinkageformulageneral}
    \end{align}
    with $\widehat{\M F}_n=\mathrm{Diag}(\widehat{\M C}_n)$ a diagonal matrix such that
    $[\widehat{\M F}_n]_{i,i}=[\widehat{\M
        C}_n]_{i,i}$, $\widetilde{\rho}_n\in[0,1]$, and $\gamma$ a factor introduced to compensate for the fact that $\widehat{\M C}$ is a biased estimate of the true (and unknown) covariance $\M C$.
        The estimator $\widetilde{\M C}_n$ defined in Eq. (\ref{eq:shrinkageformulageneral})
shrinks off-diagonal values (i.e., the covariances) of $\widehat{\M C}_n$ towards 0 (by
multiplication by the factor $1-\widetilde{\rho}_n$) and leaves diagonal values (i.e., the sample variances)
unchanged. By controlling the shrinkage amount, hyper-parameter $\widetilde{\rho}_n$ plays a critical role as it set a bias-variance trade-off.
Compared to other regularization techniques such as diagonal loading (i.e., adding a small
fraction of the identity matrix to $\widehat{\M C}_n$),
 definition (\ref{eq:shrinkageformulageneral}) is attractive because it is data-driven: it
locally adapts to the fluctuations observed in the non-stationary data and to the number of samples (in particular, we have $L_{\text{eff}} \neq L$ on the borders of the field of view). Such a shrinkage estimator
is thus well-suited to imaging systems suffering from non-stationary
perturbations.

It remains to find the optimal level of shrinkage $\widetilde{\rho}_n$ appropriate for
each patch location $n$. An optimal setting can be defined based on risk minimization
between the true covariance $\M C_n$ and its shrunk counterpart $\widetilde{\M C}_n$
\citep{ledoit2004well}. However, such an oracle estimator can not be used in practice
since $\M C_n$ is unknown.
In a recent work \citep{flasseur2024shrinkage}, we derive a practical closed-form expression for its quasi-optimal setting that asymptotically approximates the oracle (for readability, the patch index $n$ is omitted in the following equations):
\begin{equation}
  \widetilde{\rho}
  = \frac{(\gamma\nu + \epsilon - 1)\big(\Trace ( \widehat{\M C}^2) - \sum_i [ \widehat{\M C} ]_{i,i}^2\big) +  \gamma\eta \big(\Trace^2 ( \widehat{\M C}) - \sum_i [ \widehat{\M C} ]_{i,i}^2\big)}{\gamma  \nu \big(\Trace ( \widehat{\M C}^2) - \sum_i [ \widehat{\M C} ]_{i,i}^2\big)},
	\label{eq:rhoshrinkage}
\end{equation}
with:
\begin{align}
	\epsilon &= \frac{\sum_{t=1}^T \widehat{\sigma}_t^{-4}}{\left( \sum_{t=1}^T \widehat{\sigma}_t^{-2} \right)^2},\\
	\zeta &= \frac{\sum_{t=1}^T \widehat{\sigma}_t^{-6}}{\left( \sum_{t=1}^T \widehat{\sigma}_t^{-2}\right)^3},\\
	\gamma &= (1 - \epsilon)^{-1},\\
	\nu &= 1 - \epsilon - 2\,\zeta + 2\,\epsilon^2,\\
	\eta &= \epsilon - 2\,\zeta + \epsilon^2\,.
	\label{eq:epsilon_gamma}
\end{align}
This analytic solution depends solely on the sample covariance $\widehat{\M C}_n$ and patch variances $\lbrace \widehat{\sigma}_{n,t}^2 \rbrace_{t=1:T}$ introduced in the MLE estimators (\ref{eq:MLE_mu})-(\ref{eq:MLE_Cspat}) to improve robustness against outliers. In addition, formulae (\ref{eq:rhoshrinkage})-(\ref{eq:epsilon_gamma}) explicitly account for the use
of $\widehat{\V \mu}_n^\spec$ as an empirical estimate of the true unknown mean
$\V \mu_n^\spec$ \citep{flasseur2024shrinkage}. It is worth noting that the
shrinkage technique developed in this paragraph is general; it holds
  whatever the covariance structure of our problem, namely, the spatio-spectral
separability of the covariance.

In the following, the shrunk covariance is given by:
\begin{equation}
  \widetilde{\M C} = \Weight \odot \widehat{\M C},
  \label{eq:psi}
\end{equation}
where $\odot$ denotes the Hadamard (element-wise) product, and $\Weight$ is a weighting
matrix whose diagonal entries are 1 and whose off-diagonal entries are
$1 - \widetilde{\rho}$, where $\widetilde{\rho}$ is given by Eq.~\eqref{eq:rhoshrinkage}.

\subsubsection{Shrunk spatio-spectral covariance}
\label{subsubsec:loss_function_separable_model}

To introduce the shrinkage with our particular factorization of the spatio-spectral
covariance as $\M C_n^\spec \otimes \M C_n^\spat$ (see Eq.~\eqref{eq:covsep}), we propose to apply
the shrinkage on each of the components $\M C_n^\spec$ and
$\M C_n^\spat$ separately. Futhermore, following the prescription in \cite{flasseur2021rexpaco}, we estimate the skrinkage factors once at the initialization of the
reconstruction algorithm. As a consequence, in subsequent steps, the shrinkage factors depend neither on the object of interest $\V u$ nor on the nuisance statistics defined in Eqs. \eqref{eq:MLE_mu}--\eqref{eq:MLE_Cspat}. This amounts to rewriting the MLEs estimators in
Eqs.~\eqref{eq:MLE_mu}--\eqref{eq:MLE_Cspat} as:
\begin{align}
  \widetilde{\V\mu}_n^{\,\spec}
  &= \frac{\sum_{t=1}^T \widetilde\sigma_{n,t}^{-2} \, \left(\data_{n,t} - [\M M\,\widetilde{\obj}]_{n,t}\right)}%
    {\sum_{t=1}^T \widetilde\sigma_{n,t}^{-2}},
    \label{eq:moysrhink}\\
  \widetilde{\sigma}_{n,t}^2
  &=\tfrac{1}{KL}
    \left\|\V
    v_{n,t} - \widetilde{\V\mu}_{n}^{\,\spec} - [\M M\,\widetilde{\obj}]_{n,t}\right\|_{\big( \widetilde{\M
    C}_n^\spec\big)^{-1} \otimes \big( \widetilde{\M C}_n^\spat\big)^{-1}}^2,
    \label{eq:weightssrhink}\\
  \widehat{\M C}_n^\spec
  &= \tfrac{1}{TK}\sum_{t=1}^T \widetilde{\M V}_{n,t}\T
    \left(\widetilde{\sigma}_{n,t}^{2} \, \widetilde{\M C}_n^\spat\right)^{-1}\,
    \widetilde{\M V}_{n,t},
    \label{eq:covspechalfshrink}\\
  \widetilde{\M C}_n^\spec
  &= \Weight^\spec_{n} \odot \widehat{\M C}_n^\spec,
    \label{eq:covspecshrink}\\
  \widehat{\M C}_n^\spat
  &= \tfrac{1}{TL}\sum_{t=1}^T \widetilde{\M V}_{n,t}
    \left(\widetilde{\sigma}_{n,t}^{2} \, \widetilde{\M C}_n^\spec\right)^{-1}\,
    \widetilde{\M V}_{n,t}\T,
    \label{eq:covspathalfshrink}\\
  \widetilde{\M C}_n^\spat
  &= \Weight^\spat_{n} \odot \widehat{\M C}_n^\spat,
    \label{eq:covspatshrink}
\end{align}
where $\widetilde{\M V}_{n,t}$ is defined as in
Eqs.~(\ref{eq:MLE_mu})--(\ref{eq:MLE_Cspat}) but replacing $\widehat{\V\mu}_n^{\,\spec}$ by
$\widetilde{\V\mu}_n^{\,\spec}$ as well as $\widehat{\obj}$ by $\widetilde{\obj}$, and where
$\Weight^\spec_{n}$ and $\Weight^\spat_{n}$ are computed according to Eq.~\eqref{eq:psi}
for the respective sample covariances $\widehat{\M C}_n^\spec$ and
$\widehat{\M C}_n^\spat$ given by Eqs.~\eqref{eq:covspechalfshrink} and
\eqref{eq:covspathalfshrink} as estimated during the initialization stage of the
reconstruction algorithm. The sample covariances $\widehat{\M C}_n^\spec$ and
$\widehat{\M C}_n^\spat$ in Eqs.~\eqref{eq:covspechalfshrink} and
\eqref{eq:covspathalfshrink} differ from their MLEs counterparts in
Eqs.~\eqref{eq:MLE_Cspec} and \eqref{eq:MLE_Cspat} by the accounting of the shrinkage in
the whitening. The assumed separable model of the covariance now takes the form
$\widetilde{\M C}_n = \text{diag}(\widetilde{\V \sigma}_n^2) \otimes \widetilde{\M C}_n^\spec \otimes \widetilde{\M C}_n^\spat$.

\section{Reconstruction of the component of interest}
\label{sec:recons}

\subsection{Direct model}
\label{sec:directmodel}

We extend the forward model developed for ADI in \cite{flasseur2021rexpaco,flasseur2022multispectral}
by including the spectral dimension.
Since the whole ASDI sequence is acquired within a
short time (a few hours of observations during a single night), we assume the component of interest (e.g.,
circumstellar disk and potential exoplanets) does not
evolve during the observations: its proper rotation around the
host star and photometry evolution are negligible at such short time scales. The multi-spectral image of this component is
simply described by the vector $\obj\in\mathbb{R}_+^{N'L}$
of its pixel values
 and there is no temporal
dimension in this spatio-spectral reconstruction.
Due to the apparent rotation of the field of view during the ASDI sequence, the number $N'$ of pixels in each spectral band of the reconstruction should be greater than $N$ to model any part of the disk seen within the sensor field of view on at least one exposure.

The contribution of $\obj$ to the data $\data$ is modeled by $\M M\,\obj$ with the linear
operator:
\begin{align}
   \M M = \begin{pmatrix}\M M_1\\ \vdots\\\M
        M_T\\\end{pmatrix}\text{ and }\M M_t = \M S\,\M Z\,\M A\,\M B_t\,\M R_t\,,
    \label{eq:modeldirect_complet}
\end{align}
where $\M M_t$, the model for the $t$-th frame, accounts for several
instrumental effects:
\begin{itemize}
    \item a \textit{rotation} $\M R_t$ applied to all off-axis sources due to
    the
    pupil-tracking mode (the field of view rotates while the residual star
    light remains fixed),
    implemented as a sparse interpolation matrix,
    \item a \emph{blur} $\M B_t$ due to the instrumental blurring 
    modeled as a 2D
    discrete convolution by the off-axis point spread function (PSF),
    \item an \textit{attenuation} $\M A$, very strong on the optical axis,
    then
    quickly decreasing (due to the coronagraph), modeled as a diagonal matrix
    \citep{flasseur2021rexpaco},
    \item 
    the absence of measurements outside the
    spatial extension of the sensor (a non-square area due to the instrumental
    design of the integral field spectrograph), modeled as a diagonal matrix
    $\M Z$ that replaces values outside the sensor area by zeros and
    keeps other values unchanged (i.e., zero-padding).
    \item the image \textit{scaling} applied during the pre-processing step
    produces a
    last transform $\M S$ (time-invariant), corresponding to a sparse
    interpolation matrix.
\end{itemize}

With the VLT/SPHERE instrument, the
off-axis point spread function (PSF) is quite stable and its core is almost
rotation invariant, leading to the approximation $\M B_t \, \M R_t \approx \M R_t \, \M B$. The model given in Eq.~\eqref{eq:modeldirect_complet}
can thus be approximated by:
\begin{align}
    \M M \, \obj\approx\begin{pmatrix}\M F_1\\ \vdots\\\M
        F_T\\\end{pmatrix}\! \, \M B \, \obj\,,
    \label{eq:modeldirect_approche}
\end{align}
where $\M B$ is a time-invariant blurring operator and $\M F = \{\M F_t\}_{t=1:T}$
are sparse
matrices that perform rotations, scalings, and attenuations according to the
transmission of the coronagraph and the sensor field of view. The model in
Eq. (\ref{eq:modeldirect_approche}) is only approximate: it neglects
possible anisotropies or temporal evolutions of the PSF.
Thanks to this approximation, a single convolution of the multi-spectral dataset
is performed instead of $T$ convolutions, 
which leads to a dramatic acceleration of the numerical evaluation of the
forward model (by one to two orders of magnitude) which is critical to achieving
reconstructions on datasets in the order of a few hours. We
verified through numerical simulations that the impact of these approximations on the
reconstructions was negligible (less than 1\%) in practice for VLT/SPHERE data. If approximation (\ref{eq:modeldirect_approche}) does not hold (e.g., for instruments that do not produce a stable off-axis PSF or if the latter is not rotation invariant), the full model  (\ref{eq:modeldirect_complet}) can be evaluated, at each iteration of the optimization procedure (see Sect. \ref{subsec:regularized_inversion}), on a random subset of temporal frames using stochastic gradient descent (e.g., with the Adam optimizer; \cite{kingma2014adam}). The stochasticity of this procedure reduces both memory consumption and time computation and leads to an approximate solution. Based on simulated disks and off-axis PSFs, we observe a typical relative difference less than 5\% on the reconstructed flux distribution obtained with the two strategies ((i) approximate model and no stochastic optimization \textit{versus} (ii) full model and stochastic optimization). In the following, we use strategy (i) solely given that approximation (\ref{eq:modeldirect_approche}) can be made with VLT/SPHERE data.

\subsection{Regularized inversion}
\label{subsec:regularized_inversion}

We reconstruct the component of interest using a penalized maximum likelihood
approach, i.e., by solving the following numerical optimization problem:
\begin{align}
    \widehat{\obj} = \argmin_{\obj \geq \V 0}
    \Brace[\big]{\mathscr{C}(\V\Omega,\obj) \equiv \mathscr{L}\Paren*{\V\Omega, \obj} + \mathscr{R}(\obj)},
    \label{eq:minpb}
\end{align}
where
$\V\Omega = \Brace[\Big]{\V \mu_n^\spec, \V\sigma_{n}^2, \M C_n^\spec, \M C_n^\spat}_{n \in \mathbb{K}}$
represents the parameters of the statistical model of the nuisances, the co-log-likelihood
$\mathscr{L}$ is given in Eqs.~\eqref{eq:covsep}--\eqref{eq:patchcologlikelihood}, and $\mathscr{R}(\obj)$ is a
regularization term to favor plausible reconstructions $\obj$. We selected a combination
of two regularization functions applying to the same $\obj$: an edge-preserving one that
favors smooth images with sharp edges co-located at all wavelengths and a
sparsity-inducing $\text{L}^1$ norm. The regularization writes:
\begin{align}
  \mathscr{R}(\obj)
  &= \beta_{\text{smooth}}\sum_{n=1}^{N'}
    \sqrt{\tfrac{1}{L}\sum_{\ell=1}^L \Norm*{\M D_{n,\ell}\,\obj}_2^2 + \tau^2}
  \notag\\
  &\quad + \beta_{\text{sparse}}\sum_{n=1}^{N'}\sum_{\ell=1}^L |u_{n,\ell}|,
	\label{eq:regul}
\end{align}
where $\M D_{n,\ell}\,\obj \approx \M{\nabla}_{\!n}\obj_{:,\ell}$
  approximates by finite differences the 2D spatial gradient of $\obj$ at pixel $n$ in the
$\ell$-th spectral channel and with $\tau$ a parameter chosen so as to
be negligible compared to the average norm of the spatial gradient where there is a sharp
edge (the regularization then approaches an isotropic vectorial total variation;
\cite{bresson2008fast}) and similar to the gradient magnitude in smoothly-varying areas
(this prevents the apparition of the staircasing effect common with total variation;
\cite{charbonnier1997deterministic, blomgren1997total, louchet2008total}). We illustrate
qualitatively through numerical simulations in Sect.
\ref{subsec:recons_simulated_disks} that these quite classical regularization penalties in
image processing remain adapted to disks having very different morphologies, like
elliptical disks with sharp edges or spiral disks with smooth edges. Hyper-parameters
$\beta_{\text{smooth}}$ and $\beta_{\text{sparse}}$ balance the weight of each
regularization term with respect to the data-fitting term. Note that, due to the
positivity constraint in Eq. (\ref{eq:minpb}), the $\text{L}^1$ norm $\|\obj\|_1$
corresponds to the simple differentiable term $\sum_{n=1}^{N'}\sum_{\ell=1}^L u_{n,\ell}$
for any feasible object $\obj$, and thus the regularization $\mathscr{R}(\obj)$ is
differentiable for $\tau \neq 0$ (in practice, we choose $\tau = 10^{-6}$).

To solve the smooth constrained optimization problem in Eq.~\eqref{eq:minpb}, we use a
limited-memory quasi-Newton method with bound constraints, VMLM-B
\citep{thiebaut2002optimization}, which is a more efficient variant of L-BFGS-B
\citep{zhu1997algorithm}. To minimize $\mathscr{C}(\V\Omega,\obj)$ in $\obj$ given
$\V\Omega$, the VMLM-B optimizer requires to evaluate the cost
function $\mathscr{C}(\V\Omega,\obj)$ and the first derivatives
$\nabla_{\obj}\mathscr{C}(\V\Omega,\obj)$ with respect to $\obj$.
The analytic expression of these first derivatives writes, for all $\V u \ge \M 0$:
\begin{align}
  &\nabla_{\obj}\mathscr{C}\Paren*{\V\Omega,\obj} = \sum_{n\in\mathbb{K}} \underbrace{
    \M M\T \sum_{t=1}^T \frac{1}{\sigma_{n,t}^2}\,\M E_{n,t}\T\,\M\Gamma_n\,
    \Brack[\big]{\M E_{n,t}\,\M M\,\obj + \V\mu_{n} - \data_{n,t}}
  }_{\nabla_{\obj}\mathscr{L}_n\Paren*{\V\Omega_{n},\obj}\text{,
    see Eqs.~\eqref{eq:covsep}--\eqref{eq:patchcologlikelihood}}}
    \notag\\
  &\hspace*{12mm} + \underbrace{
    \beta_{\mathrm{smooth}}\sum_{n=1}^{N'}
    \frac{\M D_{n,\ell}\T \, \M D_{n,\ell}^{{\phantom{\TransposeLetter}}} \, \obj}{
      \sqrt{\frac{1}{L}\sum_{\ell=1}^L
        \Norm*{\M D_{n,\ell} \, \obj}_2^2 + \tau^2}}
    + \beta_{\mathrm{sparse}}\,\V 1}_{\nabla_{\obj}\mathscr{R}(\obj)\text{, see Eq. (\ref{eq:regul})}},
  \label{eq:gradC}
\end{align}
where $\M 1$ is an array of same size as $\obj$ filled with ones,
$\M E_{n,t}$ is the $K\,L\times K\,L\,T$ operator that extracts a multi-spectral patch at
spatial location $n$ and time frame $t$ (by extension of its definition introduced in
Sect.~\ref{subsec:patch_based_statistical_modeling}),
$\V\Omega_{n}$ denotes the subset of the statistical model parameters for
  the $n$-th patch, and $\M\Gamma_n$ is equal to
${\big( \M C_n^\spec \big)^{-1} \otimes \big( \M C_n^\spat \big)}^{-1}$.

Solving the problem in Eq.~\eqref{eq:minpb} yields an estimator $\widetilde{\obj}$ of the
object of interest given the parameters $\V\Omega$ of the statistical model. We consider
next different strategies to jointly obtain estimators of these parameters from the same
dataset.

\subsection{Joint estimation of all unknowns from the data}
\label{sec:joint-estimation}

Formally, the estimators of the object of interest $\obj$ and of the parameters $\V\Omega$
of the nuisance statistics provided by REXPACO ASDI are the ones for which
Eqs.~\eqref{eq:moysrhink}--\eqref{eq:covspatshrink} and \eqref{eq:minpb} jointly hold.
Solving this system of non-linear equations is intrinsically difficult because there is no
closed-form solution (at least due to the non-negativity constraint for $\obj$) and because
of the interdependence of the equations. In the following sub-sections, we develop
practical algorithms to iteratively solve this system of equations.

\subsubsection{Alternating strategy}
\label{sec:alternating-strategy}

Even though there is no joint closed-form solution to the set of equations
\eqref{eq:moysrhink}--\eqref{eq:covspatshrink} and \eqref{eq:minpb}, we note that each of
these equations readily provides an estimator of some unknowns when the rest of the unknowns are fixed.
This property can be exploited to solve the set of equations \eqref{eq:moysrhink}--\eqref{eq:covspatshrink} by the following alternating strategy. Given the object $\obj$, the parameters $\V\Omega$ can be estimated by repeatedly applying Eqs. \eqref{eq:moysrhink}--\eqref{eq:covspatshrink} in turn until convergence to a so-called \textit{fixed point solution}. This procedure being applied for each patch to estimate all the nuisance parameters. We denote the resulting parameters as
$\widetilde{\V\Omega}\Paren{\obj}$ in the following. A first possible algorithm to find
the solution is then:

\medskip
\noindent
\begin{tabular}{l@{\ }p{80mm}}
  1. & Let $i=0$ and initialy assume a null object $\widetilde{\obj}^{[0]} = \V0$.\\[1ex]
  2. & Estimate nuisance statistics
       $\widetilde{\V\Omega}^{[i+1]} = \widetilde{\V\Omega}\Paren[\big]{\obj^{[i]}}$
       as the fixed point solution of Eqs.~\eqref{eq:moysrhink}--\eqref{eq:covspatshrink}
       for the current estimate of the object $\obj^{[i]}$. If $i = 0$, also include
       Eq.~\eqref{eq:rhoshrinkage} in the fixed point method to determine the shrinkage
       factors $\widetilde{\rho}^{\,\spec}$ and $\widetilde{\rho}^{\,\spat}$. These
       factors define $\Weight^\spec$ and $\Weight^\spat$ for all subsequent iterations,
       i.e. for $i > 0$.\\[1ex]
  3. & Update the object
       $\widetilde{\obj}^{[i+1]} = \argmin_{\obj \geq \V 0} \mathscr{C}\Paren[\big]{\widetilde{\V\Omega}^{[i+1]}, \obj}$ by applying the
       reconstruction algorithm described in Sect.~\ref{subsec:regularized_inversion}.\\[1ex]
  4. & Let $i \gets i + 1$ and, unless estimators $\widetilde{\V\Omega}^{[i]}$ and
       $\widetilde{\obj}^{[i]}$ have converged, go to step 2.\\
\end{tabular}
\medskip
In practice, we assume the algorithm reaches convergence when the condition $\Norm[\big]{\widetilde{\V u}^{[i+1]} - \widetilde{\V u}^{[i]}} \le \eta \Norm[\big]{\widetilde{\V u}^{[i+1]}}$ is satisfied, with $\eta = 10^{-6}$.

\noindent This first algorithm implements a simple alternating strategy which is equivalent, for
non-linear equations, to the Gauss--Seidel method for solving a system of linear equations.
The alternating method converges slowly due to the need for multiple reconstructions of the object of interest, which are progressively refined in each iteration of Step 3. This process represents the primary computational bottleneck (the computational cost of estimating nuisance statistics is negligible by comparison). However, as discussed in the following subsections, the computational efficiency of this estimation strategy can be significantly improved.

\subsubsection{Partially hierarchical optimization}
\label{sec:partially-hierarchical}

Noting that the joint solution of Eqs.~\eqref{eq:moysrhink} and \eqref{eq:weightssrhink}
only depends on the object and on the spatial and spectral covariances,
we introduce the following
auxiliary cost function:
\begin{align}
  \label{eq:auxiliarycost}
  \mathscr{D}\Paren[\big]{\obj, \Brace[\big]{\M C^\spec_{n},\M C^\spat_{n}}_{n\in\mathbb{K}}}
  & = \min_{\substack{\Brace{\V\mu^\spec_{n}}_{n\in\mathbb{K}}\\
      \Brace{\sigma_{n,t}^{2}}_{n\in\mathbb{K},t\in1:T}}}
  \mathscr{C}(\V\Omega,\obj)\notag\\
  & = \mathscr{C}(\V\Omega,\obj)\,\rule[-6mm]{0.5pt}{9mm}_{
    \hspace*{-2pt}\begin{array}[b]{l}
      \scriptscriptstyle\V\mu^\spec_{n}
      \:{=}\: \widetilde{\V\mu}^{\,\spec}_{n}\Paren[\big]{\obj,\,\M C^\spec_{n},\, \M C^\spat_{n}}
      \\[1mm]
      \scriptscriptstyle\V\sigma_{n}^2
      \:{=}\: \widetilde{\V\sigma}^{2}_{n}\Paren[\big]{\obj,\,\M C^\spec_{n},\, \M C^\spat_{n}}\\
    \end{array}
    }
\end{align}
In practice, for each patch $n$ and given the object $\obj$ and the covariances
$\M C^\spec_{n}$ and $\M C^\spat_{n}$, the estimators
$\widetilde{\V\mu}^{\,\spec}_{n}\Paren[\big]{\obj,\, \M C^\spec_{n},\, \M C^\spat_{n}}$ and
$\widetilde{\V\sigma}^{2}_{n}\Paren[\big]{\obj,\, \M C^\spec_{n},\, \M C^\spat_{n}}$ are
obtained by applying Eqs.~\eqref{eq:moysrhink} and \eqref{eq:weightssrhink} iteratively
until convergence to a fixed point. Such estimators define a stationary
point of $\mathscr{C}$ with respect to the parameters $\V \mu_n$ and $\V \sigma_n$, the corresponding partial derivatives of $\mathscr{C}$ are
therefore null. Hence, by the chain rule, the derivatives of the auxiliary function
$\mathscr{D}$ in $\obj$ are simply given by $\nabla_{\obj}\mathscr{C}$ in
Eq.~\eqref{eq:gradC} evaluated at the stationary point. Thanks to this property, solving:
\begin{align}
  \widetilde{\obj} = \argmin_{\obj \geq \V 0}
  \mathscr{D}\Paren[\big]{\obj, \Brace[\big]{\widetilde{\M C}^\spec_{n},\widetilde{\M C}^\spat_{n}}_{n\in\mathbb{K}}}
  \label{eq:minhier}
\end{align}
can be done similarly to solving the constrained reconstruction problem in Eq.~\eqref{eq:minpb},
that is with a quasi-Newton method like VMLM-B \citep{thiebaut2002optimization}.

Minimizing the auxiliary function $\mathscr{D}$ instead of $\mathscr{C}$, the estimators
are obtained by the following algorithm:

\medskip
\noindent
\begin{tabular}{l@{\ }p{80mm}}
  1. & Let $i=0$, assume a null object $\widetilde{\obj}^{[0]} = \V0$, and initialize model
       statistics $\widetilde{\V\Omega}^{[0]}$ as in Step~2 of the first iteration of the
       algorithm given in Sect.~\ref{sec:alternating-strategy}.\\[2ex]
  2. & Update the object by minimizing the auxiliary cost function:\\[1ex]
     & \qquad$\widetilde{\obj}^{[i+1]} = \argmin_{\obj \geq \V 0} \mathscr{D}\Paren[\Big]{\obj,
       \Brace[\big]{\widetilde{\M C}^{\spec\,[i]}_{n},
       \widetilde{\M C}^{\spat\,[i]}_{n}}_{n\in\mathbb{K}}}$.\\[2ex]
  3. & Update the nuisance statistics:
       $\widetilde{\V\Omega}^{[i+1]} = \widetilde{\V\Omega}\Paren[\big]{\obj^{[i+1]}}$.\\[1ex]
  4. & Let $i \gets i + 1$ and, unless estimators $\widetilde{\V\Omega}^{[i]}$ and
      $\widetilde{\obj}^{[i]}$ have converged, go to step 2.\\
\end{tabular}
Like for the alternating strategy presented in Sect. \ref{sec:alternating-strategy}, we assume that  the partially hierarchical optimization scheme reaches convergence when the condition $\Norm[\big]{\widetilde{\V u}^{[i+1]} - \widetilde{\V u}^{[i]}} \le \eta \Norm[\big]{\widetilde{\V u}^{[i+1]}}$ is satisfied, with $\eta = 10^{-6}$.

\medskip

It may be noted that the estimators $\widetilde{\V\mu}^{\,\spec\,[i+1]}_{n}$ and
$\widetilde{\sigma}^{2\,[i+1]}_{n,t}$ can also be considered as a by-product of the
minimization of $\mathscr{D}$ in Step~2 of the above algorithm. Hence, Step~3 can be
modified to restrict the updating of the nuisance statistics to that of the covariances
$\widetilde{\M C}^\spec_{n}$ and $\widetilde{\M C}^\spat_{n}$
($\forall n \in \mathbb{K}$), e.g.\ by finding a fixed point of
Eqs.~\eqref{eq:covspechalfshrink}--\eqref{eq:covspatshrink}.

The hierarchical optimization in Step~2 yields estimates such that
  Eqs.~\eqref{eq:moysrhink}, \eqref{eq:weightssrhink}, and \eqref{eq:minpb} jointly hold
  for given covariance matrices. As a result, the convergence speed is improved compared
  to the Algorithm described in Sect.~\ref{sec:alternating-strategy}.

\subsubsection{Fully hierarchical approximation}
\label{sec:fully-hierarchical approximation}

In principle, all the parameters could be found by solving:
\begin{equation}
  \widetilde{\obj} = \argmin_{\obj \geq \V 0}
  \Brace[\big]{\mathscr{F}(\obj) \equiv \mathscr{C}\Paren[\big]{\widetilde{\V\Omega}(\obj),\obj}},
  \label{eq:fullhierarchical}
\end{equation}
and taking $\widetilde{\V\Omega} = \widetilde{\V\Omega}(\widetilde{\obj})$. The estimator
$\widetilde{\V\Omega}(\obj)$ of the nuisance statistics is however not truly a stationary
point of $\mathscr{C}$ for the covariance matrices $\widetilde{\M C}^\spec_{n}$ and
$\widetilde{\M C}^\spat_{n}$ although it is a stationary point for the other parameters of
the nuisance statistics. We nevertheless make the following approximation:
\begin{equation}
  \label{eq:gradientapprox}
  \nabla_{\obj}\mathscr{F}(\obj) \approx
  \left.\nabla_{\obj}\mathscr{C}\Paren{\V\Omega,\obj}\right|_{\V\Omega = \widetilde{\V\Omega}(\obj)},
\end{equation}
since, under this approximation, the constrained problem in
Eq.~\eqref{eq:fullhierarchical} can be solved by a quasi-Newton method as VMLM-B
\citep{thiebaut2002optimization}.

In practice, we verified numerically that the approximation in
Eq.~\eqref{eq:gradientapprox} holds to a numerical precision that is sufficient to achieve the convergence of the quasi-Newton method. We also verified that the fully
alternating strategy described in Sect.~\ref{sec:alternating-strategy} and the fully
hierarchical approach assuming Eq.~\eqref{eq:gradientapprox} both converge to the same
estimators. The fully hierarchical approach is however much faster than algorithms presented in Sects. \ref{sec:alternating-strategy} and \ref{sec:partially-hierarchical}. For example, the approximate fully hierarchical algorithm reduces the computational load of the alternating strategy by a factor comparable to the number of iterations $i$ required to reach convergence with the algorithm described in Sect. \ref{sec:alternating-strategy} (ranging from 30 to 100 in practice). Consequently, we exclusively employed the approximate fully hierarchical optimization strategy throughout this paper and recommend it as the preferred method for estimating parameters in REXPACO ASDI.

\subsection{Unsupervised setting of the regularization hyper-parameters}
\label{subsec:sure}

\begin{figure}
	\centering
	\includegraphics[width=0.485\textwidth]{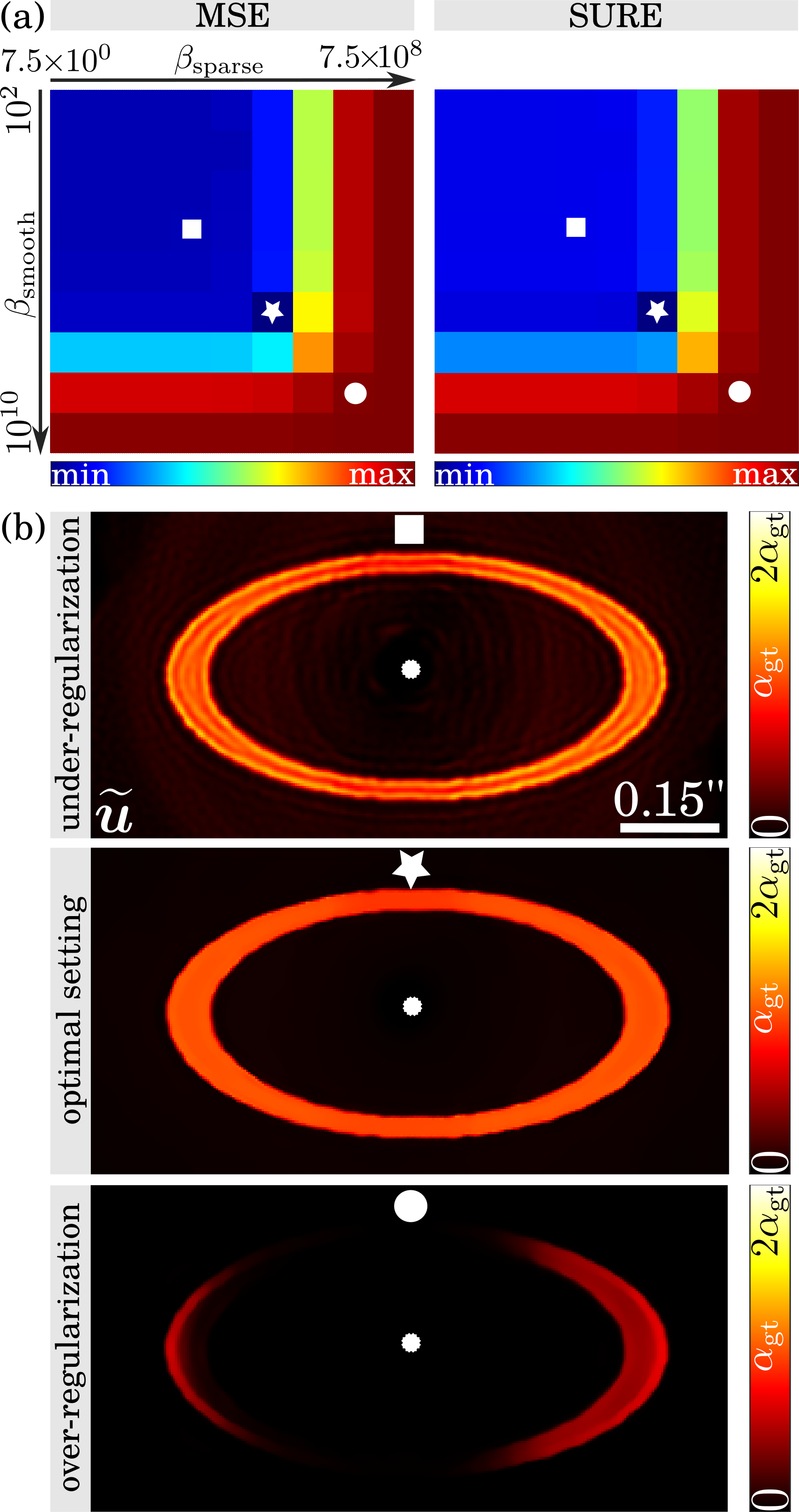}
	\caption{(figure modified) Effect of the setting of the regularization parameters on REXPACO ASDI reconstructions of a simulated elliptical disk at contrast level $\alpha_{\text{gt}} = 1 \times 10^{-5}$: (a) Comparison between the MSE (left; Eq. (\ref{eq:regularization_mse})) and the SURE criterion (right; Eq. (\ref{eq:patched_sure_criterion})). The star symbol represents the minimum of the two criteria. We observe that  it corresponds to the same setting of $\V \beta$ in both cases: $\widetilde{\beta}_{\text{sparse}}^{\text{MSE}} = \widetilde{\beta}_{\text{sparse}}^{\text{SURE}} = 7.5 \times 10^{5}$ and $\widetilde{\beta}_{\text{smooth}}^{\text{MSE}} = \widetilde{\beta}_{\text{smooth}}^{\text{SURE}} = 1 \times 10^{7}$. The square and circle symbols respectively represents examples of an under-regularization (i.e., $\V \beta < \widetilde{\V \beta}^{\text{MSE}}$) and of an over-regularization (i.e., $\V \beta > \widetilde{\V \beta}^{\text{MSE}}$). Panel (b) shows the reconstructed flux distribution $\widetilde{\obj}$ for the three settings of the regularization hyper-parameters symbolized in panel (a), see corresponding symbols on top of images. The reconstructions can be compared qualitatively to the ground truth $\obj_{\text{gt}}$ given in Fig. \ref{fig:gt_fullfig}. Given that the simulated disk has a constant contrast $\alpha_{\text{gt}}$ over the spectral channels, only the spectral mean of the spatio-spectral flux distributions reconstructed by REXPACO ASDI is displayed. Dataset: HD 172555 (2015-07-11), see Sect. \ref{subsec:datasets} for the observing parameters.}
	\label{fig:sure_convergence}
\end{figure}

As in our previous work on the REXPACO algorithm \citep{flasseur2021rexpaco}, we propose a strategy to set optimally, and in a data-driven fashion, the hyper-parameters $\V \beta = \lbrace \beta_{\text{smooth}}, \beta_{\text{sparse}} \rbrace$ involved in the regularization term $\mathscr{R}$ of Eq. (\ref{eq:regul}). These two free parameters represent the relative weights of the two combined priors on the sought flux distribution and they also set the relative weight of the priors with respect to the data-fidelity term $\mathscr{L}$ defined in Eq.~\eqref{eq:totalcologlikelihood}. In other words, the hyper-parameters $\V \beta$ set a critical bias-variance trade-off. These hyper-parameters can be tuned manually by trial and error until the reconstruction is qualitatively acceptable, but this approach relies on the user judgment and, likely, the resulting setting is not optimal. Instead, we capitalize on the large variety of methods available in the signal processing literature to set regularization hyper-parameters by minimizing a figure of merit, see e.g. \cite{craven1978smoothing,wahba1985comparison,stein1981estimation}. One of these criteria is the so-called Stein's Unbiased Risk Estimator (SURE; \cite{stein1981estimation}) that we have also selected among other metrics in our previous works dedicated to the post-processing of high-contrast observations \citep{flasseur2020paco,flasseur2021rexpaco} given its ability to approximate the mean square error (MSE) in the measurement space:
\begin{equation}
	\text{MSE}(\V \beta) = \sum\limits_{n \in \mathbb{K}} \sum\limits_{t=1}^T \left\Vert \frac{1}{\widehat{\sigma}_{n,t}^2} \M E_{n,t} \, \left( \M M \left(\obj_{\text{gt}} -
	\widetilde{\obj}_{\V \beta}(\V v)\right)\right) \right\Vert_{\widetilde{\M \Gamma}_n}^2\,,
	\label{eq:regularization_mse}
\end{equation}
with $\obj_{\text{gt}}$ the unknown ground truth flux distribution and $\widetilde{\obj}_{\V\beta}(\V v)$ the  flux distribution reconstructed from the data $\V v$ using the set of regularization hyper-parameters $\V \beta$.
	It is shown in the literature \citep{stein1981estimation} that the SURE estimator gives an unbiased estimation of $\text{MSE}(\V \beta)$ without requiring the value of the unknown ground truth flux-distribution $\obj_{\text{gt}}$ involved in the MSE  (\ref{eq:regularization_mse}).

By extending our previous work \citep{flasseur2021rexpaco} to the multi-spectral model of the nuisance and of the object components, the resulting SURE risk estimator can be numerically evaluated by:
\begin{multline}
	\text{SURE} (\V \beta)
	\approx \sum\limits_{n \in \mathbb{K}} \sum_{t=1}^{T} \left\Vert \frac{1}{\widetilde{\sigma}_{n,t}^2} \M E_{n,t}
	\left( \V v - \widetilde{\V \mu}^\spec - \M M\,\widetilde{\obj}_{\V \beta}(\V v)
	\right)\right\Vert_{\widetilde{\M \Gamma}_n}^2\\
	+ (2/\xi) \, \V b\T \, \M M \, \left[ \widetilde{\obj}_{\V \beta}(\V v + \xi \V b) -
	\widetilde{\obj}_{\V \beta}(\V v) \right] - N\,T\,L\,,
	\label{eq:patched_sure_criterion}
\end{multline}
where $\V b \in \mathbb{R}^{N'\,T\,L}$ is an independent and identically distributed pseudo-random vector of unit variance, and $\xi$ is the amplitude of this perturbation. This expression, as the MSE in Eq. (\ref{eq:regularization_mse}), tailored to our problem accounts for the structured model of the covariances of the nuisance (i.e., separable spatially and spectrally), as defined by the matrix $\M \Gamma$. It also accounts for our patch-based strategy to model the full covariance through the partition of the image into non-overlapping patches with the operator $\M E$. In addition, expression (\ref{eq:patched_sure_criterion}) is a practical approximation of the original SURE criterion that involves the computation of the Jacobian matrix of the mapping $\obj \rightarrow \widetilde{\obj}_{\V \beta}(\V v)$ with respect to the components of the data $\V v$. Given that there is no-closed-form expression for such a term, we approximate it by resorting to finite differences through a Monte-Carlo perturbation of the data, as proposed by \cite{girard1989fmcsure,ramani2012regularization}. This strategy leads to the approximate expression (\ref{eq:patched_sure_criterion}) involving the reconstruction of the two flux distributions $\widetilde{\obj}_{\V \beta}(\V v)$ and $\widetilde{\obj}_{\V \beta}(\V v + \xi \V b)$ obtained respectively from the data $\V v$ and the perturbed counterpart $\V v + \xi \V b$. The optimal setting $\widetilde{\V \beta}^{\text{SURE}}$ of the regularization hyper-parameters $\V \beta$ is obtained by minimizing the SURE score (\ref{eq:patched_sure_criterion}) with respect to $\V \beta$.

In Fig. \ref{fig:sure_convergence}, we illustrate the benefits of the proposed data-driven setting of the regularization hyper-parameters $\V \beta$ by resorting  to the numerical injection of a synthetic elliptical disk within an object-free dataset of the HD 172555 star obtained with the VLT/SPHERE-IFS instrument (see Sect. \ref{subsec:datasets} for the description of the dataset). The experiments are conducted for a disk of contrast (see definition in Sect. \ref{sec:introduction})
 $\alpha_{\text{gt}} = 1 \times 10^{-5}$ in every spectral channel. The corresponding ground truth flux distribution $\obj_{\text{gt}}$ to be reconstructed is given in Fig. \ref{fig:gt_fullfig} bottom-left. We start by comparing the SURE criterion (\ref{eq:patched_sure_criterion}) to the MSE (\ref{eq:regularization_mse}).
The tested values of the hyper-parameters are $\beta_{\text{smooth}} \in \left[ 1 \times 10^2, 1 \times 10^{10} \right]$ and $\beta_{\text{sparse}} \in \left[ 7.5 \times 10^0, 7.5 \times 10^{8} \right]$ with a regular sampling of $\log(\V \beta)$. For the computation of the SURE metric, we have to set the value of the parameter $\xi$ involved in Eq. (\ref{eq:patched_sure_criterion}), namely the strength of the perturbation $\V b$. We found  this value not to be critical, yet it should be set not too small to prevent errors due to numerical underflows in the computation of the difference $\widetilde{\V u}_{\V \beta}(\V v + \xi \V b) - \widetilde{\V u}_{\V \beta}(\V v)$ and not too large so that the approximation (\ref{eq:patched_sure_criterion}) stays valid. As in our previous work on the REXPACO algorithm \citep{flasseur2021rexpaco}, we set it empirically by $\xi = 0.1 \times \text{MAD}(\V v)$, where the median absolute deviation $\text{MAD}(\V v)=\text{median}(|\V v - \text{median}(\V v)|)$ is a robust estimator of the standard-deviation of the data $\V v$. Panel (a) of Fig. \ref{fig:sure_convergence} gives the results of the comparison between MSE and SURE. It illustrates that our custom SURE definition is an accurate proxy of the MSE: the global minimum of the two metrics is obtained for the same tested values of our grid of parameters $\V \beta$. The SURE criterion (\ref{eq:patched_sure_criterion}) can thus be safely used to approximate the MSE when facing real cases where the ground truth flux distribution $\obj_{\text{gt}}$ is not available. Panel (b) of Fig. \ref{fig:sure_convergence} completes this study by showing an example of the reconstructed flux distribution in three cases: an under-regularized reconstruction (i.e., $\widetilde{\V \beta} < \widetilde{\V \beta}^{\text{MSE}}$), the optimal regularization (i.e.,  $\widetilde{\V \beta} = \widetilde{\V \beta}^{\text{MSE}} = \widetilde{\V \beta}^{\text{SURE}}$), and an over-regularized reconstruction (i.e., $\widetilde{\V \beta} \gg \widetilde{\beta}^{\text{MSE}}$). It illustrates the benefits of the regularization with an optimal strength: the reconstructed flux distribution is very similar to the ground truth presented in Fig. \ref{fig:gt_fullfig} bottom-left. The nuisance component is well discarded, even very close to the host star, and the reconstructed disk have sharp edges matching the ground truth. An under-regularization causes a slightly worst rejection of the nuisance component (i.e., a non-null background remains in the reconstruction) and the reconstructed disk exhibits some ripples and non-homogeneous parts. In the opposite case of an over-regularization, the reconstructed flux distribution is severely biased towards zero and it results in important morphological distortions impacting the disk, in particular due to a too strong promotion of sparsity .

By construction, the optimal setting of the hyper-parameters $\V \beta$ by minimizing the SURE criterion (\ref{eq:patched_sure_criterion}) requires to perform two reconstructions ($\widetilde{\obj}_{\V \beta}(\V v)$ and $\widetilde{\obj}_{\V \beta}(\V v + \xi \V b)$) for each tested pair $\V \beta$ of hyper-parameters. Given that more than 120 individual reconstructions are presented in the following section to evaluate the performance of the proposed approach, it would have been an unreasonable computational overhead to derive $\widetilde{\V \beta}^{\text{SURE}}$ for each reconstruction. We thus chose to evaluate the optimal setting in only one case: the  disk of SA0 206462. This computation leads to $\widetilde{\beta}_{\text{sparse}}^{\text{SURE}} = 7.5 \times 10^{4}$ and $\widetilde{\beta}_{\text{smooth}}^{\text{SURE}} = 1 \times 10^{6}$. These values are not too far from the optimal ones derived in the numerical experiments performed in Fig. \ref{fig:sure_convergence} of this section on a totally different dataset. When facing a new dataset, we thus simply weight these pre-computed values according to the number of frames within the target dataset with respect to the dataset of SAO 206462 in order to keep a constant relative weighting between the regularization and the data fidelity terms. We found that this setting was qualitatively acceptable in all our experiments, i.e. no significant artifact was ever observed either in terms of a bad rejection of the nuisance component or in terms of non-physical discontinuities in the disk structures. We recommend to use this strategy when facing the processing of a large number of datasets. A careful data-dependent and data-driven setting of the hyper-parameters $\V \beta$ with SURE can be reserved to specific cases where the setting seems to be more critical (e.g., in the case of a very faint disk) or to refine the reconstruction obtained with the pre-computed and scaled values of the regularization hyper-parameters.

\section{Results}
\label{sec:results}

\subsection{Datasets description}
\label{subsec:datasets}

\begin{table*}
		\caption{Summary of the main observational parameters for the VLT/SPHERE datasets analyzed in this paper. The columns include: target name, ESO survey ID, observation date, number ($L$) of spectral channels, spectral filter band ($\Delta_{\lambda}$), number ($T$) of available temporal frames, total apparent field of view rotation ($\Delta_{\text{par}}$), number of sub-integration exposures (NDIT), individual exposure time (DIT; Detector Integration Time), average coherence time ($\tau_0$), average seeing, and the first publication reporting an analysis of the same data. All observations were conducted using the apodized Lyot coronagraph \citep{carbillet2011apodized} on the VLT/SPHERE instrument. $^{\text{(a)}}$The contribution of the three known exoplanets (HR 8799 c, d, e), which are within the SPHERE-IFS field of view, was masked.
$^{\text{(b)}}$While the IRDIS dataset from the same epoch (recorded simultaneously using the IRDIFS-EXT mode of SPHERE) was analyzed in \citep{boccaletti2021investigating}, no reconstruction from the IFS dataset was reported in that study. $^{\text{(c)}}$The first value is the real amplitude of the parallactic rotation, while the second corresponds to the simulated parallactic rotation used in our experiments with synthetic disk simulations (see Sect. \ref{subsec:recons_simulated_disks}).}
		\label{tab:dataset_logs}
		\centering
		\begin{tabular}{ccccccccccccc}
				\toprule
				Target & ESO ID & Obs. date & $L$ & $\Delta_{\lambda}$ & $T$ & $\Delta_{\text{par}}$ & NDIT & DIT & $\tau_0$ & Seeing & Related paper\\
				& & & & ($\micro\meter$) & & (°) & & (s) & (ms) & ('') & \\
				\midrule
				\multicolumn{12}{c}{\textit{SPHERE-IFS data used for validation of the statistical model, see Sect. \ref{subsec:validation_model}}}\\
				\midrule
				HR 8799$^{\text{(a)}}$ & 095.C-0298(C) & 2015-07-04 & 39 & 0.96-1.64 & 46 & 16.4 & 4 & 64 & 2.3 & 0.94 & \cite{langlois2021sphere}\\
				\midrule
				\multicolumn{12}{c}{\textit{SPHERE-IFS data used for qualitative analysis by reconstructing known real disks, see Sect. \ref{subsec:recons_real_disks}}}\\
				\midrule
				HR 4796 & 095.C-0298(H) & 2015-02-03 & 39 & 0.96-1.33 & 56 & 48.2 & 4 & 64 & 13.7 & 0.67 & \cite{milli2017near}\\
				SAO 206462 & 095.C-0298(A) & 2015-05-15 & 39 & 0.96-1.64 & 63 & 63.7 & 4 & 64 & 8.9 & 0.59 & \cite{maire2017testing}\\
				MWC 758 & 1100.C-0481(K) & 2018-12-17 & 39 & 0.96-1.33 & 63 & 29.2 & 4 & 96 & 8.3 & 0.98 & \cite{boccaletti2021investigating}\\
				PDS 70 & 1100.C-0481(D) & 2018-02-24 & 39 & 0.96-1.64 & 87 & 93.4 & 3 & 96 & 7.5 & 0.66 & \cite{mesa2019vlt}\\
				HD 163296 & 1100.C-0481(G) & 2018-05-07 & 39 & 0.96-1.64 & 48 & 14.2 & 3 & 96 & 2.6 & 1.04 & \cite{mesa2019determining}\\
				AB Aurigae & 104.20V7.001 & 2020-01-18 & 39 & 0.96-1.64 & 51 & 38.5 & 2 & 64 & 5.6 & 0.71 & this paper$^{\text{(b)}}$\\
				\midrule
				\multicolumn{12}{c}{\textit{SPHERE-IFS data used for quantitative analysis by reconstructing synthetic disks, see Sect. \ref{subsec:recons_simulated_disks}}}\\
				\midrule
				HD 172555 & 095.C-0192 & 2015-07-11 & 39 & 0.96-1.33 & 62 & 12.9//30.0$^{\text{(c)}}$ & 8 & 32 & 3.9 & 1.20 & \cite{flasseur2020paco}\\
				\midrule
				\multicolumn{12}{c}{\textit{SPHERE-IRDIS data used to compare ADI and ASDI post-processing, see Sect. \ref{subsec:importance_spectral_processing}}}\\
				\midrule
				SAO 206462 & 095.C-0298(A) & 2015-05-15 & 2 & 2.11-2.25 & 63 & 63.7 & 4 & 64 & 8.9 & 0.59 & \cite{maire2017testing}\\
				\bottomrule
		\end{tabular}
\end{table*}

For our comparisons, we selected eight datasets from the SPHERE-IFS instrument, acquired under diverse observing conditions.

First in Sect. \ref{subsec:validation_model}, we consider a dataset of HR 8799 to assess the relevance of the statistical model proposed in this paper. This emblematic star hosts four known exoplanets, all detected by direct imaging  \citep{marois2008direct,marois2010images}. Three of which fall within the SPHERE-IFS field of view. After masking the contribution of these point-like sources within the data, we conduct a model ablation analysis to show that it is critical to accurately model the correlations of the nuisance component.

Then in Sects. \ref{subsec:validation_model} and \ref{subsec:recons_real_disks}, we consider six additional datasets from stars with previously imaged circumstellar disks. These datasets are used to qualitatively assess  the benefits of the proposed algorithm on real disks in comparison to baseline methods. The selected disks are at different evolution stages and have very diverse morphologies. The stars included in the analysis are:\\
-- HR 4796A, which is the primary member of a
				binary system within the TW Hydrae association with an age of about 12 Myr
				\citep{bell2015self}. Located at about 72.8 pc
				\citep{van2007validation}, HR 4796A harbors a debris disk observable in
				a face-on configuration, initially imaged by the Hubble Space Telescope
				\citep{schneider1999nicmos}. Subsequently, its morphology and spectroscopy have been studied
				intensively by direct imaging
				\citep{milli2017near,milli2019optical}. The disk showcases a slender ring and a high surface brightness  hinting at the potential presence of exoplanets, though no companion has been detected yet.\\
-- SAO 206462, which is located within the Upper Centaurus Lupus constellation, has an estimated age of about 9 Myr \citep{muller2011hd}. Located at about 157 pc \citep{brown2016gaia}, it hosts a nearly face-on transition disk imaged both in thermal emission \citep{doucet2006mid} and in scattered light \citep{grady2009revealing}. It includes two discernible spiral arms, several asymmetric features, and an inner cavity. High-contrast and high-resolution observations suggest that the observed structures may be attributed to the presence of low-mass exoplanets located within the spiral arms or within the inner cavity \citep{maire2017testing}.\\
-- MWC 758, which is located within the Taurus association, has as estimated age of about 3.5 Myr. Located at about 156 pc \citep{brown2021gaia}, it hosts a protoplanetary disk in the form of a spiral with (at least) three arms \citep{reggiani2018discovery}. Recently, two candidate protoplanets have been proposed based on the post-processing of VLT/SPHERE and LBTI/LMIRCam observations by algorithms dedicated to the detection of point-like sources \citep{reggiani2018discovery,wagner2023direct}. The first one is interior to the spiral (angular separation about 0.11'') and the second one is exterior to the Southern arm (angular separation about 0.62''). According to numerical models, each of these two massive candidate exoplanets would be able to generate the observed spiral arm \citep{wagner2019thermal}. However, the real existence of the spotted candidate exoplanets remains uncertain given the presence of disk material at the location of the candidate exoplanets, that could also lead to misinterpret disk features as point-like sources.\\
-- PDS 70, which is located within the Scorpius-Centaurus association, has an an estimated age of about 5 Myr \citep{muller2018orbital}. Located at about 113 pc \citep{brown2016gaia}, this star is notable for hosting a protoplanetary disk within which two confirmed exoplanets, PDS 70 b and PDS 70 c, are in the process of formation. The exoplanet PDS 70 b was directly imaged using the VLT/SPHERE instrument in near-infrared \citep{keppler2018discovery}, while PDS 70 c was unveiled through observations with the VLT/MUSE instrument in $\text{H}_\alpha$ \citep{haffert2019two}. A third additional candidate exoplanet was also recently detected using JWST observations in the near and mid-infrared \citep{christiaens2024minds}. By harboring multiple nascent exoplanets, this system stands as a unique case. Several structures such as arcs, outer and inner gaps, and potential spiral arms, particularly on the north side of the outer disk were also resolved by direct imaging \citep{riaud2006coronagraphic,keppler2018discovery,mesa2019vlt,juillard2022analysis}.\\
-- HD 163296, which is located within the Sagittarius association, has an estimated age of about 5 Myr. Located at about 101.5 pc \citep{gaia2018gaia}, it hosts a protoplanetary disk with a diameter larger than 1000 au \citep{isella2007millimeter,tilling2012gas,muro2018dust}. Sub-millimeter observations have shown that this disk harbors multiple rings whose structure are due to variations in the gas pressure \citep{teague2018kinematical}. Moreover, multiple asymmetries in the continuum emission have been observed, which supports the hypothesis of the existence of (yet undetected) sub-stellar companions \citep{isella2018disk}. Near infrared observations with VLT/SPHERE allowed to put mass limits of about 3-4 M\textsubscript{Jup} at 30 au, 6-7 M\textsubscript{Jup} between 30 and 80 au, and 2-4 M\textsubscript{Jup} beyond 200 au for such plausible exoplanets \citep{mesa2019determining}.\\
-- AB Aurigae, which is located within the Auriga association, has as estimated age of about 4 Myr. Located at about 163 pc \cite{brown2016gaia}, it hosts a protoplanetary disk with complex spiral features \citep{boccaletti2020possible}. Recently, three candidate point-like sources were identified within the circumstellar environment. Two of them were identified from VLT/SPHERE observations \citep{boccaletti2020possible}. The first one appears very elongated and is embedded within the Southern spiral arm. The second one is located exterior to the Northern spiral arm and is more similar to a point-like source (while being detectable only from SPHERE-IRDIS data and not from SPHERE-IFS data recorded simultaneously). In addition, these two features are detectable both in polarimetry and total intensity, which suggests that they are more likely due to scattering dust particles \citep{boccaletti2020possible}. A third candidate protoplanet was identified by \cite{currie2022images} from SUBARU/SCExAO data. It behaves as a bright emission source at an angular separation of about 0.59'', interior to a dust ring resolved in millimeter observations. However, given that the candidate exoplanet would be at its first stage of formation, likely still accreting material from the disk, it does not appear as a point-like source, but rather as a very elongated pattern, which makes the detection difficult to confirm. Nevertheless, its location and estimated SED would be compatible with model predictions as a driver of the observed spiral arms \citep{currie2022images}.

In Sect. \ref{subsec:recons_simulated_disks}, we quantitatively assess the performance of the proposed algorithm against baseline methods of the field. To this end, we resort to numerical injections of synthetic disks of various morphologies into a real SPHERE-IFS dataset of the HD 122555 star \citep{schutz2005mid, lisse2009abundant}. To the best of our knowledge, no off-axis objects (either point-like sources or disk) have ever been imaged around this star within the SPHERE-IFS field of view \citep{nielsen2008constraints, nielsen2010uniform}. We also generate a synthetic vector of parallactic angles (linearly distributed between 0° and 30°) differing from the experimental value, to vanish out any potential signal from (unknown) real objects.

Finally in Sect. \ref{subsec:importance_spectral_processing}, we consider an additional dataset from the Infrared Dual-band Imager and Spectrograph (IRDIS; \cite{dohlen2008infra}) of the SPHERE instrument. Its dual band mode allows simultaneous imaging at two distinct spectral channels for each individual exposure \citep{vigan2014sphere}. The selected dataset corresponds to the observation of the star SAO 206462. The IRDIS and IFS data of this star were collected simultaneously using the IRDIFS-EXT mode of the SPHERE instrument \citep{beuzit2019sphere}. In our previous work with the ADI version of the REXPACO algorithm \citep{flasseur2021rexpaco}, we processed this dataset but with a mono-spectral approach. In Sect. \ref{subsec:importance_spectral_processing}, we revisit this data with the proposed REXPACO ASDI algorithm to illustrate the benefits of a joint spectral processing.

\begin{figure*}
	\centering
	\includegraphics[width=0.70\textwidth]{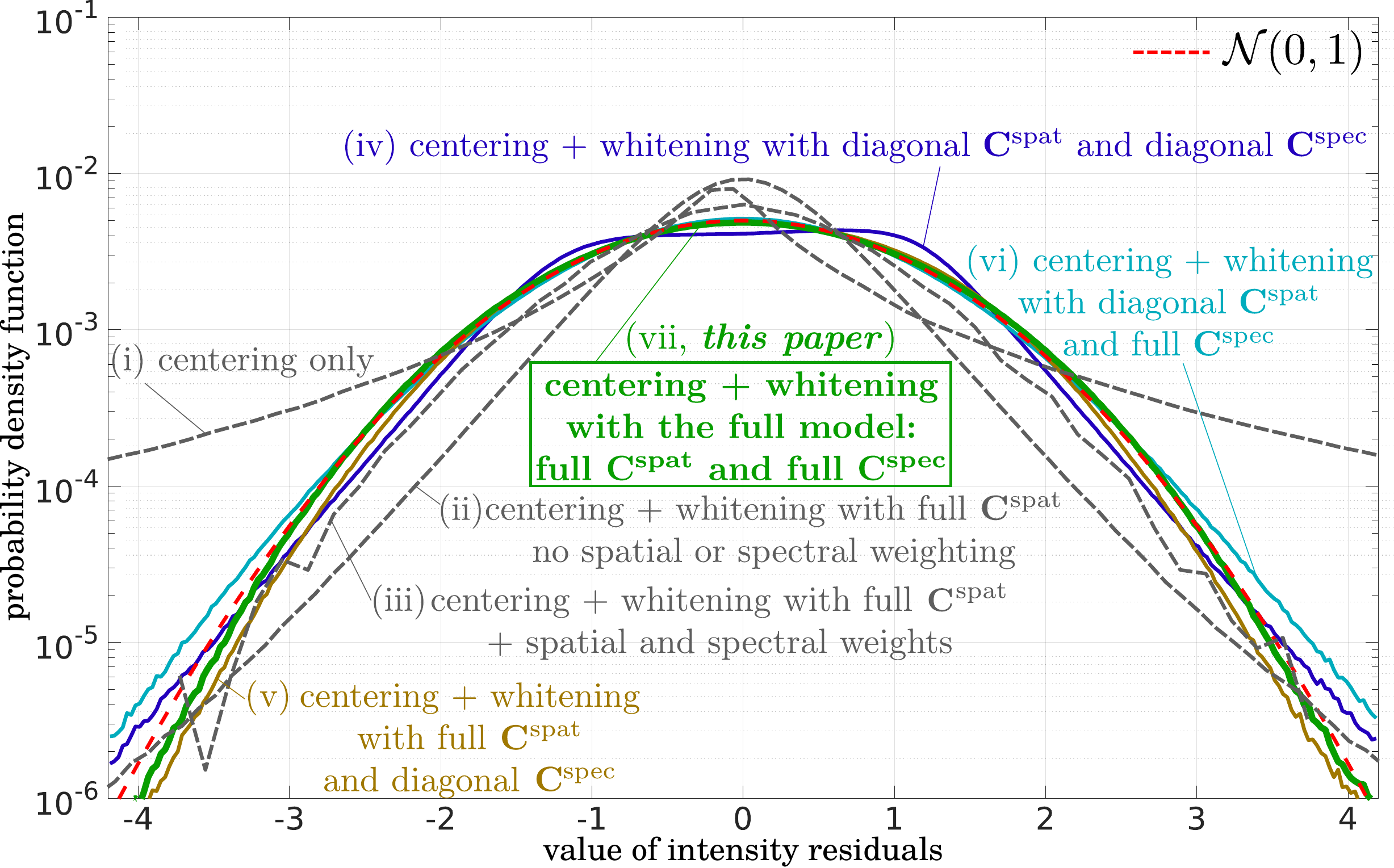}
	\caption{Empirical distributions of the centered and whitened patches averaged over the whole field of view for different covariance models. Data used in this figure contain only the contribution from the nuisance component (i.e., no exoplanet or disk). Dataset: HR 8799 (2015-07-04) with the three known exoplanets masked out, see Table \ref{tab:dataset_logs} for the observation parameters.}
	\label{fig:statval}
\end{figure*}

\begin{figure*}
	\centering
	\includegraphics[width=\textwidth]{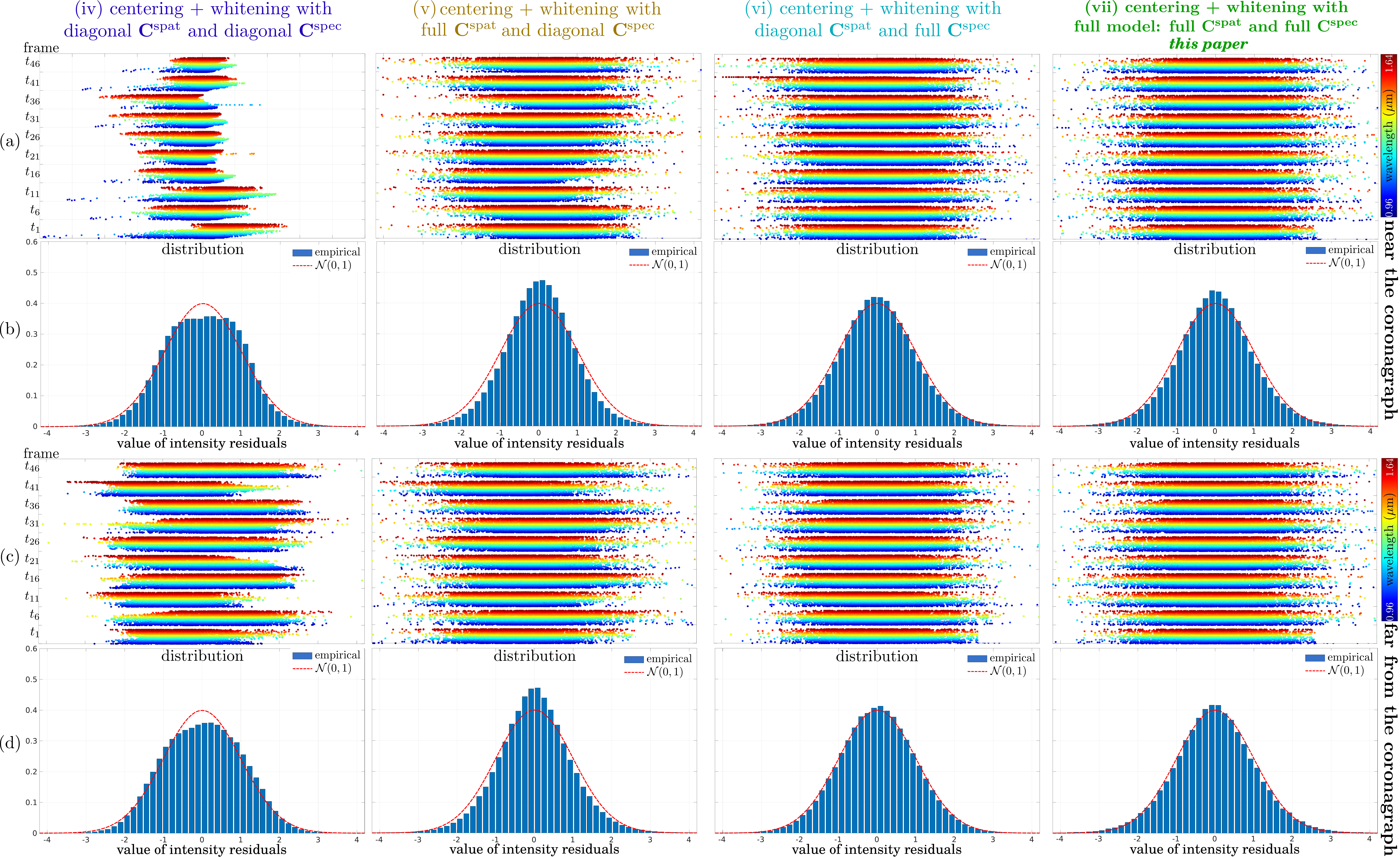}
	\caption{(figure modified) Empirical distributions of the centered and whitened patches for different covariance models: diagonal spatial and spectral covariances in the first column; full spatial covariance, diagonal spectral covariance, and temporal weighting in the second column; full spectral covariance, diagonal spatial covariance, and temporal weighting in the third column; and finally in the last column the full spatio-spectral separable model introduced in this work. The reported empirical distribution are computed locally: a location selected at a small angular separation in rows (a) and (b); a location selected at a larger angular separation in rows (c) and (d). The corresponding global empirical distributions computed over the whole field of view are given in Fig. \ref{fig:statval}. Patches represented in this figure contain only the contribution from the nuisance component (i.e., no exoplanet or disk). Dataset: HR 8799 (2015-07-04) with the three known exoplanets masked out, see Table \ref{tab:dataset_logs} for the observation parameters.}
	\label{fig:statisticalval_big}
\end{figure*}

\medskip

\noindent All datasets were calibrated and assembled from SPHERE raw data using the pre-reduction and handling pipeline of the SPHERE consortium \citep{pavlov2008sphere}. During this step, background, flat-field, bad pixels, registration, true-North, wavelength and astrometric calibrations are performed. These standard pre-processing steps are followed by additional refinements implemented at the SPHERE Data Center \citep{delorme2017sphere}, aimed at reducing cross-talk, enhancing bad pixel correction, and mitigating spectral cross-talk effects.

Table \ref{tab:dataset_logs} summarizes the main observation parameters associated to each dataset.

\subsection{Validation of the statistical model of the nuisance component}
\label{subsec:validation_model}

Before evaluating the reconstruction method on high-contrast observations of circumstellar disks, we aim to show that our statistical model of the nuisances is relevant. We use the same ASDI dataset (HR 8799, 2015-07-04) as in \cite{flasseur2020paco} with the three known exoplanets within the SPHERE-IFS field of view masked out so that the resulting data correspond only to the nuisance term. Following a similar analysis as in \cite{flasseur2020paco}, Fig. \ref{fig:statval} displays the empirical distribution of all patches in the field of view after performing different post-processing.
If random vectors $\V v_n$ are accurately modeled by a Gaussian distribution with mean $\V\mu_n$ and covariance $\M C_n$, as described in Eq. (\ref{eq:gauss_patch}), the centered and whitened vectors $\M C_n^{-1/2}(\V v_n-\V\mu_n)$ should follow $\mathcal{N}(\V 0,\M I)$, corresponding to the red dashed line in Fig. \ref{fig:statval}. We thus compare a standard Gaussian distribution with the empirical marginal distribution of
$\M C_n^{-1/2}(\V v_n-\V\mu_n)$ for several covariance models: the three models considered in \cite{flasseur2020paco}, drawn in gray dashed-lines: (i) no covariance ($\M C_n=\M I$); (ii) only spatial covariances; (iii) spatial covariances plus  temporal and spectral weighting; and four additional models: (iv) diagonal spatial and spectral covariances (i.e., spatial, spectral, and temporal weighting via a separable model); (v) full spatial covariance, diagonal spectral covariance, and temporal weighting; (vi) full spectral covariance, diagonal spatial covariance, and temporal weighting; and finally (vii) the full separable model introduced in this paper, see Eq. (\ref{eq:covmodel}).
As shown by Fig. \ref{fig:statval}, the full spatio-spectral separable model (green curve) provides the best fit to the empirical distribution (i.e., the green curve closely matches the red dashed line of the standard Gaussian distribution). This justifies the use of the full spatio-spectral separable model in our loss function $\mathscr{L}_n$.
As an average trend over the field of view, we also observe that neglecting spatial covariances is more detrimental than ignoring spectral covariances, as model (v) better approximates $\mathcal{N}(\V 0, \M I)$ than model (vi). Figure \ref{fig:statisticalval_big} completes this study with a more localized examination of the empirical distribution of patches for models (iv)-(vii) across two nuisance regimes: (1) a regime near the star where speckles dominate, and (2) a regime at larger separations where stochastic noise prevails. A similar representation was provided in Fig. 4 of
\cite{flasseur2020paco} on this dataset for three additional models considered
in our previous work for exoplanet detection in angular and spectral
differential imaging: (i) no covariance; (ii) spatial covariances only; and (iii)
spatial covariances with temporal and spectral weighting. Based on this analysis, the full spatio-spectral separable model introduced in this paper is the most effective at statistically describing the fluctuations of the nuisance component across both noise regimes (i.e., regardless of the distance to the star). Notably, in both models, the empirical distributions of centered and whitened patches more closely follow a Gaussian law with zero mean and unit variance far from the star than near the star. This is to be expected, given that the nuisance is stronger, more correlated, and fluctuates more in the vicinity of the star than farther away; see Fig. \ref{fig:covmat}.

\begin{figure*}
	\centering
	\includegraphics[width=\textwidth]{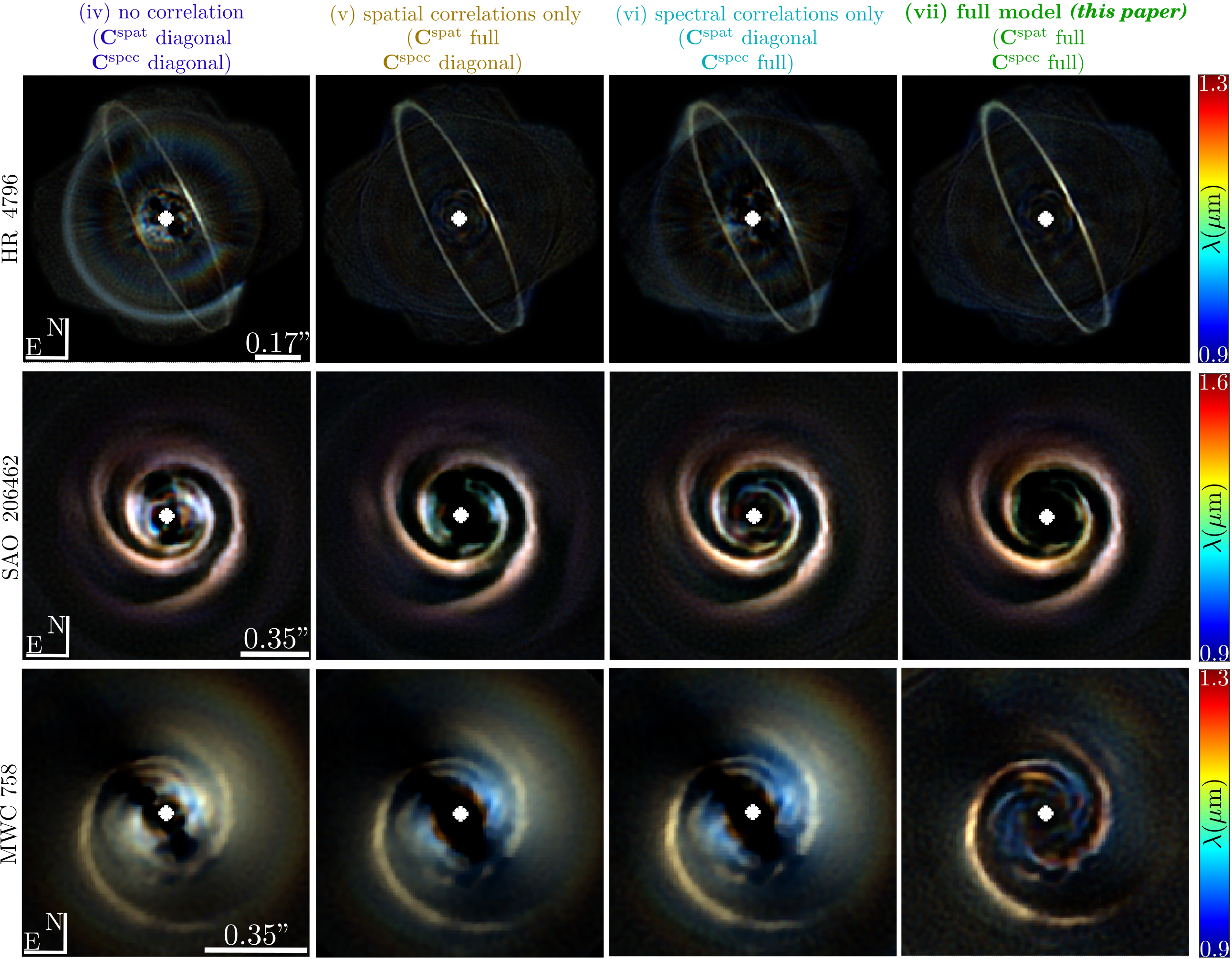}
	\caption{Ablation study: neglecting spatial and/or spectral correlations of the nuisance component significantly degrades reconstruction quality. The displayed images represent the deconvolved reconstructions $\widetilde{\obj}$. Pseudo-color images are shown with colors selected to cover the infrared spectrum according to the colormaps on the right. Datasets: HR 4796 (2015-02-03), SAO 206462 (2015-05-15) and MWC 758 (2018-12-17), see Table \ref{tab:dataset_logs} for the observation parameters.}
	\label{fig:ablationstudy}
\end{figure*}

We conclude this ablation study by showing how the reconstruction results are impacted if simpler covariance models are considered rather than the full model of Eq. (\ref{eq:covmodel}). Figure \ref{fig:ablationstudy} displays examples on real data of the reconstructed disk component for the same four nuisance models as in Fig. \ref{fig:statisticalval_big} (i.e., models (iv)-(vii)). The datasets of HR 4796 and MWC 758 suffer from a strong nuisance component. Ignoring the spatial correlations leads to severe artifacts: a ghost circular structure is reconstructed and contaminates a large fraction of the field of view. For SAO 206462 and PDS 70, close inspection of the central region reveals spurious structures in all reconstructions except those obtained with the full model (vii) of the nuisance.
While ignoring spectral correlations is also harmful (e.g., a bright nuisance halo remains around the MWC 758 disk), its effect is less pronounced compared to omitting spatial correlations, aligning with the findings in Fig. \ref{fig:statval}, where empirical residual distributions were analyzed across the whole field of view.
These qualitative observations emphasize again the value of accurately modeling the nuisance's spatial and spectral correlations to improve the reconstruction quality.

\medskip

\begin{figure}
	\centering
	\includegraphics[width=0.5\textwidth]{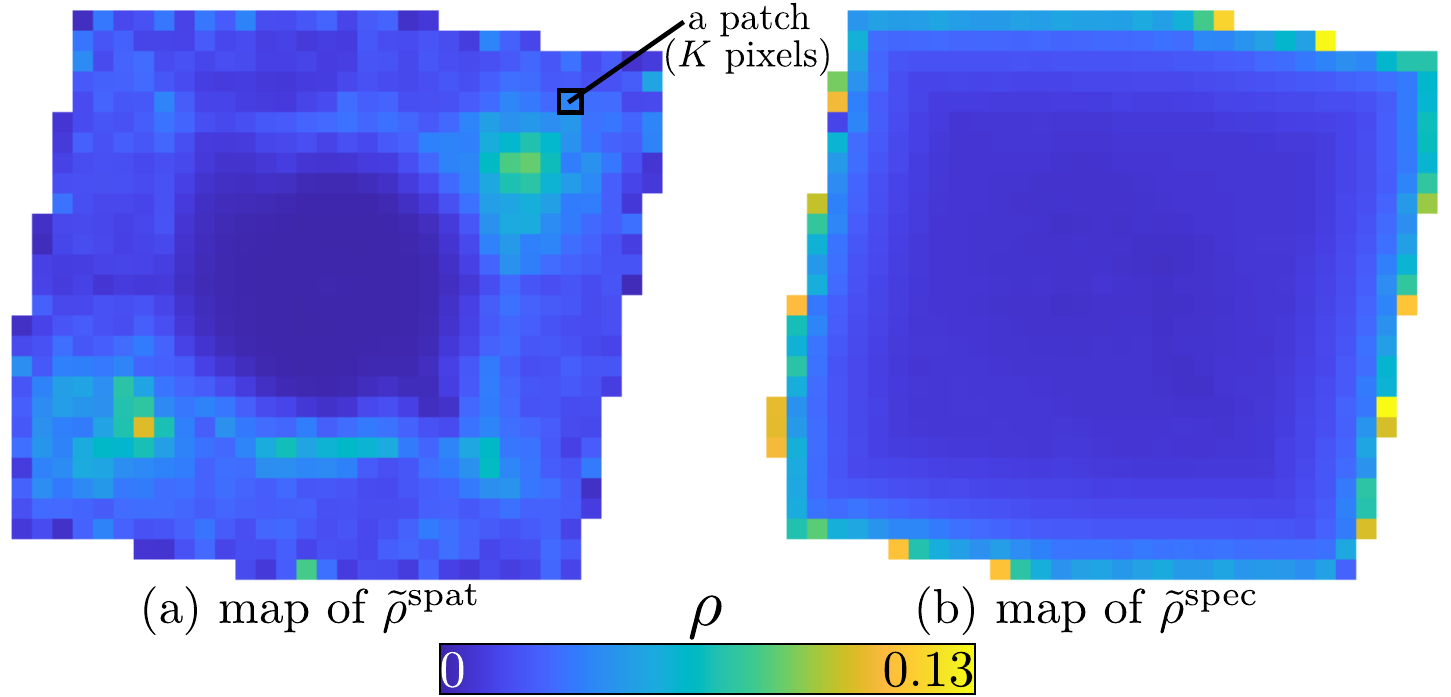}
	\caption{Spatial distribution of the spatial and spectral shrinkage parameters. Dataset: HR 8799 (2015-07-04), see Table \ref{tab:dataset_logs} for the observation parameters.}
	\label{fig:shrinkmaps}
\end{figure}

\noindent The shrinkage parameters $\widetilde{\rho}_n^\spat$ and $\widetilde{\rho}_n^\spec$ can significantly influence the statistical model. We monitor their values by displaying maps of the spatial and spectral shrinkage parameters in Fig. \ref{fig:shrinkmaps} for the dataset shown in Fig. \ref{fig:data}. Values of $\widetilde{\rho}_n^\spat$ and $\widetilde{\rho}_n^\spec$ remain relatively low (below 0.13), suggesting a moderate bias towards zero and indicating that the off-diagonal sample covariances are only slightly attenuated by the shrinkage. Spectral shrinkage intensifies at the edges of the field of view, where fewer samples are available due to spectral scaling (i.e., $L_{\text{eff}} \le L$). Conversely, spatial shrinkage is stronger at some locations of the field of view for this dataset, illustrating that a uniform shrinkage value across the entire field of view would be sub-optimal.

\subsection{Qualitative analysis: reconstruction of disks from SPHERE-IFS data}
\label{subsec:recons_real_disks}

\begin{table}
	\centering
	\caption{Number of modes optimized for PCA ASDI reconstructions.}
	\begin{tabular}{cccc}
				\toprule
				& \multicolumn{3}{c}{\textit{Known real disks, see Sect. \ref{subsec:recons_real_disks}}}\\
				\midrule
				\midrule
				HR 4796 & \multicolumn{3}{c}{18}\\
				SAO 206462 & \multicolumn{3}{c}{6}\\
				MWC 758 & \multicolumn{3}{c}{4}\\
				PDS 70 & \multicolumn{3}{c}{14}\\
				HD 163296 & \multicolumn{3}{c}{42}\\
				AB Aurigae & \multicolumn{3}{c}{20}\\
				\midrule
				\midrule
				& \multicolumn{3}{c}{\textit{Synthetic disks, see Sect. \ref{subsec:recons_simulated_disks}}}\\
				 & $\alpha_{\text{gt}} = 1 \times 10^{-6}$ & $\alpha_{\text{gt}} = 5 \times 10^{-6}$ & $\alpha_{\text{gt}} = 1 \times 10^{-5}$\\
				\midrule
				\midrule
				elliptical disk & 14 & 4 & 4\\
				circular disk & 26 & 10 & 4\\
				spiral disk & 26 & 12 & 4\\
				\bottomrule
	\end{tabular}
	\label{tab:pca_number_of_modes}
\end{table}

\begin{figure*}
	\centering
	\includegraphics[width=\textwidth]{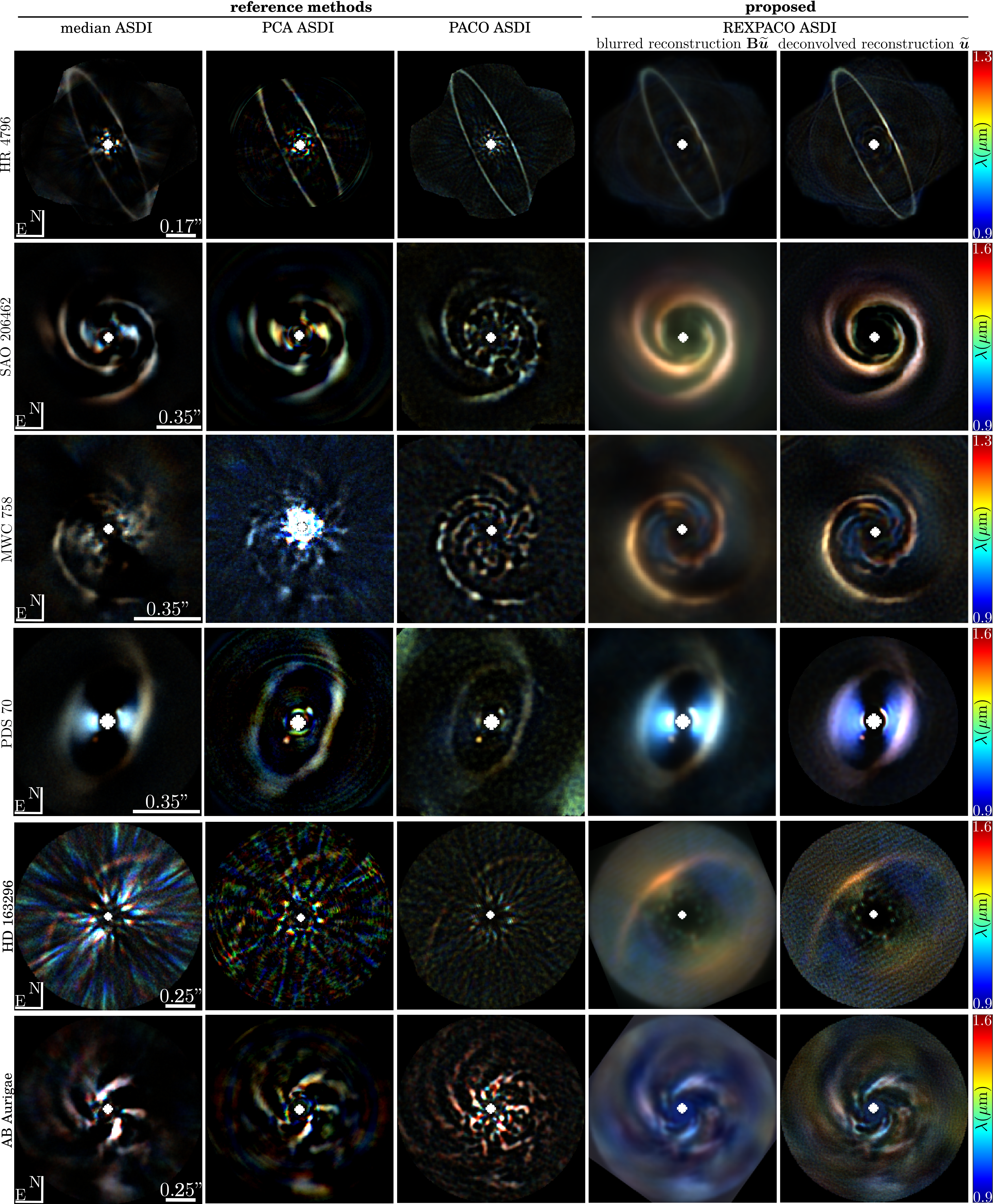}
	\caption{Disk reconstruction results from real data obtained with the SPHERE-IFS instrument. The images in the fourth column show the outputs of the proposed REXPACO ASDI method, which have been re-blurred for direct comparison with the reference methods (i.e., these correspond to $\M B \, \widetilde{\obj}$). The deblurred reconstructions $\widetilde{\obj}$ are displayed in the last column (corresponding to the last column of Fig.  \ref{fig:ablationstudy}). Pseudo-color images are displayed as in Fig. \ref{fig:ablationstudy}. Datasets: HR 4796 (2015-02-03), SAO 206462 (2015-05-15), MWC 758 (2018-12-17), PDS 70 (2018-02-24), HD 163296 (2018-05-07) and AB Aurigae (2020-01-18), see Table \ref{tab:dataset_logs} for the observation parameters.}
	\label{fig:realdisks}
\end{figure*}

Having established the benefits of the proposed statistical model, we now apply it to the six SPHERE-IFS datasets presented in Sect. \ref{subsec:datasets}, which correspond to observations of
stars hosting known circumstellar disks with diverse morphological structures. These include SAO 206462 (already shown in Fig. \ref{fig:data}) and MWC 758, both featuring a spiral disk; HR 4796, which hosts a thin elliptical disk; and PDS 70, AB Aurigae and HD 163296, each hosting a protoplanetary disk of complex shape and several candidate or confirmed exoplanets
 in formation within the surrounding gas and dust material.
Figure \ref{fig:realdisks} presents reconstructions produced by various reference
methods alongside those obtained with our method. As the other
methods do not perform a deconvolution, we show in the fourth column of Fig.
\ref{fig:realdisks} our reconstruction \textit{re-blurred} at the resolution of the
instrument (i.e., $\M B\widetilde{\obj}$ instead of $\widetilde{\obj}$).
Based on code availability,
 three standard methods
were selected for comparison: (i) median ASDI \citep{sparks2002imaging,marois2006angular,thatte2007very} which estimates the
nuisance component by temporally and spectrally stacking the observations using medians, (ii) PCA ASDI \citep{soummer2012detection,amara2012pynpoint,christiaens2019separating}
 which employs principal component analysis to
remove the nuisance component, and (iii) PACO ASDI \citep{flasseur2020paco}
originally developed for exoplanet detection from ASDI datasets but also capable of
partially reconstructing thin disks, see Sect. \ref{sec:introduction}. For median ASDI and PCA ASDI, we used the Vortex Image Processing (VIP; \cite{gonzalez2017vip,christiaens2023vip}) package\footnote{See \href{https://github.com/vortex-exoplanet/VIP}{https://github.com/vortex-exoplanet/VIP}.}, whereas we employed our unsupervised
pipeline\footnote{See \href{http://doi.org/10.5281/zenodo.3679426}{http://doi.org/10.5281/zenodo.3679426} for a frozen implementation.} for PACO ASDI \citep{flasseur2020paco}. The number of modes in PCA ASDI has been manually optimized, with the selected value being constant across all angular separations (i.e., we applied so-called \textit{full frame} PCA ASDI).
In practice, we evaluated all possible mode numbers (in increments of two). For experiments involving a synthetic disk with a known ground truth flux distribution $\V u_{\text{gt}}$, we selected the number of modes that minimizes the MSE between the estimate $\widetilde{\V u}$ and the ground truth $\V u_{\text{gt}}$. For real disks, we visually selected the optimal number of modes to best preserve fine structures while effectively removing most of the stellar leakage. Table \ref{tab:pca_number_of_modes} summarizes the number of modes for PCA ASDI reconstructions of the disks analyzed in this paper. For the other hyper-parameters of median ASDI and PCA ASDI, we used default values provided within VIP.
The reconstructions obtained
with the reference methods all suffer from noticeable artifacts, particularly at the
center of the field of view where the reduced angular diversity makes it challenging to disentangle the components. In comparison, both the blurred and the deblurred
reconstructions shown in the last two columns of Fig. \ref{fig:realdisks} are
far more satisfactory. The REXPACO ASDI reconstruction of the HR 4796 disk displays a near-continuous elliptical structure and a flux asymmetry on the West side of the ring, consistent with the predictions of  intensity scattering models, see \cite{milli2017near}. For the SAO 204642, the REXPACO ASDI reconstruction exhibits two main spiral arms whose overall morphology and spatial extent are in good agreement with radiative transfer and hydro-dynamical models  of transitional disks shaped by giant planets, which are responsible for sculpting multiple spiral arms, see e.g. \cite{bae2016planetary, maire2017testing}.
Additionally, the REXPACO ASDI
reconstructions of HR 4796 (respectively, SAO 206462, MWC 758, PDS 70, HD 163296) can be qualitatively compared
 with the reconstructions in Fig. 4 of \cite{milli2017near} (respectively, Fig. 1
bottom-left of \cite{maire2017testing}, Fig. A.1 second line of \cite{boccaletti2021investigating}, Fig. 1 first line-second row of \cite{mesa2019vlt}, Fig. 1 right of \cite{mesa2019determining}). These results were derived from custom routines of respectively median ASDI, RDI ADI, median ASDI, PCA ASDI, and PCA ASDI applied to the same datasets.
 The REXPACO ASDI reconstructions exhibit significantly fewer
artifacts, such as non-physical discontinuities in the disk structures and
residuals stellar leakages near the star.
The deconvolution step in the
proposed method also enhances the spatial resolution of thin disk structures. In contrast, baseline methods like median ASDI and PCA ASDI tend to subtract part of the disk
component when removing the nuisance term. This leads to substantial flux biases and
a high-pass filtering effect. PACO ASDI, being optimized for point-like detections, manages to recover parts of the disks in large gradient areas. It is much
more successful on the thin disk of HR 4796 and on the extended disk of PDS 70 than on the thicker spiral disk of
SAO 206462 and of MWC 758.
Finally, the multi-spectral REXPACO ASDI reconstructions in Fig. \ref{fig:realdisks} can
be compared to the mono-spectral reconstructions produced by the REXPACO ADI
algorithm \citep{flasseur2021rexpaco} (see Fig. 11 of \citep{flasseur2021rexpaco})
on mono-spectral datasets of the same target stars (excepted MWC 758, HD 163296, AU Aurigae). These mono-spectral
datasets were recorded using the InfraRed Dual Imaging Spectrograph (IRDIS) of
the SPHERE instrument, operating simultaneously with the IFS but in a different
spectral band and resolution. The joint multi-processing leads to a better rejection of the
nuisance component, thereby reducing non-physical reconstruction artifacts such as
discontinuities, especially within spiral arms. These comparisons illustrate that joint processing of multi-spectral datasets is
particularly beneficial for disks having a circular symmetry, such as SAO 206462 or MWC 758,
as it helps to disentangle the disk light from the stellar light. This is because these two components do not always superimpose  due to the chromatic scaling of speckles induced by ASDI. The advantages of joint spectral processing are further explored and discussed in Sect. \ref{subsec:importance_spectral_processing}.

\begin{figure*}
	\centering
	\includegraphics[width=0.75\textwidth]{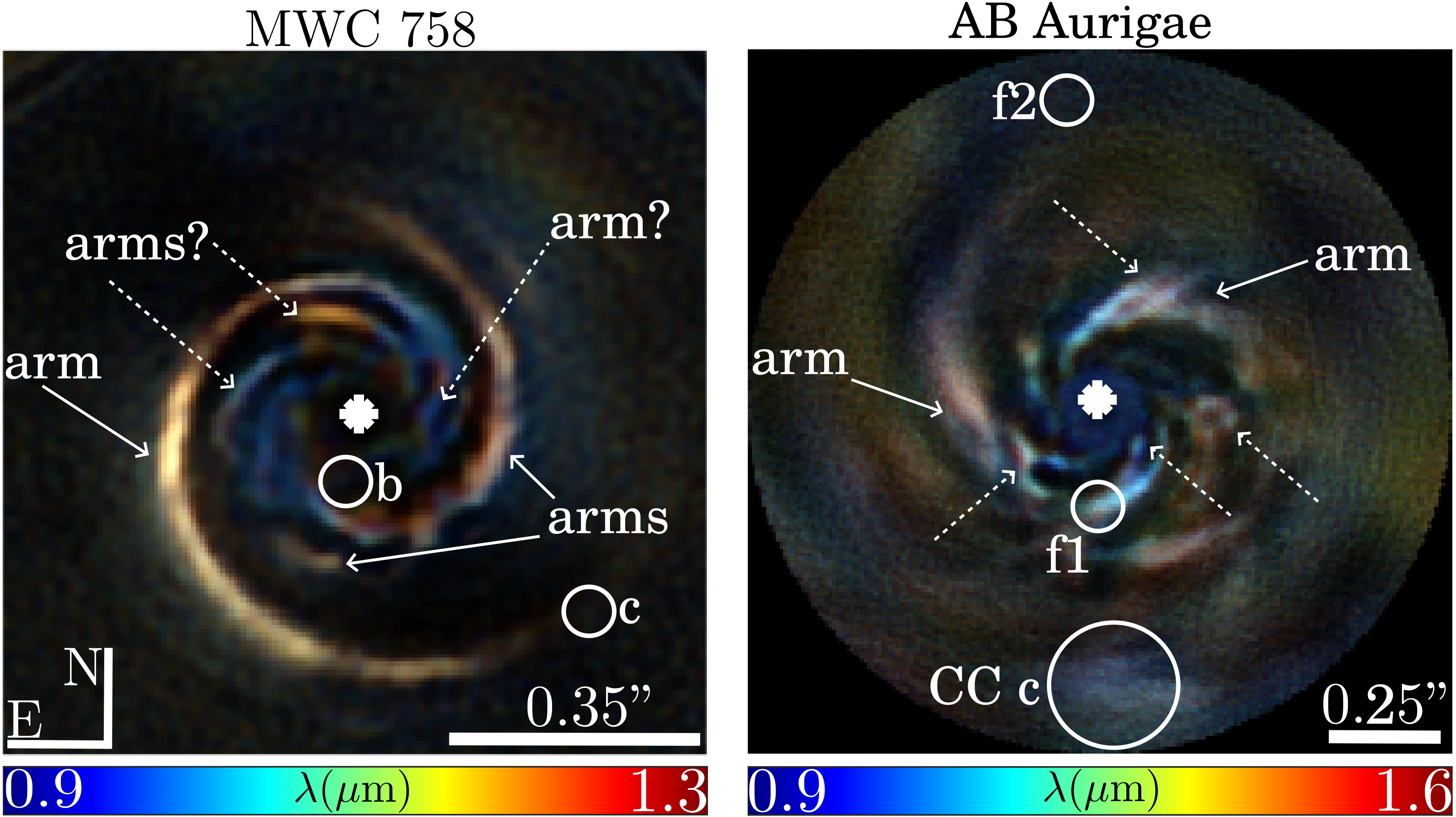}
	\caption{Disk reconstructions $\widetilde{\obj}$ obtained with REXPACO ASDI: same results as in Fig. \ref{fig:realdisks} (last column) for the protoplanetary disks MWC 758 and AB Aurigae. The main disk features reported in the literature are overlaid with straight lines and (candidate) point-like sources also identified in the literature (not only based on data of the same instrument) are highlighted with circles. Newly identified candidate disk features from our reconstructions are indicated with dashed lines, see text. Datasets: MWC 758 (2018-12-17) and AB Aurigae (2020-01-18), see Table \ref{tab:dataset_logs} for the observation parameters.}
	\label{fig:realdisks_rexpaco_asdi}
\end{figure*}

Figure \ref{fig:realdisks_rexpaco_asdi} focuses on protoplanetary disks MWC 758 and AB Aurigae reconstructed with the proposed REXPACO ASDI algorithm. Known disk features and (candidate) point-like sources reported in the literature, as well as new disk features identified through our reconstructions, are overlaid.
For MWC 758, the three spiral arms identified by \cite{wagner2019thermal} (highlighted with solid arrows) are well reconstructed by REXPACO ASDI. We also reconstruct two additional elongated structures interior to the Northern main spiral arm. These features could be interpreted as additional spiral arms and they appear connected to the main spiral arms by material bridges. None of the two point-like sources (b and c) identified by \cite{reggiani2018discovery,wagner2023direct} are detected in our reconstruction. This may be due to the VLT/SPHERE-IFS observations being taken in the Y-J spectral band, whereas the two exoplanets were discovered using Keck/NIRC2 and LBTI/LMIRCam observations in the L' and M' bands, where contrast for such candidate sources is more favorable.
For AB Aurigae, REXPACO ASDI reconstructs the two main spiral arms previously identified by \cite{boccaletti2020possible}. We also identify additional complex structures such as gaps and splittings within the main spiral arms. Consistent with \cite{boccaletti2020possible}, we detect a bright emission source (f1) embedded within the Southern spiral arm, though it appears very extended, suggesting that it is part of the disk. Like \cite{boccaletti2020possible}, we do not detect the Northern point-like source (f2) from this SPHERE-IFS dataset. It can be also noted that point-like source f2 were identified by \cite{boccaletti2020possible} at the same epoch, but from a dataset obtained with the SPHERE-IRDIS instrument, operating simultaneously to SPHERE-IFS. We also clearly detect the Northern bright emission source (CC c) identified by \cite{currie2022images} from SUBARU/SCExAO data. However, CC c does not appear as a point-like source in our reconstruction, likely because this candidate exoplanet, if real, would be at its first stage of formation, still accreting material from the disk. The SPHERE-IFS wavelengths being shorter than on SUBARU/SCExAO, is it also possible that the point sources are beyond reach at these wavelengths with SPHERE-IFS.

\subsection{Quantitative analysis: reconstruction of synthetic disks injected into SPHERE-IFS data}
\label{subsec:recons_simulated_disks}

\begin{figure*}
	\centering
	\includegraphics[width=\textwidth]{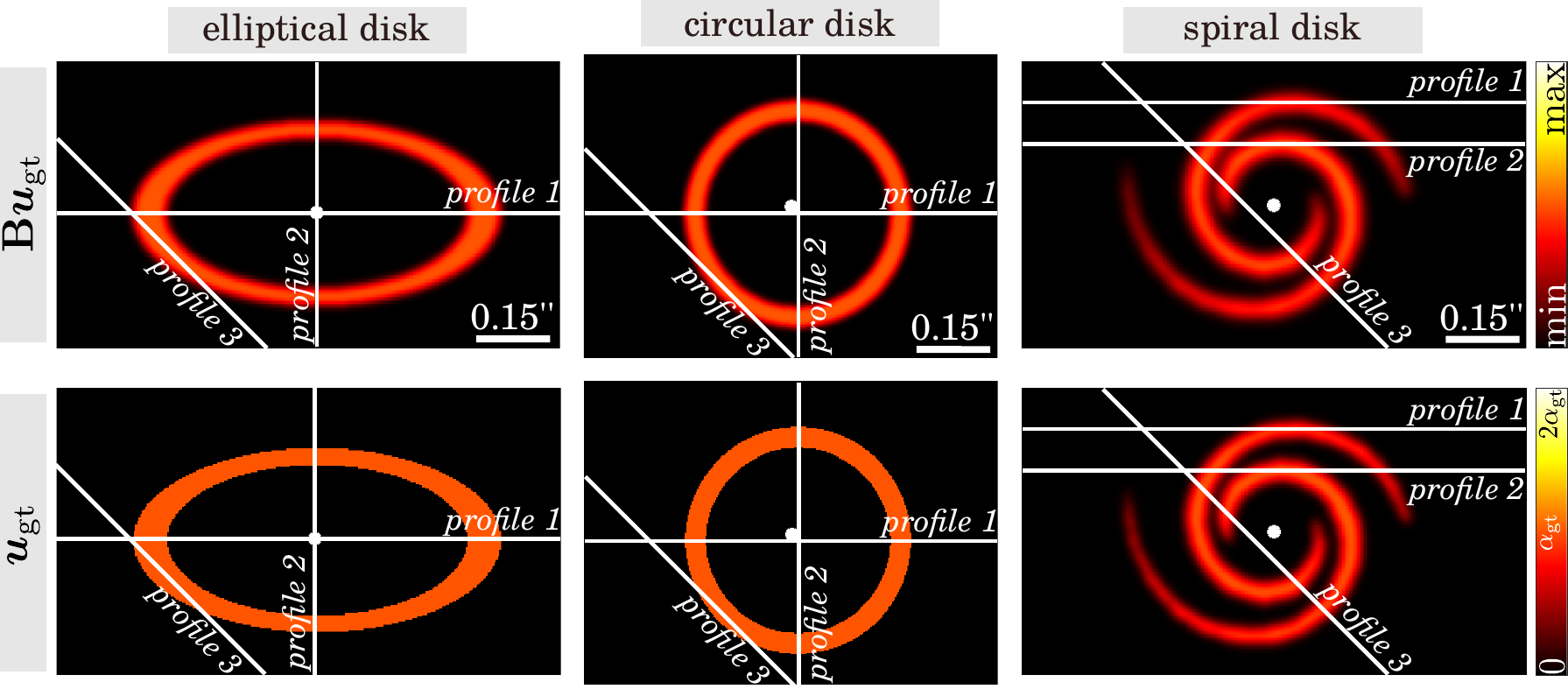}
	\caption{Ground truth images for three synthetic disks: an elliptical disk, a circular disk, and a spiral disk. The first line gives the contribution $\M B \, \obj_{\text{gt}}$ of the disks (i.e., blurred by the off-axis PSF to be at the same spatial resolution than the instrument) that are injected within the data. The second line gives the flux distribution $\obj_{\text{gt}}$ free from the blur introduced by the off-axis PSF. For each disk, three slice-cuts are defined (denoted by \textit{profile 1, 2, 3}) for 1D visualization of the reconstructed flux performed in Figs. \ref{fig:ellipse_cuts_fullfig}, \ref{fig:circle_cuts_fullfig}, \ref{fig:spiral_cuts_fullfig}, and  \ref{fig:adi_vs_asdi_synthetic_disks_fullfig_cuts_only} of Appendix \ref{app:additional_results}.}
	\label{fig:gt_fullfig}
\end{figure*}

In this section, we quantitatively assess the performance of the proposed approach in comparison to three baseline methods: median ASDI, PCA ASDI and PACO ASDI. The general principles of these approaches are outlined in Sect. \ref{sec:introduction}, and their specific settings are detailed in \ref{subsec:recons_real_disks}.

We consider three simulated disks representative of common morphologies in high-contrast observations: (i) a spatially centered elliptical disk with sharp edges and with an eccentricity of about 0.80; (ii) a circular disk with sharp edges and whose center is shifted by five pixels from the star center in the two spatial dimensions; (iii) a spiral disk exhibiting two arms with smooth edges. Figure \ref{fig:gt_fullfig} illustrates the ground truth flux distribution for for each of these disk types used in this analysis.

While these toy models were not generated using physics-based simulators (e.g., modeling the hydrodynamics and radiative transfer), cases (i) and (ii) typically correspond to debris disks while case (iii) resembles a particular instance of transition or protoplanetary disks. Additionally, it can be noted that these synthetic disks resemble the real circumstellar disks reconstructed in Fig. \ref{fig:realdisks} so that these simulations can help to assess the quality of the reconstructions of these real circumstellar disks: the elliptical disk (i) has a spatial extent similar to the HR 4796 disk, and the spiral disk (iii) has similar spatial extent and morphology to the SAO 206462 disk.

Each simulated disk is injected into the HD 172555 dataset (which contains no known off-axis source), at three different contrast levels $\alpha_{\text{gt}} \in \lbrace 1\times 10^{-6}, 5\times 10^{-6}, 1\times 10^{-5} \rbrace$. For our simulations, we consider \textit{gray objects}, meaning the contrast is constant across the spectral band, resulting in an identical flux distribution across all spectral channels. Consequently, all reconstructions presented in this section are averaged over the whole spectral band. A total of 90 semi-synthetic datasets have been generated: for each disk type and contrast level $\alpha_{\text{gt}}$, the simulated disk has been injected at ten different orientations relative to the nuisance component (which remains the same for all simulations). This simulation protocol allows us to evaluate the mean and variance of the reconstructions.

\begin{figure*}
	\centering
	\includegraphics[width=\textwidth]{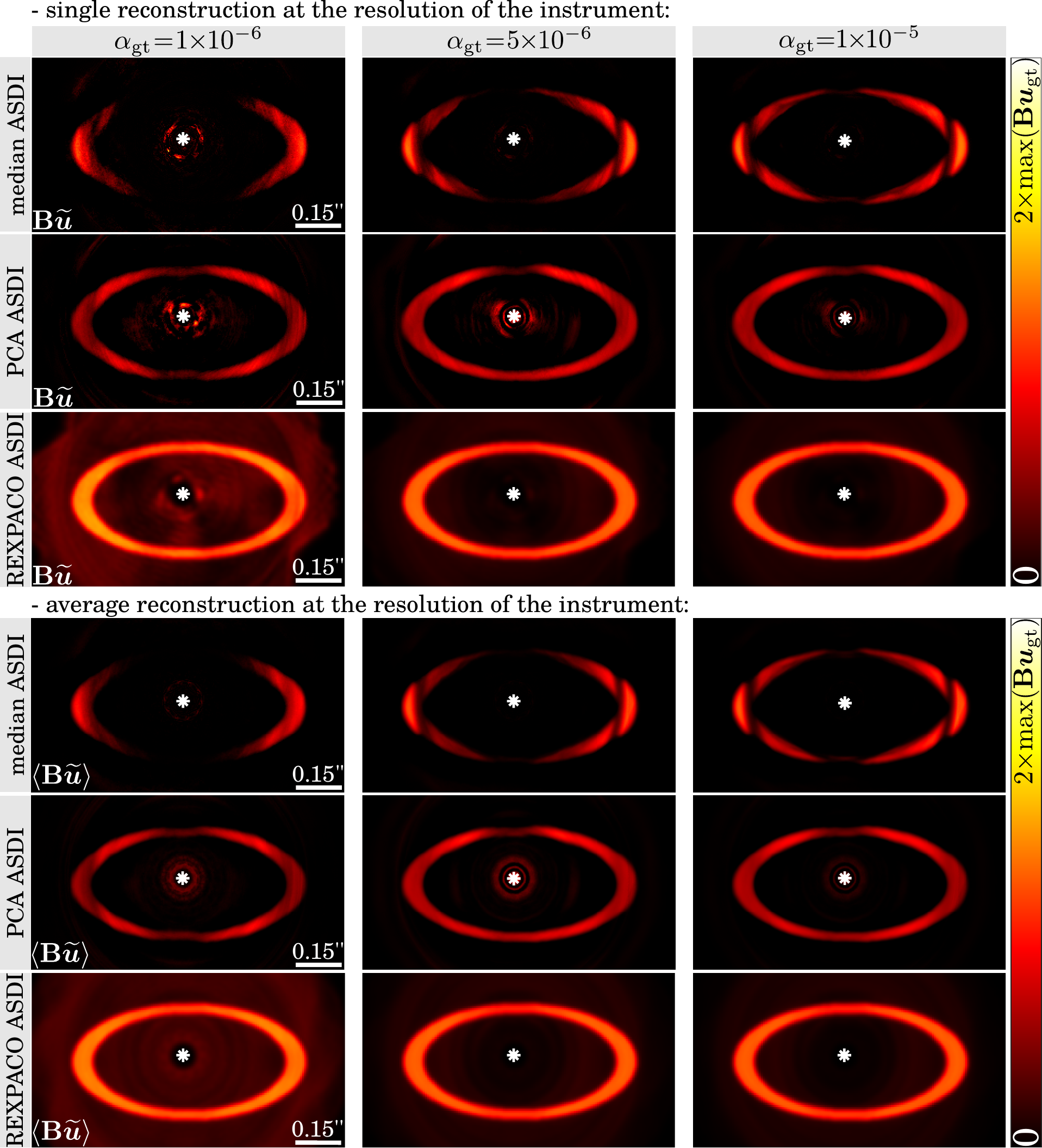}
	\caption{Reconstructions of simulated elliptical disks: comparisons between median ASDI, PCA ASDI and re-blurred REXPACO ASDI reconstructions. Single reconstructions (associated to a selected orientation of the disk with respect to the nuisance) and the average reconstructions (over ten different injections of the same disk but with various orientations with respect to the nuisance) are displayed. The three columns correspond to the three considered levels of contrast: $\alpha_{\text{gt}} \in \lbrace 1\times 10^{-6}, 5\times 10^{-6}, 1\times 10^{-5} \rbrace$. Dataset: HD 172555 (2015-07-11), see Table \ref{tab:dataset_logs} for the observation parameters.}
	\label{fig:ellipse_blurred_fullfig}
\end{figure*}

\begin{figure*}
	\centering
	\includegraphics[width=\textwidth]{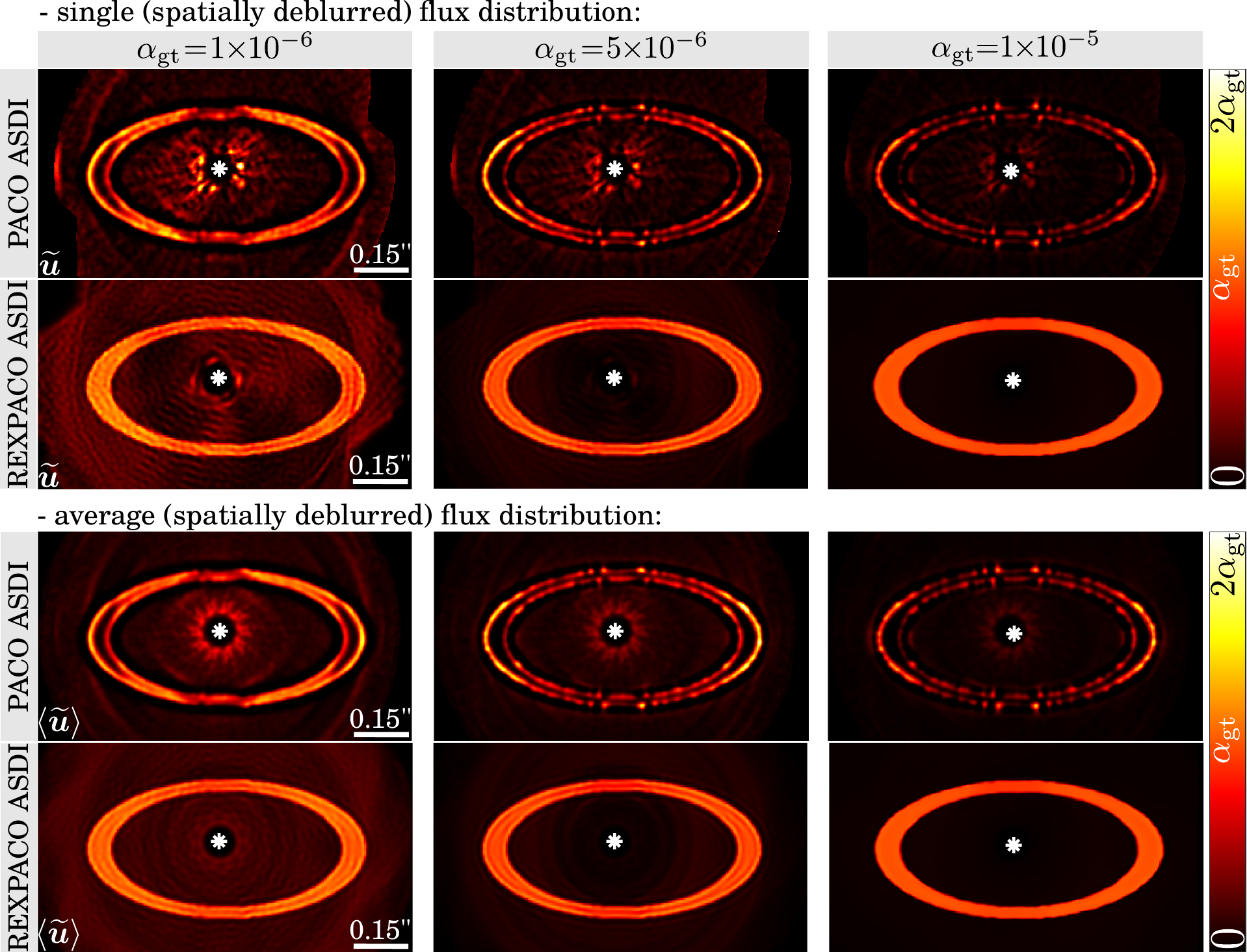}
	\caption{Reconstructions of simulated elliptical disks: comparisons between PACO ASDI and deconvolved REXPACO ASDI reconstructions. Single reconstructions (associated to a selected orientation of the disk with respect to the nuisance) and the average reconstructions (over ten different injections of the same disk but with various orientations with respect to the nuisance) are displayed. The three columns correspond to the three considered levels of contrast: $\alpha_{\text{gt}} \in \lbrace 1\times 10^{-6}, 5\times 10^{-6}, 1\times 10^{-5} \rbrace$. Dataset: HD 172555 (2015-07-11), see Table \ref{tab:dataset_logs} for the observation parameters.}
	\label{fig:ellipse_deblurred_fullfig}
\end{figure*}

\begin{figure*}
	\centering
	\includegraphics[width=0.92\textwidth]{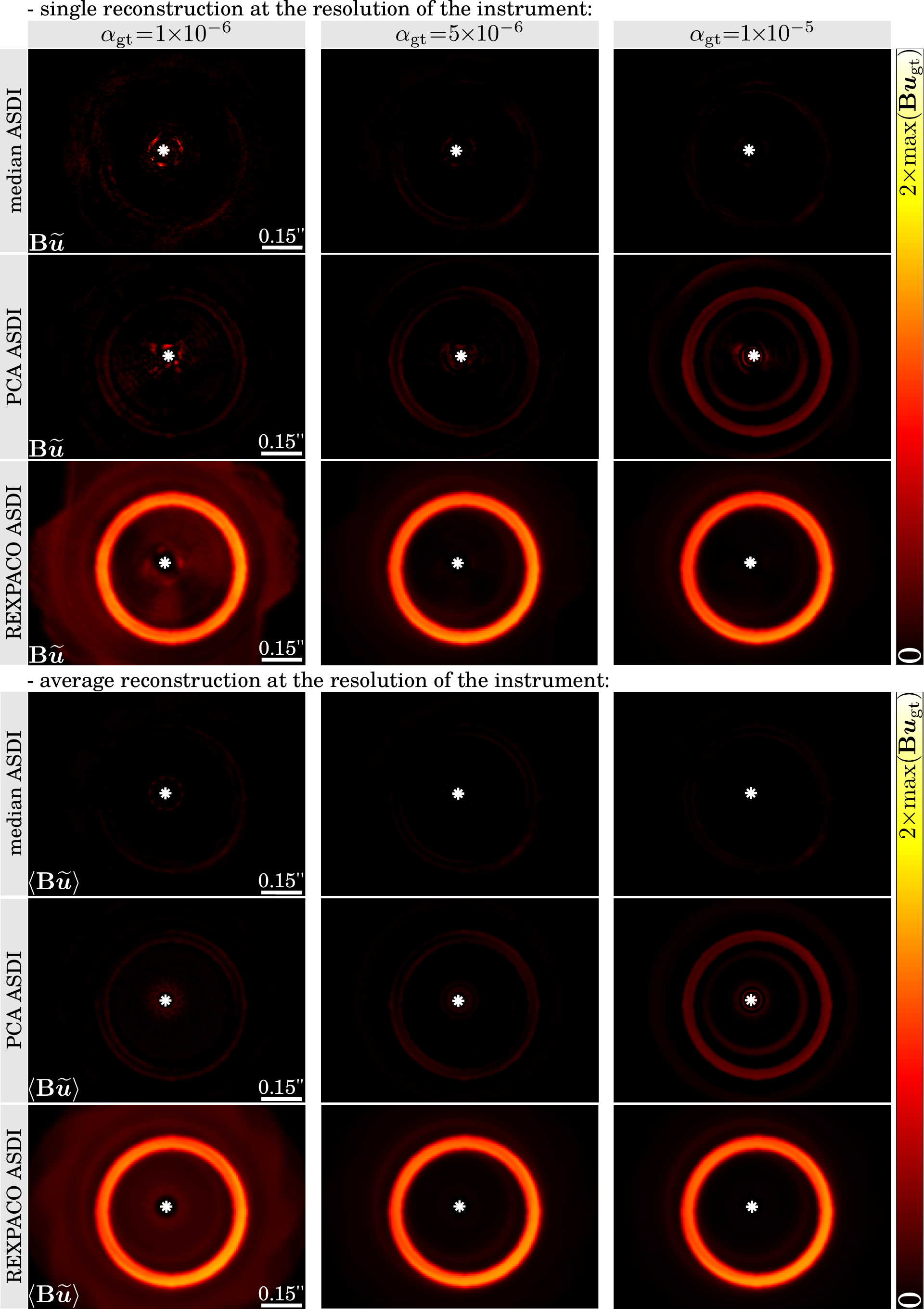}
	\caption{Same than Fig. \ref{fig:ellipse_blurred_fullfig} for synthetic circular disks.}
	\label{fig:circle_blurred_fullfig}
\end{figure*}

\begin{figure*}
	\centering
	\includegraphics[width=0.92\textwidth]{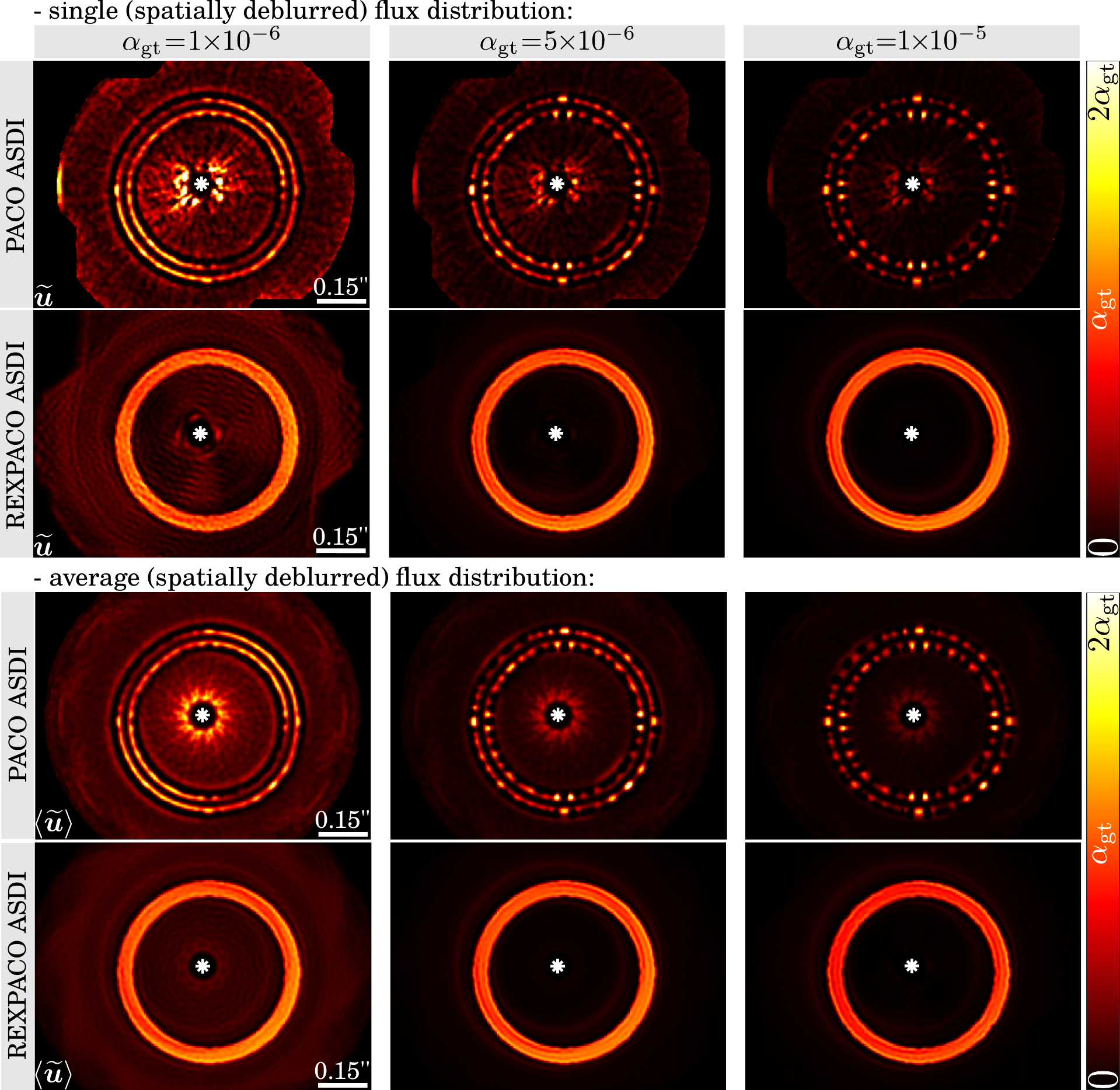}
	\caption{Same than Fig. \ref{fig:ellipse_deblurred_fullfig} for synthetic circular disks.}
	\label{fig:circle_deblurred_fullfig}
\end{figure*}

\begin{figure*}
	\centering
	\includegraphics[width=\textwidth]{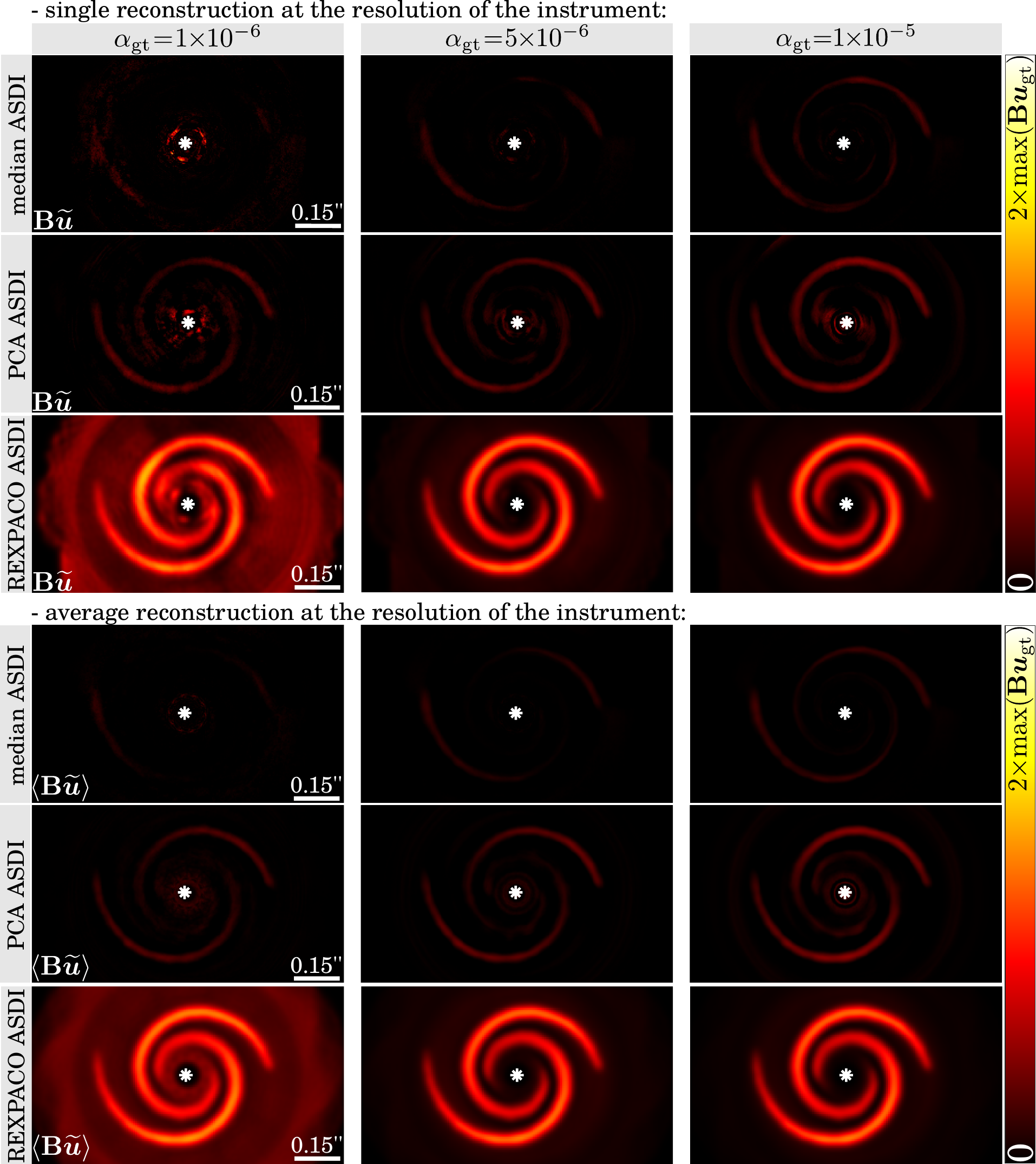}
	\caption{Same than Fig. \ref{fig:ellipse_blurred_fullfig} for synthetic spiral disks.}
	\label{fig:spiral_blurred_fullfig}
\end{figure*}

\begin{figure*}
	\centering
	\includegraphics[width=\textwidth]{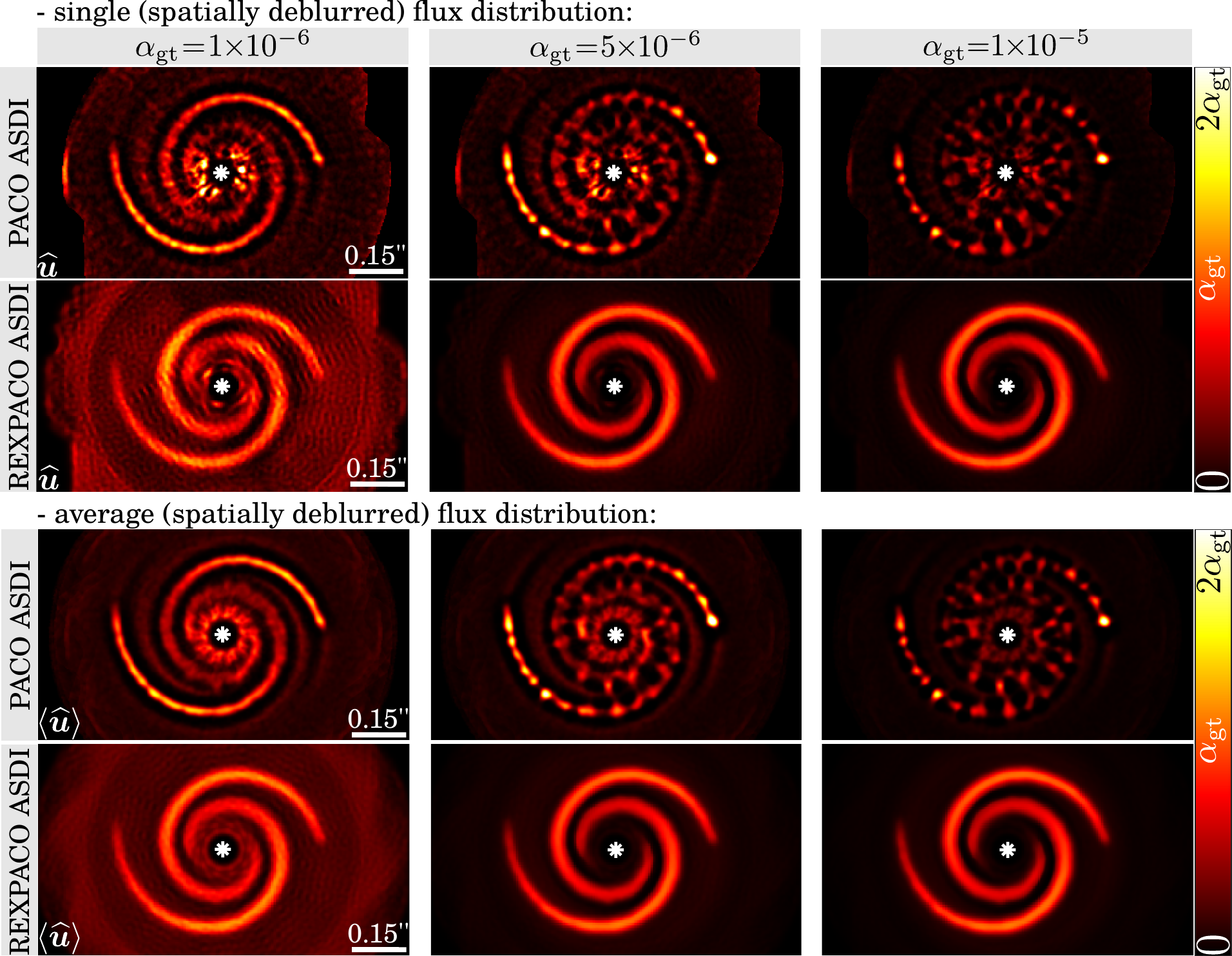}
	\caption{Same than Fig. \ref{fig:ellipse_deblurred_fullfig} for synthetic spiral disks.}
	\label{fig:spiral_deblurred_fullfig}
\end{figure*}

\begin{table*} %
			\caption{Quantitative assessment of the reconstruction quality on synthetic disks. N-RMSE as defined in
				Eq.~\eqref{eq:n_rmse} is reported for the reconstructions displayed in
				Figs. \ref{fig:ellipse_blurred_fullfig}-\ref{fig:spiral_deblurred_fullfig} and \ref{fig:ellipse_cuts_fullfig}-\ref{fig:spiral_cuts_fullfig}. The N-RMSE is
				also computed
				on the restrictions $\mathcal{D}(\obj_{\text{gt}})$ and
				$\mathcal{D}(\widetilde{\obj})$ to the area actually covered by the simulated disks.
				The best scores are highlighted in bold fonts.}
			\centering%
			\begin{tabular}{ccccc} \toprule Score & Algorithm &
				$\alpha_{\text{gt}} = 1\times 10^{-6}$ & $\alpha_{\text{gt}} = 5\times
				10^{-6}$ & $\alpha_{\text{gt}} = 1\times 10^{-5}$ \\
				\midrule
				\midrule
				& & \multicolumn{3}{c}{---\textit{ Elliptical disk, see Figs. \ref{fig:ellipse_blurred_fullfig}, \ref{fig:ellipse_deblurred_fullfig} and \ref{fig:ellipse_cuts_fullfig} }---} \\
				$\text{N-RMSE}\left(\obj_{\text{gt}},\widetilde{\obj}\right)$ & PACO ASDI & 0.52 & 0.53 & 0.58\\
				$\text{N-RMSE}\left(\obj_{\text{gt}},\widetilde{\obj}\right)$ & \textbf{REXPACO ASDI} & \textbf{0.12} & \textbf{0.11} & \textbf{0.10}\\
				\midrule
				$\text{N-RMSE}\left(\mathcal{D}(\obj_{\text{gt}}),\mathcal{D}(\widetilde{\obj})\right)$ & PACO ASDI & 0.26 & 0.41 & 0.53\\
				$\text{N-RMSE}\left(\mathcal{D}(\obj_{\text{gt}}),\mathcal{D}(\widetilde{\obj})\right)$ & \textbf{REXPACO ASDI} & \textbf{0.10} & \textbf{0.06} & \textbf{0.04}\\
				\midrule
				$\text{N-RMSE}\left(\M B \, \obj_{\text{gt}},\M B \,\widetilde{\obj}\right)$ & median ASDI & 0.68 & 0.56 & 0.46\\
				$\text{N-RMSE}\left(\M B \, \obj_{\text{gt}},\M B \,\widetilde{\obj}\right)$ & PCA ASDI & 0.40 & 0.30 & 0.30\\
				$\text{N-RMSE}\left(\M B \, \obj_{\text{gt}},\M B \, \widetilde{\obj}\right)$ & \textbf{REXPACO ASDI} & \textbf{0.13} & \textbf{0.06} & \textbf{0.05}\\
				\midrule
				$\text{N-RMSE}\left(\M B \, \mathcal{D}(\obj_{\text{gt}}),\M B \, \mathcal{D}(\widetilde{\obj})\right)$ & median ASDI & 0.66 & 0.54 & 0.45\\
				$\text{N-RMSE}\left(\M B \, \mathcal{D}(\obj_{\text{gt}}),\M B \, \mathcal{D}(\widetilde{\obj})\right)$ & PCA ASDI & 0.39 & 0.29 & 0.27\\
				$\text{N-RMSE}\left(\M B \, \mathcal{D}(\obj_{\text{gt}}),\M B \, \mathcal{D}(\widetilde{\obj})\right)$ & \textbf{REXPACO ASDI} & \textbf{0.12} & \textbf{0.03} & \textbf{0.01}\\
				\midrule
				\midrule
				& & \multicolumn{3}{c}{---\textit{ Circular disk, see Figs. \ref{fig:circle_blurred_fullfig}, \ref{fig:circle_deblurred_fullfig} and \ref{fig:circle_cuts_fullfig} }---}\\
				$\text{N-RMSE}\left(\obj_{\text{gt}},\widetilde{\obj}\right)$ & PACO ASDI & 0.74 & 0.71 & 0.77\\
				$\text{N-RMSE}\left(\obj_{\text{gt}},\widetilde{\obj}\right)$ & \textbf{REXPACO ASDI} & \textbf{0.14} & \textbf{0.12} & \textbf{0.11}\\
				\midrule
				$\text{N-RMSE}\left(\mathcal{D}(\obj_{\text{gt}}),\mathcal{D}(\widetilde{\obj})\right)$ & PACO ASDI & 0.51 & 0.60 & 0.71\\
				$\text{N-RMSE}\left(\mathcal{D}(\obj_{\text{gt}}),\mathcal{D}(\widetilde{\obj})\right)$ & \textbf{REXPACO ASDI} & \textbf{0.10} & \textbf{0.06} & \textbf{0.04}\\
				\midrule
				$\text{N-RMSE}\left(\M B \, \obj_{\text{gt}},\M B \, \widetilde{\obj}\right)$ & median ASDI & 0.97 & 0.97 & 0.97\\
				$\text{N-RMSE}\left(\M B \, \obj_{\text{gt}},\M B \, \widetilde{\obj}\right)$ & PCA ASDI & 0.91 & 0.87 & 0.65\\
				$\text{N-RMSE}\left(\M B \, \obj_{\text{gt}},\M B \, \widetilde{\obj}\right)$ & \textbf{REXPACO ASDI} & \textbf{0.15} & \textbf{0.08} & \textbf{0.07}\\
				\midrule
				$\text{N-RMSE}\left(\M B \, \mathcal{D}(\obj_{\text{gt}}),\M B \, \mathcal{D}(\widetilde{\obj})\right)$ & median ASDI & 0.97 & 0.96 & 0.96\\
				$\text{N-RMSE}\left(\M B \, \mathcal{D}(\obj_{\text{gt}}),\M B \, \mathcal{D}(\widetilde{\obj})\right)$ & PCA ASDI & 0.90 & 0.87 & 0.63\\
				$\text{N-RMSE}\left(\M B \, \mathcal{D}(\obj_{\text{gt}}),\M B \, \mathcal{D}(\widetilde{\obj})\right)$ & \textbf{REXPACO ASDI} & \textbf{0.12} & \textbf{0.04} & \textbf{0.02}\\
				\midrule
				\midrule
				& & \multicolumn{3}{c}{---\textit{ Spiral disk, see Figs. \ref{fig:spiral_blurred_fullfig}, \ref{fig:spiral_deblurred_fullfig} and \ref{fig:spiral_cuts_fullfig} }---}\\
				$\text{N-RMSE}\left(\obj_{\text{gt}},\widetilde{\obj}\right)$ & PACO ASDI & 0.63 & 0.64 & 0.69\\
				$\text{N-RMSE}\left(\obj_{\text{gt}},\widetilde{\obj}\right)$ & \textbf{REXPACO ASDI} & \textbf{0.60} & \textbf{0.39} & \textbf{0.38}\\
				\midrule
				$\text{N-RMSE}\left(\mathcal{D}(\obj_{\text{gt}}),\mathcal{D}(\widetilde{\obj})\right)$ & PACO ASDI & 0.25 & 0.40 & 0.60\\
				$\text{N-RMSE}\left(\mathcal{D}(\obj_{\text{gt}}),\mathcal{D}(\widetilde{\obj})\right)$ & \textbf{REXPACO ASDI} & \textbf{0.06} & \textbf{0.05} & \textbf{0.03}\\
				\midrule
				$\text{N-RMSE}\left(\M B \, \obj_{\text{gt}},\M B \, \widetilde{\obj}\right)$ & median ASDI & 0.99 & 0.96 & 0.91\\
				$\text{N-RMSE}\left(\M B \, \obj_{\text{gt}},\M B \, \widetilde{\obj}\right)$ & PCA ASDI & 0.82 &  0.80 & 0.70\\
				$\text{N-RMSE}\left(\M B \, \obj_{\text{gt}},\M B \, \widetilde{\obj}\right)$ & \textbf{REXPACO ASDI} & \textbf{0.58} & \textbf{0.36} & \textbf{0.35}\\
				\midrule
				$\text{N-RMSE}\left(\M B \, \mathcal{D}(\obj_{\text{gt}}),\M B \, \mathcal{D}(\widetilde{\obj})\right)$ & median ASDI & 0.99 & 0.96 & 0.91\\
				$\text{N-RMSE}\left(\M B \, \mathcal{D}(\obj_{\text{gt}}),\M B \, \mathcal{D}(\widetilde{\obj})\right)$ & PCA ASDI & 0.82 & 0.80 & 0.69\\
				$\text{N-RMSE}\left(\M B \, \mathcal{D}(\obj_{\text{gt}}),\M B \, \mathcal{D}(\widetilde{\obj})\right)$ & \textbf{REXPACO ASDI} & \textbf{0.14} & \textbf{0.05} & \textbf{0.04}\\
				\bottomrule
			\end{tabular}
			\label{tab:nrmse_photometry}
\end{table*}

Figures \ref{fig:ellipse_blurred_fullfig}-\ref{fig:ellipse_deblurred_fullfig}, \ref{fig:circle_blurred_fullfig}-\ref{fig:circle_deblurred_fullfig} and \ref{fig:spiral_blurred_fullfig}-\ref{fig:spiral_deblurred_fullfig} report the reconstruction results for the circular, elliptical and spiral disks, respectively. Figures \ref{fig:ellipse_cuts_fullfig}, \ref{fig:circle_cuts_fullfig}, and \ref{fig:spiral_cuts_fullfig} complement these reconstruction results with a slice-cuts analysis along the three profiles defined in Fig. \ref{fig:gt_fullfig}.

Because median ASDI and PCA ASDI do not perform a deconvolution, the comparisons are performed at the resolution of the instrument, as in Sect. \ref{subsec:recons_real_disks}. The deconvolved flux distributions $\widetilde{\obj}$ estimated by REXPACO ASDI are thus re-blurred by the off-axis PSF so that the quantity $\M B\,\widetilde{\obj}$ can be directly compared with the median ASDI and PCA ASDI images in Figs.  \ref{fig:ellipse_blurred_fullfig}, \ref{fig:spiral_blurred_fullfig} and \ref{fig:circle_blurred_fullfig}. REXPACO ASDI reconstructions $\widetilde{\obj}$ deconvolved from the off-axis PSF are more specifically compared to PACO ASDI flux distribution maps in Figs. \ref{fig:ellipse_deblurred_fullfig}, \ref{fig:circle_deblurred_fullfig} and \ref{fig:spiral_deblurred_fullfig}.

Overall, significant errors both in terms of morphology distortions and photometry under-estimations are made on the sought objects by the three comparative techniques, regardless of the type of disk.
These errors are more pronounced when the diversity induced by ASDI is the most limited to disentangle the nuisance from the off-axis objects. As an illustration, the circular disk and arms of the spiral disk are barely visible near the star in the median ASDI and PCA ASDI images, even for the brightest cases, which is the sign that an important self-subtraction occurs.
In addition, some stellar leakages remain, especially near the star due to the absence of explicit modeling of the correlations of the nuisance. Flux distributions estimated by PACO ASDI are also affected by significant artifacts: continuous structures manifest as a series of point sources due to assumptions made in the model regarding the target objects. Unlike other tested algorithms, this effect worsen when the contrast improves. In addition, only gradient of smooth structures are (approximately) recovered by PACO ASDI.

In comparison, reconstructions produced by REXPACO ASDI seem much closer to the ground truth, even for the lowest level of contrast $\alpha_{\text{gt}} = 1 \times 10^{-6}$, with an improved object fidelity and a better rejection of the star light. Unlike median ASDI or PCA ASDI reconstructions, which display non-physical negative values, REXPACO ASDI flux distributions are consistently non-negative (see slice-cuts profiles in Figs. \ref{fig:ellipse_cuts_fullfig}-\ref{fig:spiral_cuts_fullfig}), owing to the explicit non-negativity constraint imposed in the minimization problem~\eqref{eq:minpb}. In addition, an important result is the ability of REXPACO ASDI to reconstruct disks having a quasi-circular symmetry (that are especially challenging to reconstruct due to the lack of angular diversity), without the need of additional diversity complementing ASDI, e.g. leveraging multiple datasets as done in RDI techniques (see Sect. \ref{sec:introduction}).
	The deblurred reconstructions of REXPACO ASDI shown in Figs. \ref{fig:ellipse_deblurred_fullfig}, \ref{fig:circle_deblurred_fullfig}, \ref{fig:spiral_deblurred_fullfig} and \ref{fig:ellipse_cuts_fullfig}-\ref{fig:spiral_cuts_fullfig} are in good agreement with the ground truth. As expected, the reconstruction fidelity is
		higher when the disk is brighter: more spurious fluctuations are visible in
		the deblurred reconstruction at $\alpha_{\text{gt}} = 1 \times 10^{-6}$ than at $\alpha_{\text{gt}} = 1 \times 10^{-5}$. Moreover, the spatial resolution is also significantly improved by the deconvolution process. However, some discrepancies can be noted in the deblurred line profiles, such as a slight Gibbs effect (i.e., signal ripples) near sharp edges induced by the edge-preserving regularization (even if it is beneficial in overall) and a residual bias on the photometry for some parts of the spiral disk (even the overall morphology is preserved). We discuss the latter phenomenon in Sect. \ref{subsec:importance_spectral_processing} dedicated to the comparison between ADI and ASDI processing.

\medskip

After this qualitative analysis, we now compare, as done in \cite{flasseur2021rexpaco}, the reconstruction quality of median ASDI, PCA ASDI, PACO ASDI and REXPACO ASDI by reporting the normalized root mean square error (N-RMSE, the lower the higher reconstruction fidelity):
\begin{equation}
	\text{N-RMSE}(\obj_{\text{gt}}, \widetilde{\obj}) = \frac{|| \obj_{\text{gt}} - \widetilde{\obj}||_2}{||\obj_{\text{gt}}||_2}\,.
	\label{eq:n_rmse}
\end{equation}
Table~\ref{tab:nrmse_photometry} reports the N-RMSE for two regions of the reconstructed flux distribution: (i) the entire image, and (ii) the disk area only. In the latter case, Eq. (\ref{eq:n_rmse}) is modified to account solely for disk regions. This metric shows a clear improvement brought by REXPACO ASDI compared to the other tested algorithms, with error reduction exceeding a factor 10 for more challenging configurations (e.g., circular or spiral disks). A more modest error reduction is obtained for configurations (i.e., morphology and contrast) leading to an easier separation of the disk from the nuisance contribution, like for the elliptical disk.

This study also provides valuable insights for interpreting the reconstructions of real disks presented in Fig. \ref{fig:realdisks}, as both the simulations and real data share comparable angular and spectral diversity (i.e., similar amounts of parallactic rotation, same number and spreading of the spectral channels).  Additionally, the simulated disks possess morphologies closely resembling the real disks.
Consequently, this study suggests that the reconstructed flux distribution of HR 4796 can be confidently interpreted as having an elliptical morphology. Similarly, the outer disk of HD 163296 has roughly the same morphology, allowing for confidence in the reconstructed structures on the Northern side, though the quality of the reconstruction is strongly limited by the low disk contrast (lower than $5 \times 10^{-7}$) on the Southern side. SAO 206462, MWC 758 and AB Aurigae, all of which exhibit spiral arms with a spatial extent quite similar to the simulated spiral disk studied in this section. We can thus expect that the morphology of these three real disks are well reconstructed with, likely, a slight photometric bias on some structures in the vicinity of the host star. The case of PDS 70 is more challenging due to its intricate structures including a smooth flux distribution near the star in the shortest wavelengths. While no non-physical discontinuities are observed in the outer disk, dedicated hydro-dynamical simulations of this object are needed to identify the areas impacted by potential artifacts. Such a study is out of the scope of this paper and is left for a future work dedicated to the re-analysis of multi-epochs and multi-instruments observations of PDS 70.

\subsection{On the importance of a joint spectral processing}
\label{subsec:importance_spectral_processing}

In this section, we aim to illustrate the benefits of joint spectral processing, incorporating fine modeling of correlations between spectral channels, compared to mono-spectral processing that does not leverage the apparent chromatic displacement of the speckle field induced by ASDI (see Sect. \ref{sec:introduction}).

\begin{figure*}
	\centering
	\includegraphics[width=0.93\textwidth]{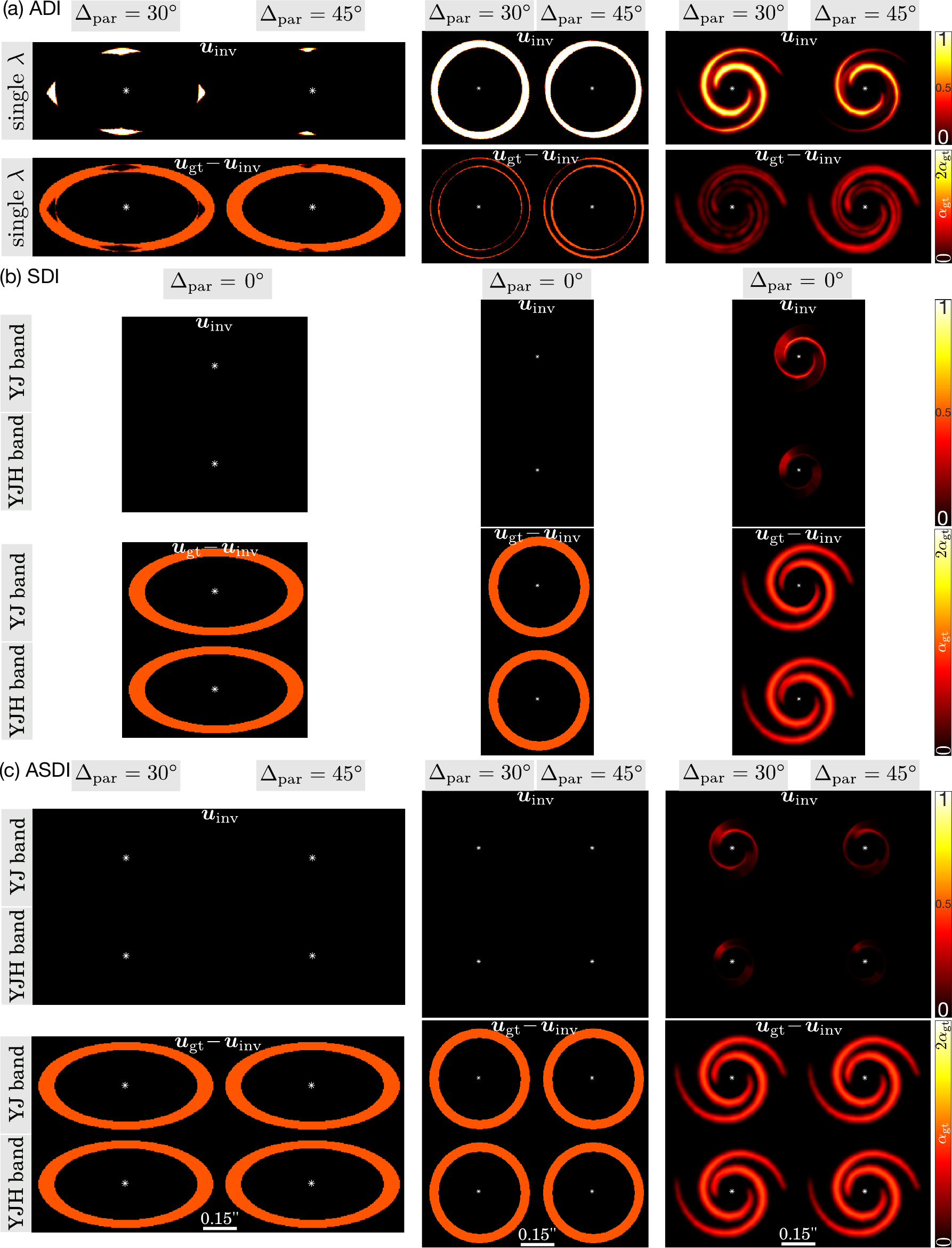}
	\caption{Nuisance-free study on the importance of a joint spectral processing. Invariant flux distribution $\obj_{\text{inv}}$ as defined in Eq. (\ref{eq:invariant_flux}) is reported on the first line of panels (a), (b) and (c) for the three synthetic disks (i.e., elliptical, circular, and spiral) whose ground truth flux distribution $\obj_{\text{gt}}$ are represented in Fig. \ref{fig:gt_fullfig}. The second line of panels (a), (b) and (c) gives the difference $\obj_{\text{gt}} - \obj_{\text{inv}}$. Two total amounts $\Delta_{\text{par}}$ of parallactic rotation and spectral bands are considered (if applicable) in each case. Panel (a) is for ADI (i.e., each spectral channel is considered independently in Eq. (\ref{eq:invariant_flux})), panel (b) is for SDI (i.e., assuming the parallactic angle is equal for each temporal exposure), and panel (c) is for ASDI (i.e., all temporal frames and spectral channels are processed jointly in Eq. (\ref{eq:invariant_flux})).}
	\label{fig:bias_fullfig}
\end{figure*}

On the latter point, we start by identifying parts of disks that are expected to suffer the most from the self-subtraction effect for different disk morphologies, spectral bands and total amounts of parallactic rotation. For that purpose, we consider the three synthetic disk morphologies studied in Sect. \ref{subsec:recons_simulated_disks}, and we assume a null nuisance component to evaluate solely the influence of limited angular and spectral diversity on reconstruction quality. As done by \cite{juillard2023inverse} for ADI, given a ground truth flux distribution $\obj_{\text{gt}} \in \mathbb{R}^{N'\times L}$, we define the spectrally aggregated flux $\obj_{\text{inv}} \in \mathbb{R}^{N'}$ which is invariant both from the apparent rotation induced by ADI and from the homothetic spectral motion of the speckle field induced by SDI as:
\begin{equation}
	{\left[ \obj_{\text{inv}} \right]}_n = \text{min}_{t=1:T,\, \ell=1:L} \left[ \M F \, \obj_{\text{gt}} \right]_{n,t,\ell} \,, \forall n \in \llbracket 1; N' \rrbracket\,,
	\label{eq:invariant_flux}
\end{equation}
with $\M F$ the sparse operator performing rotations, scalings, and attenuations as defined for the forward image formation model in Sect. \ref{sec:directmodel}.
Taking the minimum intensity value (operator min) across the temporal and spectral dimensions in Eq. (\ref{eq:invariant_flux}) enables the identification of ASDI-invariant flux regions. The output is 0 for non-invariant regions and 1 for areas of the disk that are fully affected by angular and/or spectral invariance.
We also consider the quantity $\obj_{\text{gt}} - \obj_{\text{inv}}$ representing the expected reconstructed flux distribution if the invariant component $\obj_{\text{inv}}$ can not be disentangle from the nuisance component (i.e., the angular and spectral diversity are not sufficient to perform signal unmixing).
 Fig. \ref{fig:bias_fullfig} represents these two quantities in ADI, SDI and ASDI for the three typical morphologies considered in Sect. \ref{subsec:recons_simulated_disks}, for a simulated total amount of parallactic rotation $\Delta_{\text{par}} = \lbrace 30\degree, 45\degree \rbrace$, and for simulated spectral bands YJ (i.e., $\lambda \in \left[  0.96-1.33 \right]\, \micro \meter$) or YJH (i.e., $\lambda \in \left[ 0.96-1.64 \right]\, \micro \meter$). In ADI, i.e. in the absence of a joint spectral processing, we observe that a large fraction of the circular and spiral disks remain invariant with respect to the background. The elliptical disk is less affected by this phenomenon, even if it is not negligible, especially near the ellipse handles along its minor axis. This lack of diversity translates into a partial attenuation and distortion of the reconstructed disk, due to object self-subtraction, see e.g. \cite{milli2012impact, pairet2019iterative,juillard2023inverse} for related studies in ADI. Moreover, as expected the total amount of parallactic rotation brings only a limited diversity at short angular separations: the angular-invariant flux distribution only slightly decreases when $\Delta_{\text{par}}$ evolves from 30° to 45°, regardless of the disk morphology. SDI effectively eliminates most signal ambiguities caused by object invariances. It leads to no invariant flux for elliptical and circular disks.  Joint spectral processing with ASDI further improves the unmixing capability of post-processing algorithms as only a very slight fraction of the spiral disk remains invariant for the setting $\Delta_{\text{par}} = 30\degree$ in YJ band. It results in a slight object self-subtraction that could explain the observed photometric bias in Figs. \ref{fig:spiral_deblurred_fullfig} and \ref{fig:spiral_cuts_fullfig} on the reconstructed spiral disk for similar settings (in terms of disk morphology, parallactic rotation, and spectral band). Increasing the spectral width towards the H band and the total parallactic rotation towards 45° leads to a negligible invariant flux distribution, that would allow to reconstruct the underlying off-axis object without self-subtraction with REXPACO ASDI, and without the need to leverage a database archive as in RDI techniques.

\begin{figure*}
	\centering
	\includegraphics[width=\textwidth]{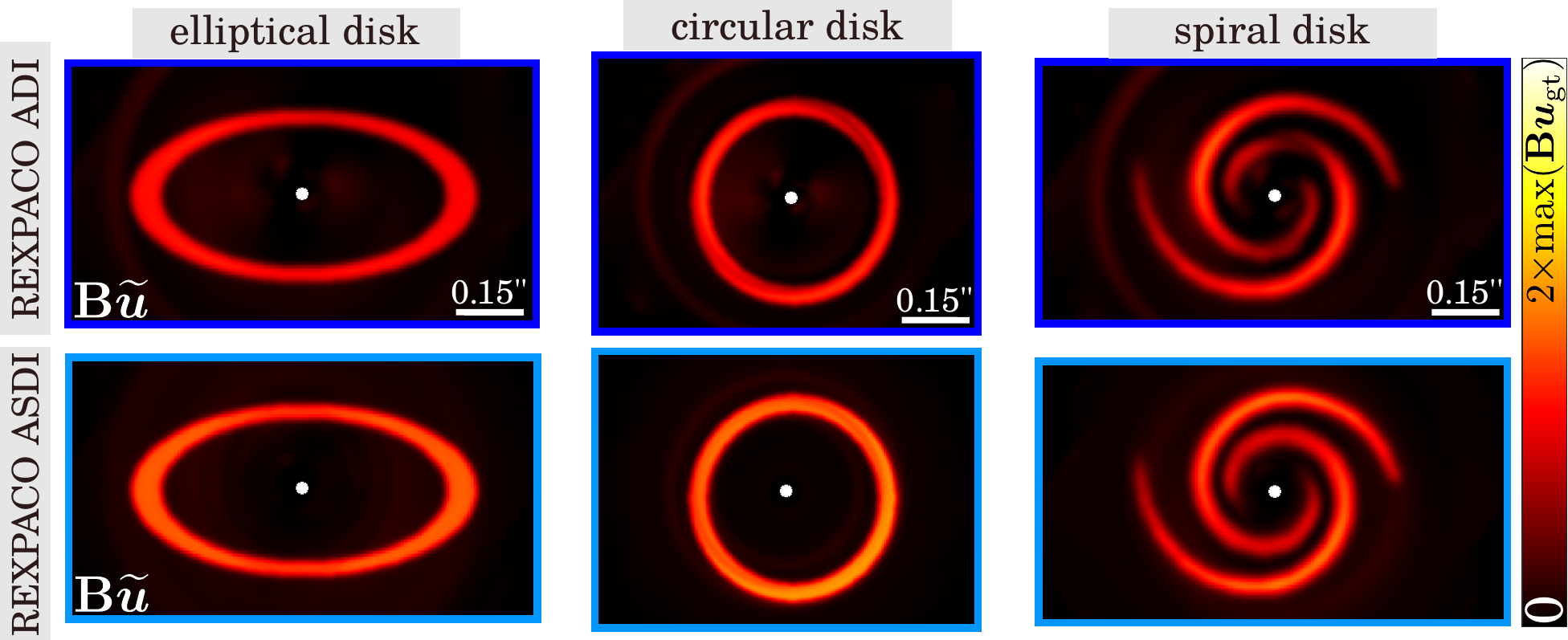}
	\caption{
	Comparison between ADI and ASDI on the reconstruction of synthetic disks. The considered elliptical, circular, and spiral disks are injected ($\alpha_{\text{gt}} = 1 \times 10^{-5}$, $\Delta_{\text{par}} = 30\degree$, YJ band), within a real SPHERE-IFS dataset and processed with the mono-spectral algorithm REXPACO ADI \citep{flasseur2021rexpaco} and its multi-spectral version REXPACO ASDI proposed in this paper. The reconstructions $\M B\, \widetilde{\obj}$ are re-blurred by the off-axis PSF, and REXPACO ASDI results are similar to the ones presented in the third column of Figs. \ref{fig:ellipse_blurred_fullfig}, \ref{fig:circle_blurred_fullfig}, and \ref{fig:spiral_blurred_fullfig}. Dataset: HD 172555 (2015-07-11), see Table \ref{tab:dataset_logs} for the observation parameters.
	}
	\label{fig:adi_vs_asdi_synthetic_disks_fullfig}
\end{figure*}

Figure \ref{fig:adi_vs_asdi_synthetic_disks_fullfig} completes this study by comparing a post-processing relying on ADI only (here, with the REXPACO ADI algorithm) to a post-processing leveraging also on the spectral diversity brought by ASDI (here, with the REXPACO ASDI algorithm) for three particular configurations of the extensive simulations performed in Sect. \ref{subsec:recons_simulated_disks}. Figure \ref{fig:adi_vs_asdi_synthetic_disks_fullfig_cuts_only} complements results displayed in Fig. \ref{fig:adi_vs_asdi_synthetic_disks_fullfig} with a slice-cuts analysis along the three profiles defined in Fig. \ref{fig:gt_fullfig}.

In all cases, the same total amount of information is used, i.e. all spectral channels are considered in ADI but they are processed individually instead of jointly as in ASDI. The conclusions derived from the nuisance-free simulations in Fig. \ref{fig:bias_fullfig} directly translates on the reconstruction quality: REXPACO ADI leads to a bias (respectively, by up to 20 \% and 60\%) on the reconstructed flux distribution of the elliptical and circular disks, respectively. This bias is almost null for the elliptical disk reconstructed with the proposed REXPACO ASDI algorithm. For the spiral disk, a bias up to 20\% can remain on some parts of the ASDI reconstruction, even though it is significantly smaller than for ADI. This residual bias can be attributed to a still insufficient angular and spectral diversity, as shown by the nuisance-free study for a simulated spiral disk in the YJ band with a total parallactic rotation $\Delta_{\text{par}}$ of 30°.

\begin{figure}
	\centering
	\includegraphics[width=0.45\textwidth]{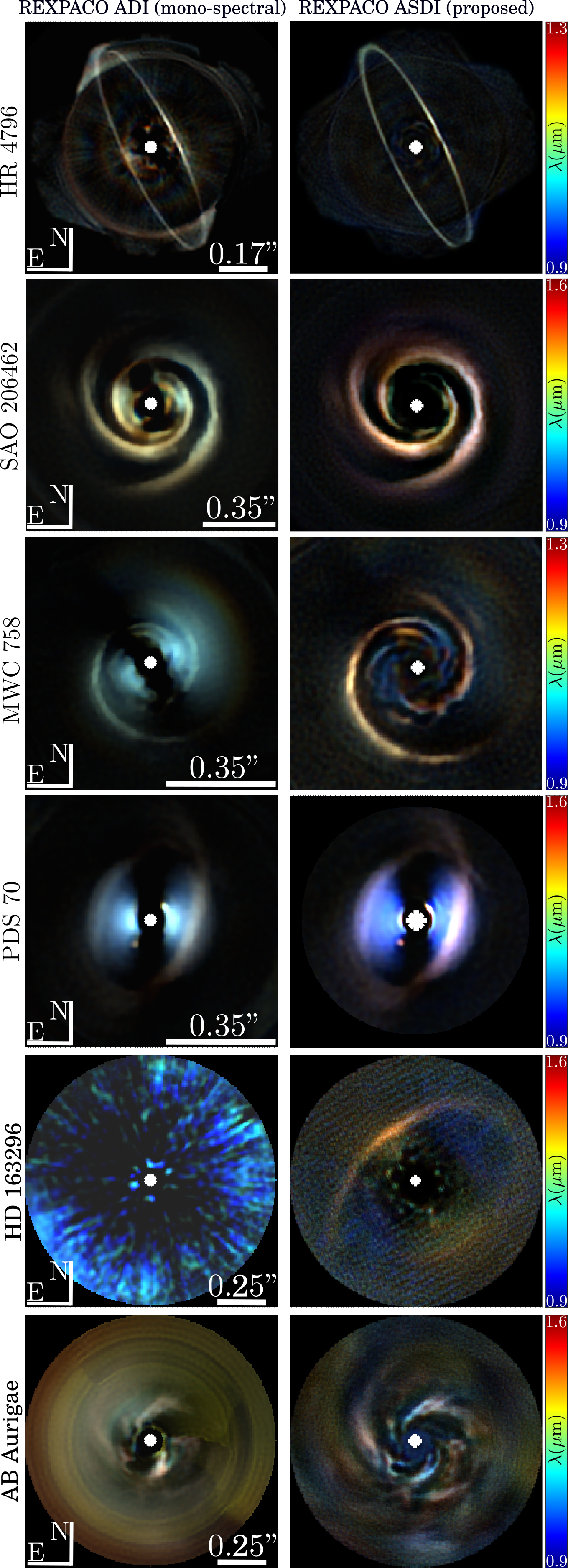}
	\caption{Comparison of reconstructions $\widetilde{\obj}$ obtained on SPHERE-IFS data with the mono-spectral REXPACO algorithm (left) and the proposed multi-spectral REXPACO ASDI algorithm (right, similar results as in last column of Figs. \ref{fig:realdisks}). Pseudo-color images are displayed as in Figs. \ref{fig:ablationstudy}. Datasets: same as in Fig. \ref{fig:realdisks}.}
	\label{fig:real_disk_ifs_adi_fullfig}
\end{figure}

Conversely, on the same real data of Sects. \ref{subsec:validation_model}-\ref{subsec:recons_real_disks}, we perform a model ablation study complementary to the one presented in Sect. \ref{subsec:validation_model}. Unlike in Sect. \ref{subsec:validation_model}, we consider here the spatial covariances when estimating the nuisance component but we process each spectral channel individually with REXPACO ADI instead of jointly with REXPACO ASDI as done in Sects. \ref{subsec:validation_model}-\ref{subsec:recons_real_disks}. Figure \ref{fig:real_disk_ifs_adi_fullfig} displays the resulting reconstructed flux distributions compared to the corresponding REXPACO ASDI reconstructions. The absence of joint spectral processing is detrimental on three aspects. First, important residual star light remains in the ADI reconstructions, in particular for HR 4796 and AB Aurigae. Their typical signatures in \textit{rainbow pattern} is due to the absence of modeling of the spectral correlations of the nuisance. Second, the sensitivity is lowered due to the absence of explicit exploitation of the spectral diversity, even if the same total amount of data is processed. As an illustration, the HD 163296 disk is almost invisible in the ADI reconstruction. Third, important non-physical artifacts and discontinuities on the disk features are present on the ADI reconstructions, especially for disk having a circular symmetry like SAO 206462, MWC 758 and PDS 70. This latter effect is due to the lack of diversity between the sought off-axis objects and the nuisance component in ADI, as discussed in the previous paragraph.

\begin{figure}
	\centering
	\includegraphics[width=0.47\textwidth]{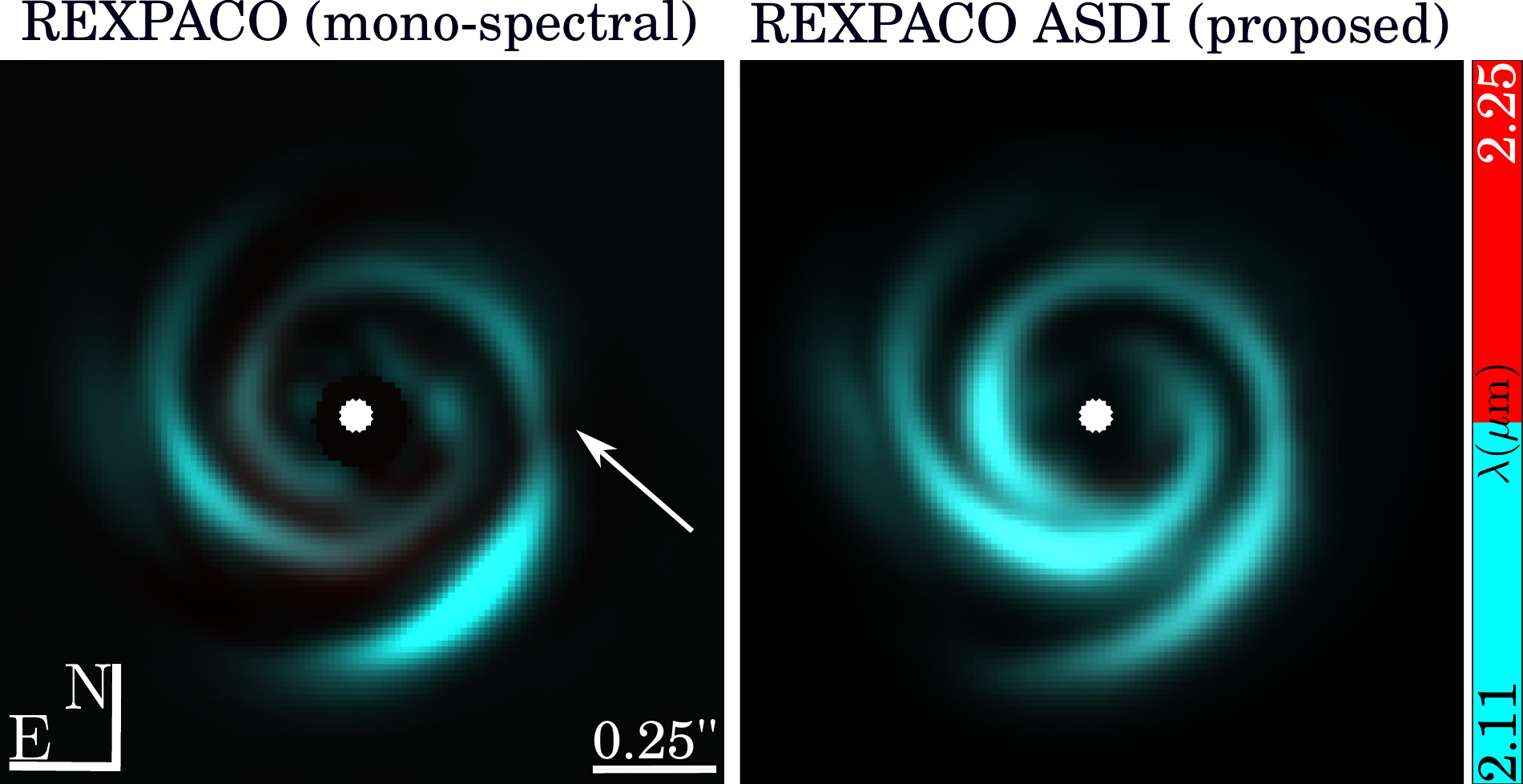}
	\caption{Comparison of reconstructions obtained on SPHERE-IRDIS data with the mono-spectral REXPACO ADI algorithm (left) and the proposed multi-spectral REXPACO ASDI algorithm. The white arrow points out a part of West spiral arm severely impacted by reconstruction artifacts with a post-processing based on ADI solely, see Fig. 11 of \citep{flasseur2021rexpaco}. Pseudo-color images are displayed as in Figs. \ref{fig:ablationstudy}. Dataset: SAO 206462 (2015-05-15), see Table \ref{tab:dataset_logs} for the observation parameters.}
	\label{fig:irdis_asdi_fullfig}
\end{figure}

Now that we have established the causes of the limitations of ADI and emphasized the benefits of ASDI to produce faithful reconstructions of the circumstellar environment from IFS data, we illustrate that even a limited spectral diversity can be useful to improve the quality of the reconstructions. In our previous work on the REXPACO algorithm designed for ADI \citep{flasseur2021rexpaco}, we considered datasets from the SPHERE-IRDIS imager in its dual band configuration (i.e., producing simultaneously datasets on $L=2$ spectral channels). In \cite{flasseur2021rexpaco}, we have shown that REXPACO ADI is able to produce disk reconstructions with a significantly improved quality compared to standard post-processing methods like median ADI, PCA ADI, PACO ADI. We also notice that some plausible artifacts can remain due to the lack of diversity between the disk and the nuisance component. Here, we re-visit with the proposed REXPACO ASDI algorithm a SPHERE-IRDIS dataset (SAO 206462) considered in \cite{flasseur2021rexpaco} and for which the reconstruction seems the most impacted by residual artifacts. Figure \ref{fig:irdis_asdi_fullfig} compares our new reconstruction obtained by a joint spectral processing with REXPACO ASDI to the REXPACO ADI reconstruction. 
Notably, we identified in our ADI reconstruction a spurious reconstruction effect on the West spiral arm, taking the form of a flux discontinuity (see white arrow in Fig. \ref{fig:irdis_asdi_fullfig}). This likely artifact is effectively mitigated in the ASDI reconstruction, primarily due to the joint spectral processing of both available spectral channels. Furthermore, the disk appears significantly fainter in the second channel compared to the first, leading to better separation between the disk and the nuisances. In this case, the second channel serves almost like a reference channel, nearly free from the signal of the target object.
Overall, the morphology of SAO 206462 extracted with REXPACO ASDI from the IRDIS dataset exhibits structures very similar to these in the IFS reconstruction presented in Fig. \ref{fig:realdisks}. This example illustrates qualitatively that even a very limited spectral diversity (in the present case, $L=2$ spectral channels, and a band width $\Delta_\lambda < 0.15 \, \micro \meter$) is sufficient to improve significantly the reconstruction quality by reducing morphological distortions and flux attenuations.

\medskip

\noindent This study yields two main conclusions. First, ASDI post-processing should be favored over ADI and SDI, as it significantly mitigates ambiguities due to object invariances. This finding supports the choices made in Sects. \ref{subsec:recons_real_disks} and \ref{subsec:recons_simulated_disks} regarding the application of comparative algorithms that exploit jointly ASDI diversities. Second, while ASDI offers a theoretical advantage in diversity, this benefit fully  translates into improved reconstruction fidelity only when appropriate models of the data are employed. As an illustration, all median ASDI reconstructions (i.e., based on an overly simplistic and empirical model of the nuisance) shown in Sects. \ref{subsec:recons_real_disks} and \ref{subsec:recons_simulated_disks} display strong artifacts, despite the method jointly exploiting both ADI and SDI diversities.

\section{Unmixing point-like sources from extended features}
\label{sec:unmixing_ps_disk}

\subsection{Alternate unmixing}
\label{subsec:alternate_unmixing}

\begin{algorithm}[t]
	\SetAlgoVlined \DontPrintSemicolon\SetSideCommentLeft
	\caption{Alternating REXPACO ASDI and PACO ASDI\newline(unmixing disk and point-like sources).}
	\label{alg:rexpaco_paco}%
	\begin{minipage}{0.98\columnwidth}
		\KwIn{ASDI sequence $\V v$.}
		\KwIn{Forward operator $\M M$.}
		\KwIn{Relative precision $\eta \in (0,1)$, $\eta=10^{-3}$ in practice.}
		\smallskip
		\KwOut{Flux distribution $\widetilde{\obj}$ of disk.}
		\KwOut{Flux distribution $\widehat{\V \alpha}$ of point-like sources.}
		\KwOut{S/N of detection of point-like sources.}
		\KwOut{Astrometry (separation $\widehat{\rho}$, angle $\widehat{\theta}$) of point-like sources.}
		\smallskip
		\AlgoStep{1}{Initialization.}
		$i \leftarrow 0$ \Comment*[r]{iteration counter}
		$\widetilde{\obj}^{[i]} \leftarrow \text{REXPACO ASDI}(\V v)$ \Comment*[r]{apply REXPACO on data}
		$\text{S/N}^{\left[i\right]}, \widehat{\V \alpha}^{\left[i\right]} \leftarrow \text{PACO ASDI}(\V v)$ \Comment*[r]{apply PACO on data}
		\smallskip
		\AlgoStep{2}{User identification of (candidate) point-like sources.}
		$P > 0$ \Comment*[r]{set number of sources}
		$\widehat{\rho}^{[i]}_{1:P}, \widehat{\theta}^{[i]}_{1:P}$ \Comment*[r]{set rough astrometry}
		\smallskip
		\AlgoStep{3}{Main iteration loop.}
		\Do{ $\Norm[\big]{\estim{\V \alpha}^{[i]} - {\estim{\V \alpha}}^{[i-1]}} >
					\eta\,\Norm[\big]{\estim{\V \alpha}^{[i]}}$ }{
			$i \leftarrow i + 1$ \Comment*[r]{update iteration counter}
			\smallskip
			$\widetilde{\obj}^{[i]} \leftarrow \text{REXPACO ASDI}(\underbrace{\V v - \M M \, \widehat{\V \alpha}^{[i-1]}}_{\text{PACO residuals}})\hfill$ 
			\smallskip
			$\text{S/N}^{\left[i\right]}, \widehat{\V \alpha}^{\left[i\right]}, \widehat{\rho}^{[i]}_{1:P}, \widehat{\theta}^{[i]}_{1:P} \leftarrow \text{PACO ASDI}(\underbrace{\V v - \M M \, \widetilde{\obj}^{[i-1]}}_{\text{REXPACO residuals}})\hfill$ 
		\smallskip
			}
	\end{minipage}
\end{algorithm}

\begin{figure*}
	\includegraphics[width=\textwidth]{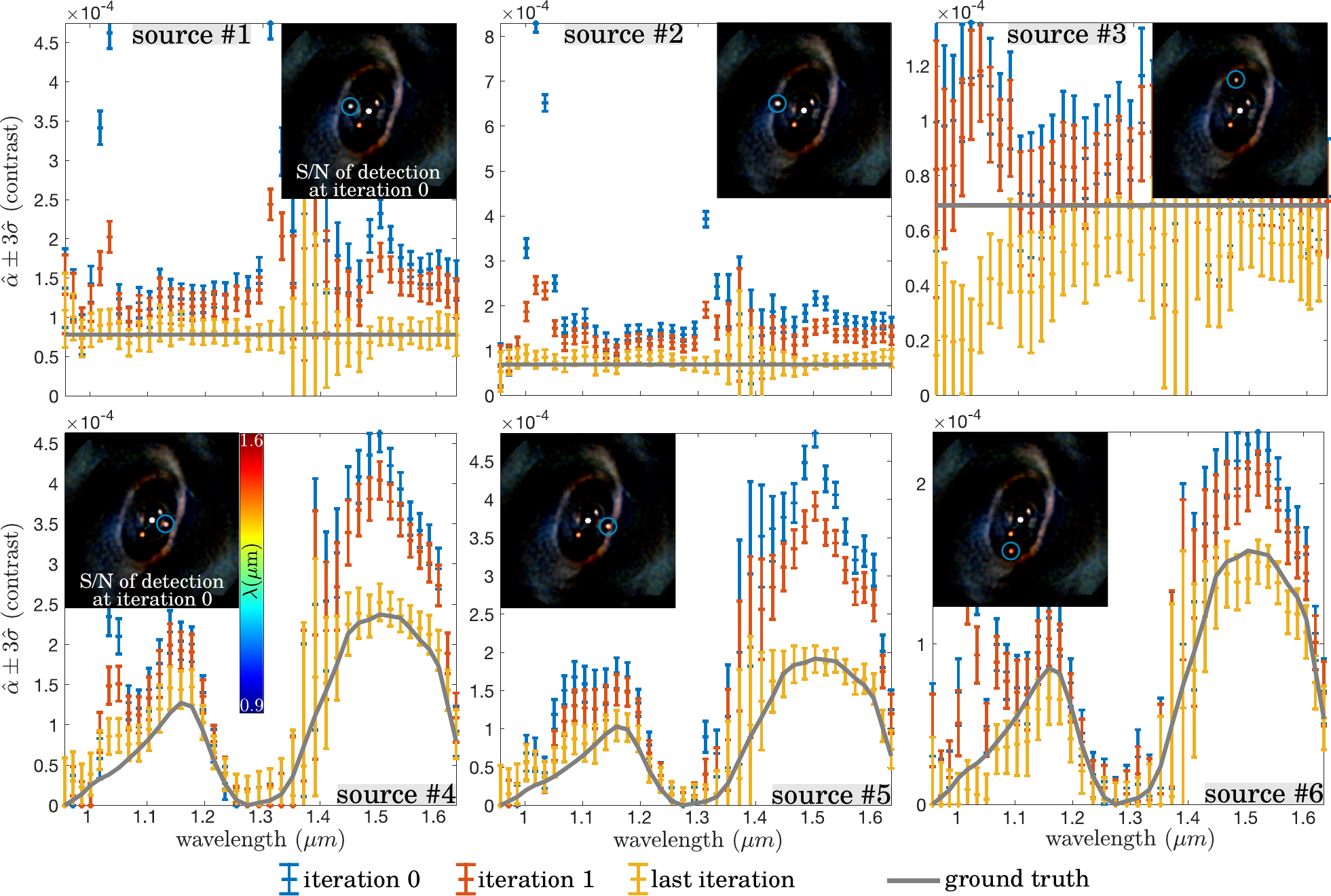}
	\caption{Unmixing synthetic sources from the circumstellar environment. Each sub-graph corresponds to a specific fake source (numbered \#1 to \#6). Sources \#1, \#2, and \#3 (top) exhibit a flat SED in contrast units (i.e., same spectrum as the star), while sources \#4, \#5, and \#6 (bottom) have a bi-modal SED. The SEDs estimated by Algorithm \ref{alg:rexpaco_paco} are shown with error bars, where the color represents the iteration. These estimates can be compared to the simulated ground truth SED indicated by gray straight lines. The S/N maps obtained with PACO ASDI at  iteration 0 of Algorithm \ref{alg:rexpaco_paco} are also included as insets, highlighting simulated point-like sources within the circumstellar environment (see circles).}
	\label{fig:estimated_spectrum}
\end{figure*}

In this section, we investigate the unmixing of the contribution of point-like sources embedded in spatially extended structures like circumstellar disks. The approach we propose is an extension to multi-spectral observations of the unmixing strategy described in our previous work \cite{flasseur2021rexpaco} for ADI data. It consists in combining REXPACO ASDI with PACO ASDI \citep{flasseur2020paco}; the former being dedicated to the reconstruction of disks while the latter being dedicated to the detection and to the sub-pixel characterization of point-like sources. In our experiments, this alternated strategy proved to be more satisfactory than a joint and regularized reconstruction of both a sparse component (for point-like sources) and of a smooth component (for the disk). One of the main peculiarities of the proposed alternated strategy is the ability to select manually the number and the rough location of candidate point-like sources to unmix from the disk material. In contrast, a joint reconstruction of both a sparse component and a smooth component leads either to many more nonzero values in the sparse component than the actual number of point-like sources, or misses the faintest sources, depending on the relative weights given to the sparsity and smoothness regularizations.

The proposed unmixing procedure works as follows: (i) REXPACO ASDI and PACO ASDI are applied independently on a target ASDI observation; (ii) based on the spatio-spectral S/N maps obtained with PACO ASDI and on the spatio-spectral flux distribution obtained with REXPACO ASDI, candidate point-like sources to unmix from the disk material are identified manually by the user; (iii) REXPACO ASDI and PACO ASDI are iteratively applied until convergence of the two retrieved components. During step (iii), the astrometry and photometry of the selected point-like sources are refined with sub-pixel accuracy by PACO ASDI within a $3\times3$ pixels box, based on the residual data obtained after subtraction of the disk contribution as currently reconstructed by REXPACO ASDI. Similarly, the spatio-spectral flux distribution of the disk is refined by REXPACO ASDI on updated residuals obtained after subtraction of the refined point-sources contribution estimated by PACO ASDI. This procedure is summarized by Algorithm \ref{alg:rexpaco_paco}.

\subsection{Case study on the PDS 70 system}
\label{subsec:pds70_case_study}

\begin{figure*}
	\centering
	\includegraphics[width=\textwidth]{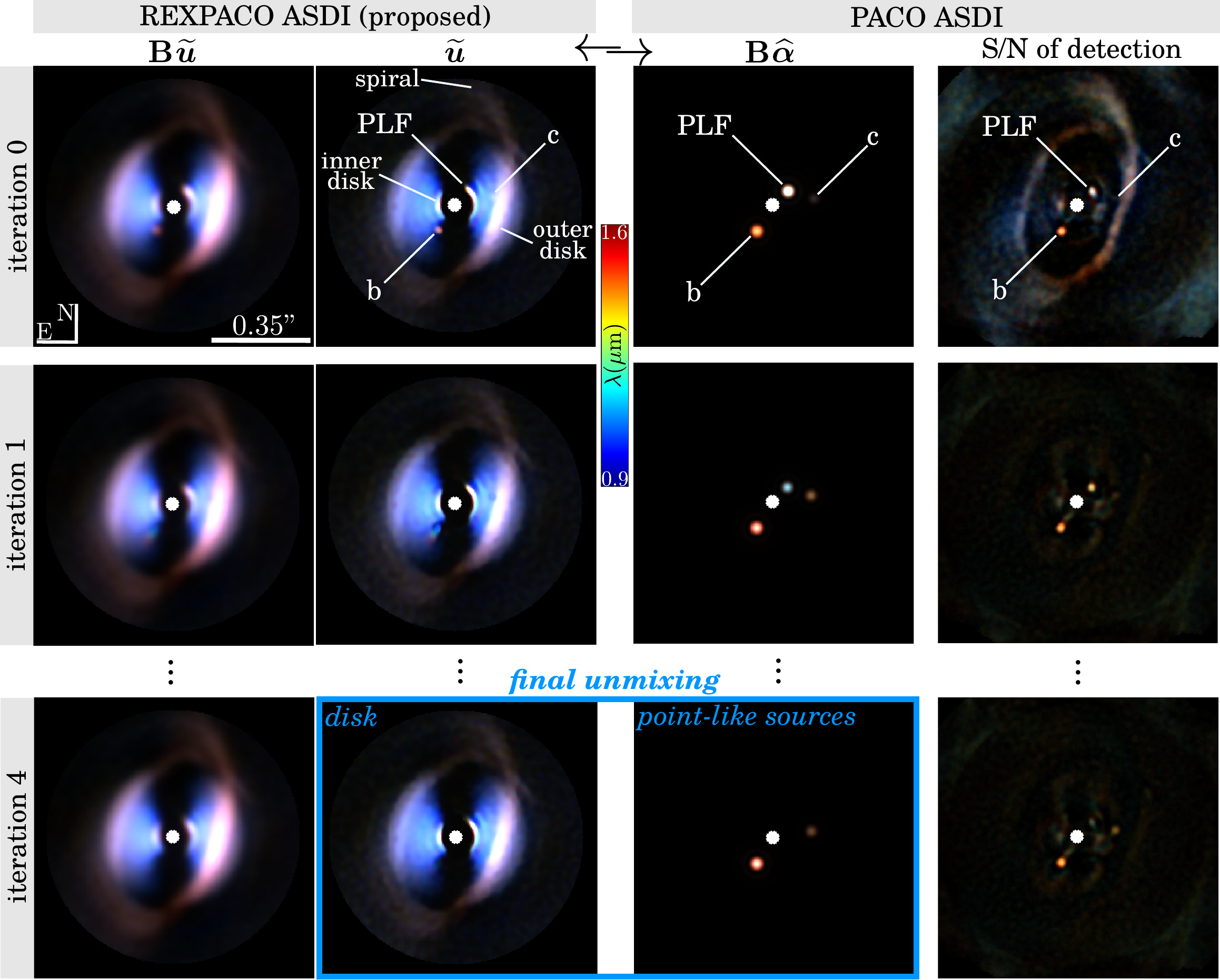}
	\caption{Unmixing disk and point-like components by combining REXPACO ASDI and PACO ASDI. The two algorithms are applied independently at iteration 0 of Algorithm \ref{alg:rexpaco_paco}, while they are applied on the current residual data (i.e., subtraction to the data of the current contribution estimated by the algorithm lastly applied) for the other iterations. Concerning the reconstruction of the disk component, the spatio-spectral flux distribution $\widetilde{\obj}$ and its re-blurred version $\M B \, \widetilde{\obj}$ obtained with REXPACO ASDI are displayed for each iteration. Concerning the estimation of the point-like contributions, the spatio-spectral S/N maps of detection and the estimated flux contribution maps $\M B \, \widehat{\V \alpha}$ obtained with PACO ASDI are reported for each iteration. As the reported S/N comes from a detection algorithm, it should not be interpreted as a proper image of the multi-spectral flux distribution. The flux maps are non-null only at the locations of the three characterized point-like sources. Dataset: PDS 70 (2018-02-24), see Sect. \ref{subsec:datasets} for the observation parameters.}
	\label{fig:pds70_fullfig}
\end{figure*}

\noindent We first evaluate the unmixing ability of the proposed algorithm through numerical experiments on a SPHERE-IFS dataset of PDS 70. We injected (not simultaneously) six faint point-like sources, and we disregarded the unmixing of the real known exoplanets, focusing solely on separating the synthetic sources from the circumstellar environment. Figure \ref{fig:estimated_spectrum} compares the estimated SED of the synthetic sources across various iterations of Algorithm \ref{alg:rexpaco_paco}. It shows that estimation errors decrease over iterations, generally converging towards zero (except for the first spectral channels of source \#3 that display a remaining discrepancy with the ground truth). The errors are larger when the SED of the point-like sources closely resembles that of the star (i.e., for sources \#1, \#2 and \#3), as the disk material shares spectral similarities with it, making unmixing more ambiguous. Overall, these results demonstrate the capability of the proposed approach to effectively disentangle the signal from point-like sources, even when they are partially buried into disk material.

As a case-study, we apply Algorithm \ref{alg:rexpaco_paco} on the same SPHERE-IFS dataset of PDS 70, focusing now on unmixing real point-like sources. We recall that PDS 70 hosts two known exoplanets \citep{keppler2018discovery,haffert2019two} in accretion phase within a protoplanetary disk \citep{Isella2019}, see also Sect. \ref{subsec:recons_real_disks}.
Based on the processing of the same dataset, \cite{mesa2019vlt} also identified a point-like feature (PLF) with several post-processing algorithms dedicated to the detection of point-like sources. As in \cite{mesa2019vlt}, the independent application of PACO ASDI allows to identify a PLF in the spatio-spectral S/N maps produced with PACO ASDI, see iteration 0 in Fig. \ref{fig:pds70_fullfig}. Exoplanet PDS 70 c cannot be detected in the same S/N maps, likely due to its proximity to the disk material which is over-subtracted by PACO ASDI, since it is not specifically designed to preserve extended structures. Scrutinizing the REXPACO ASDI reconstruction allows to detect PDS 70 b and c, appearing as red point-like sources, even though they are embedded within the disk material. The outer and inner structures of the disks, as well as the spiral feature identified by \cite{juillard2022analysis} from SPHERE-IRDIS observations are also reconstructed. 
At the first application of REXPACO ASDI, the PLF seems to be more likely a part of the inner disk hosted by the star, see iteration 0 in Fig. \ref{fig:pds70_fullfig}. Iterating between REXPACO ASDI and PACO ASDI leads to several remarks. First, the extractions of PDS 70 b and PDS 70 c improve along the iterations. As a qualitative illustration, the REXPACO ASDI reconstruction obtained after a single iteration with Algorithm \ref{alg:rexpaco_paco} 
exhibits a discontinuous footprint within the disk material at the location of the two exoplanets, as a sign of an overestimation of their contribution by PACO ASDI. At convergence of the proposed unmixing scheme, the disk component appears smooth and continuous at the locations of the two exoplanets without any residual signature of PDS 70 b and c. Across the iterations, the contribution of the PLF in the sparse component decreases since it is increasingly  explained by the disk component. At convergence of the iterative procedure, the residual spatio-spectral S/N maps from PACO ASDI are almost free from the disk contribution and the signature of the PLF is significantly attenuated with respect to the initial S/N maps at iteration 0. These results also support the conclusions of \cite{mesa2019vlt} likely attributing the PLF as a part of the disk, based on its estimated photometry (its SED being very similar to the disk one).

\begin{figure}
	\centering
	\includegraphics[width=0.48\textwidth]{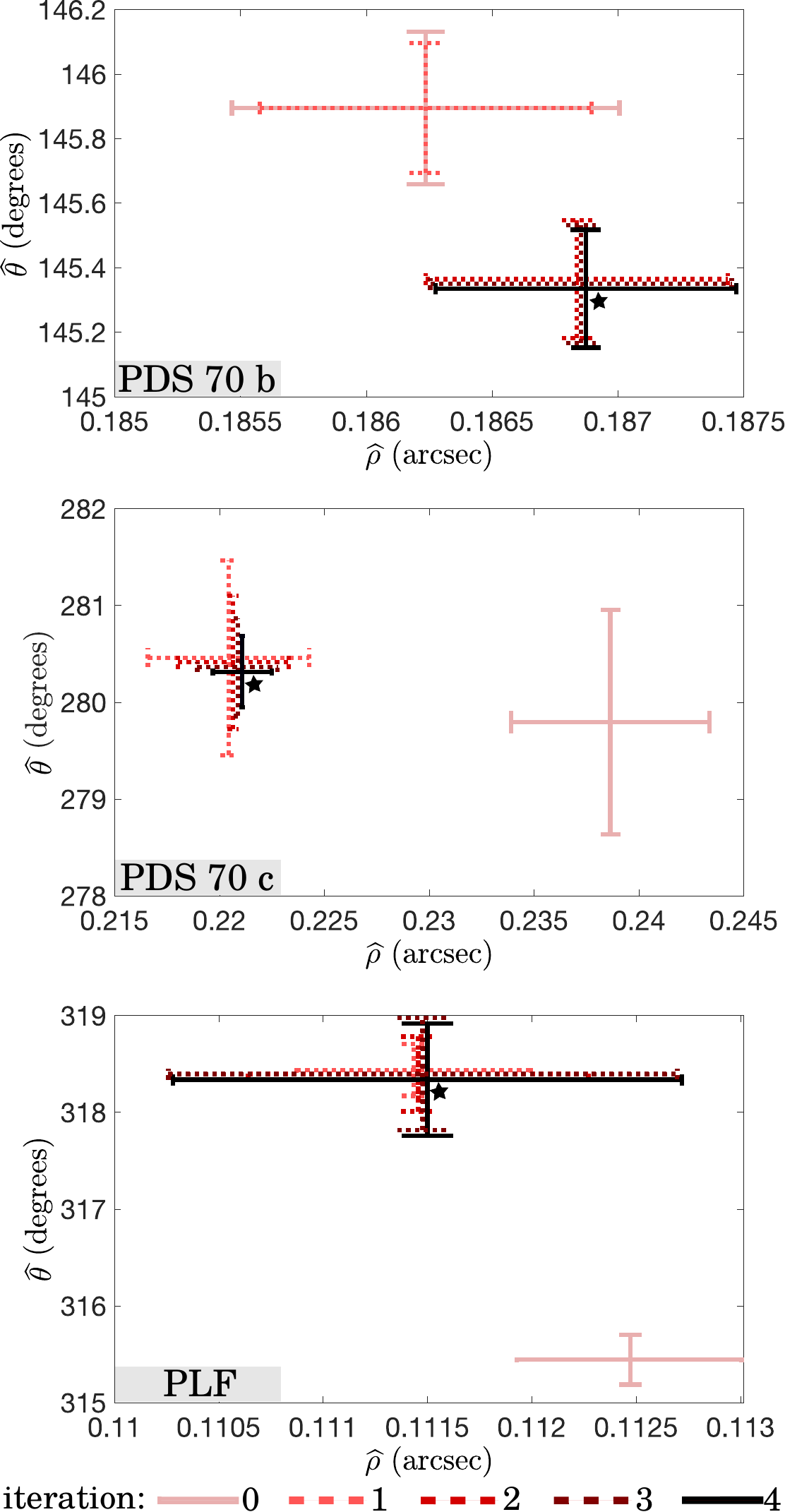}
	\caption{Unmixing disk and point-like components by combining REXPACO ASDI and PACO ASDI. The estimated astrometry along the iterations are reported for the three considered point-like sources (PDS 70 b, PDS 70 c and the PLF). When estimations are similar from one iteration to the other, the error-bars are slightly shifted artificially to better see the evolution of the estimation accuracy. These cases are marked by a star symbol and the common estimated astrometry is given by the highest iteration. Dataset: PDS 70 (2018-02-24), see Sect. \ref{subsec:datasets} for the observation parameters.}
	\label{fig:pds70_astrometry}
\end{figure}

\begin{table} %
			\caption{Estimated astrometry $(\widehat{\rho},\,\widehat{\theta})$ of PDS 70 b and c as well as the candidate PLF. Values obtained with our unmixing scheme combining REXPACO ASDI and PACO ASDI are compared to the values reported in the literature on data from the same instrument taken at the same observation date.}
			\centering%
			\begin{tabular}{cccc}
				\toprule
				Source & $\widehat{\rho}$ (mas) & $\widehat{\theta}$ (degrees) & reference\\
				\midrule
				PDS 70 b & 192.2 $\pm$ 8.0 & 146.8 $\pm$ 2.4 & \cite{muller2018orbital}\\
				PDS 70 b & 186.8 $\pm$ 0.2 & 145.4 $\pm$ 0.1 & this paper\\
				\midrule
				PDS 70 c & 209 $\pm$ 13 & 281.2 $\pm$ 0.5 & \cite{mesa2019vlt}\\
				PDS 70 c & 211.0 $\pm$ 0.5 & 280.3 $\pm$ 0.1 & this paper\\
				\midrule
				PLF & 118 $\pm$ 4 & 316.8 $\pm$ 0.5 & \cite{mesa2019vlt}\\
				PLF & 111.5 $\pm$ 0.3 & 318.4 $\pm$ 0.1 & this paper\\
				\bottomrule
			\end{tabular}
			\label{tab:astrometry_real_sources}
\end{table}

Figure \ref{fig:pds70_astrometry} completes this study by showing the estimated astrometry of PDS 70 b and c, as well as of the PLF along the iterations of the unmixing method. It shows, that the estimated astrometry of the three sources evolves during the iterations. The estimated angular separation $\widehat{\rho}$ evolves up to 15 mas (i.e., 2 pixels) and the estimated parallactic angle $\widehat{\theta}$ evolves up to 0.5 degree for PDS 70 c (located very near the outer disk arm), which is an illustration of the impact of the disk material on the characterization and orbital parameters estimation of point-like sources embedded within. The estimation shift both in angular separation and in parallactic angle is even more important for the PLF since its sparse contribution gets fainter during the iterations and does not resemble to a point-source anymore. In addition, the accuracy of astrometry improves (i.e., the error bars get smaller) for PDS 70 b and c while it degrades (i.e., the error bars get larger) for the PLF. This observation is in agreement with the qualitative results presented in Fig. \ref{fig:pds70_fullfig} attributing preferentially the PLF as part of the disk component. Table \ref{tab:astrometry_real_sources} reports our final astrometric measurements obtained for the considered sources. The retrieved values are compared to the most accurate measurements available in the literature using direct imaging for these three sources and at the same observation date. Overall, our estimations are compatible (within two times the standard-deviation, at most) with the values reported in the literature. However, our estimations are much more accurate: the uncertainties are decreased by a factor between 5 and 40. If the astrometric estimations we derived are confirmed (e.g., based on a multi-epochs analysis), they could be significant corrective factors of the orbit of the exoplanet PDS 70 b and c. 

\medskip

\noindent Beyond the benefits of the proposed iterative approach to unmix point-like sources from the circumstellar environment, this study illustrates that applying a post-processing algorithm not specifically designed for the recovery of extended sources can lead to critical artifacts and biases. In particular, it can lead to misinterpret a disk feature for a point-like source. These observations could encourage to revisit systems where candidate point-like sources embedded in disk material were recently identified via a post-processing of the data by algorithms not tailored to reconstruct extended features and even less to unmix disk and point-like components.

\section{Discussion and conclusion}
\label{sec:conclusion}

In this paper, we introduced REXPACO ASDI, a new algorithm for reconstructing circumstellar environments from high-contrast observations in pupil-tracking mode. Our approach utilizes spectral diversity inherent in ASDI data. REXPACO ASDI combines a tailored statistical model of non-stationary nuisances with a forward image formation model of the off-axis sources. These models are jointly used to solve a reconstruction task in a regularized inverse problem framework. This method, specifically designed for extended sources, is the first to leverage jointly angular and spectral diversity introduced by ASDI for reconstructing the spatio-spectral flux distribution of circumstellar environments.

\medskip

\noindent On the methodological side, we employ a local modeling approach to capture spatial and spectral correlations of nuisances for a more accurate statistical description of the data. This model utilizes a spatio-spectral separable approximation to reduce the large number of free parameters needed to model full covariances. For similar reasons, the model is local, i.e. its parameters differ with the location in the field of view and are estimated at the scale of small patches. Our model can thus be interpreted as a block-diagonal approximation of the full spatio-spectral covariance. Tailored estimators of model parameters, based on covariance shrinkage, are developed to reduce estimation uncertainty and improve robustness. We illustrate on real data that this approximate statistical model effectively captures most nuisance correlations. Ablation study reveals that jointly accounting for spatio-spectral correlations directly from the data is crucial for capturing accurately the statistics of ASDI observations, outperforming methods that first model spatial correlations from spatio-temporo-spectral data and then spectral correlations from reduced quantities, as in our previous work dedicated to exoplanet detection from similar ASDI observations \citep{flasseur2020paco}.

We proposed a specific reconstruction strategy to refine jointly the statistical model of the nuisance and the reconstructed flux distribution of the circumstellar environment. This hierarchical estimation strategy derives estimators of the nuisance component mostly unbiased from the contamination of the sought off-axis objects. This method also prevents iterating between the characterization of the nuisance and the reconstruction task, thus leading to an algorithm that scales to the size of typical datasets recorded with the ASDI technique, both in terms of computational burden and memory storage.
We apply regularization to the spatio-spectral flux distribution using suitable penalties. These penalties improve both rejection of residual starlight and fidelity of reconstructed features. We demonstrate the versatility of these priors in recovering various structures within the circumstellar environment, such as sharp edges and smooth transitions.

REXPACO ASDI operates in a fully unsupervised manner, allowing optimal estimation of all hyper-parameters from the dataset itself, without relying on prior knowledge about the disk properties or requiring trial and error reconstructions. Among the free hyper-parameters, the patch size is set based on the full width at half maximum of the off-axis PSF. The spatially adaptive regularization of noisy covariances through shrinkage is obtained via a derived closed-form expression, minimizing estimation risk for the statistical nuisance model. Hyper-parameters that determine the relative weights of reconstruction regularization can be estimated quasi-optimally by minimizing Stein's unbiased risk estimator. However this process is time-consuming because it requires multiple reconstructions with different penalty weights. 
As this setting is not the most critical, it can be approximated from the optimal setting obtained on a standard dataset by scaling regularization parameters with respect to the acquired number of frames.

\medskip

\noindent We tested the proposed algorithm using injection of synthetic disks with different morphologies, orientations, and contrast levels. While these simulations could be complemented and refined by even more extensive experiments, they allowed to identify the key capabilities and benefits of REXPACO ASDI. We showed that the proposed method is very versatile since it is able to reconstruct faithful spectral images of the considered disks for contrasts up to $10^{-6}$. One of our major result is the ability of REXPACO ASDI to reconstruct disks being partly rotation-invariant, i.e. whose morphology makes the unmixing of the disk and the speckles particularly difficult when only leveraging the ADI diversity.
These disks are known to be especially challenging to reconstruct without an additional source of diversity in the data, for instance provided by multiple observations as in RDI. Unlike this latter category of methods, the unmixing capability of REXPACO ASDI is achieved from a single ASDI dataset, i.e. the model of the nuisance is dataset-dependent. Using simulated flux distributions, we also illustrated that the theoretical fraction of flux lost due to unmixing ambiguities is negligible if the different spectral channels are processed jointly. This property of ASDI is due to the chromatic scaling of the speckle field caused by the diffraction.

By resorting to a model ablation, applied both on synthetic and real disks, we illustrated that the joint spectral processing of REXPACO ASDI efficiently unmixes disk features from the nuisance and requires an accurate model of spatio-spectral correlations that are very strong in ASDI observations. Ignoring these correlations in the statistical model of the nuisance is particularly detrimental to the quality of the reconstruction. 

As a proof of concept, we analyzed real SPHERE-IFS datasets containing six known circumstellar disks with various morphologies, including challenging features like spiral arms. Despite the absence of ground truth for these real objects, we observed that our method outperformed median ASDI, PCA ASDI, PACO ASDI, and the mono-spectral version of REXPACO in rejecting nuisances. Our approach significantly reduced non-physical artifacts, such as discontinuities from partial self-subtraction. Additionally, the reconstructed flux distribution showed improved spatial resolution compared to the original data, as we accounted for the blur introduced by the off-axis PSF through deconvolution in the forward image formation model. We also processed a dual-band dataset obtained with the IRDIS imager of the SPHERE instrument. Although spectral diversity was limited in this dataset, we illustrated that our approach enhanced reconstruction quality compared to the mono-spectral REXPACO algorithm designed for ADI observations.

Given the complementary capabilities of REXPACO ASDI, we can expect that it will be helpful to unveil new disks, to improve the spatio-spectral interpretation of their flux distribution, and thus to better understand the phenomena governing the formation of planetary systems like the intricate interactions between exoplanets and the disk material. In particular, we illustrated that the latter goal can be achieved by combining REXPACO ASDI with the detection algorithm PACO ASDI to unmix point-like sources from the circumstellar material.
As initialization step, this latter strategy only requires the rough locations (typically, with pixel-level accuracy) of candidate point-like sources to be unmixed from the disk material. Based on numerical experiments, we illustrated that this combined approach can reduce significantly the photometry bias occurring during characterization of point-like sources embedded within disk material. As a case-study, we applied this strategy on a dataset of PDS 70. Our results illustrated the ability of the proposed approach to identify components being more likely disk features than point-like sources, even when they are mistaken as point-sources at initialization step.

\medskip

\noindent As future work, we plan to improve the fidelity of the model of the nuisance, especially in the vicinity of the star where the model is slightly inaccurate.
 Disk reconstruction is very challenging in this area due to large stellar leakages that could be more accurately captured by accounting for the spatial correlations at a larger spatial scale than a patch of a few pixels. Complementary to that, even the spectral diversity is very useful to retrieve faithful flux distribution, a distortion can remains in some cases.
 This limitation could be tackled by building a more complex model leveraging deep learning techniques to model the nuisance distribution from multiple archival data.

Beyond the specific field of application of the proposed algorithm, its statistical modeling of the spatio-spectral correlations of the nuisance component and the estimation strategy of the underlying parameters are very general approaches. These methodological developments could be specialized to other large-scale reconstruction problems encountered in other imaging modalities such as microscopy or remote sensing. These fields often involve multi-spectral measurements, where signals of interest are faint and affected by multi-correlated and non-stationary nuisances.

\section*{Acknowledgements}

We thank the anonymous Referee for her/his careful reading of the manuscript as well as her/his insightful comments and suggestions.

This project was funded in part by the French National Research Agency (ANR) under the project DDISK (grant ANR-21-CE31-0015) and by the Région Auvergne-Rhône-Alpes under the project DIAGHOLO. This work was also supported by the ANR under the France 2030 program (PEPR Origins, reference ANR-22-EXOR-0016), by the French National Programs (PNP and PNPS), and by the Action Spécifique Haute Résolution Angulaire (ASHRA) of CNRS/INSU co-funded by CNES.

OF, LD, and {\'E}T conceived and designed the method presented in this paper. OF developed, tested, and implemented the algorithm. OF selected the raw data. ML pre-reduced them through the SPHERE Data Center. OF performed the analysis of the data. OF, LD, {\'E}T, and ML wrote the manuscript.

\section*{Data Availability}

The raw data used in this article are freely available on the ESO archive facility at \href{http://archive.eso.org/eso/eso\_archive\_main.html}{http://archive.eso.org/eso/eso\_archive\_main.html}. They were pre-reduced with the SPHERE Data Centre, jointly operated by OSUG/IPAG (Grenoble), PYTHEAS/LAM/CESAM (Marseille), OCA/Lagrange (Nice), Observatoire de Paris/LESIA (Paris), and Observatoire de Lyon/CRAL (Lyon, France). The resulting pre-processed datasets will be shared based on reasonable request to the corresponding author.



\bibliographystyle{mnras}
\bibliography{rexpaco_asdi_mnras_2024_archive_version} 

\begin{thebibliography}{}
\makeatletter
\relax
\def\mn@urlcharsother{\let\do\@makeother \do\$\do\&\do\#\do\^\do\_\do\%\do\~}
\def\mn@doi{\begingroup\mn@urlcharsother \@ifnextchar [ {\mn@doi@}
  {\mn@doi@[]}}
\def\mn@doi@[#1]#2{\def\@tempa{#1}\ifx\@tempa\@empty \href
  {http://dx.doi.org/#2} {doi:#2}\else \href {http://dx.doi.org/#2} {#1}\fi
  \endgroup}
\def\mn@eprint#1#2{\mn@eprint@#1:#2::\@nil}
\def\mn@eprint@arXiv#1{\href {http://arxiv.org/abs/#1} {{\tt arXiv:#1}}}
\def\mn@eprint@dblp#1{\href {http://dblp.uni-trier.de/rec/bibtex/#1.xml}
  {dblp:#1}}
\def\mn@eprint@#1:#2:#3:#4\@nil{\def\@tempa {#1}\def\@tempb {#2}\def\@tempc
  {#3}\ifx \@tempc \@empty \let \@tempc \@tempb \let \@tempb \@tempa \fi \ifx
  \@tempb \@empty \def\@tempb {arXiv}\fi \@ifundefined
  {mn@eprint@\@tempb}{\@tempb:\@tempc}{\expandafter \expandafter \csname
  mn@eprint@\@tempb\endcsname \expandafter{\@tempc}}}

\bibitem[\protect\citeauthoryear{Aharon, Elad  \& Bruckstein}{Aharon
  et~al.}{2006}]{aharon2006k}
Aharon M.,  Elad M.,   Bruckstein A.,  2006, IEEE Transactions on Signal
  Processing, 54, 4311

\bibitem[\protect\citeauthoryear{Amara \& Quanz}{Amara \&
  Quanz}{2012}]{amara2012pynpoint}
Amara A.,  Quanz S.~P.,  2012, Monthly Notices of the Royal Astronomical
  Society, 427, 948

\bibitem[\protect\citeauthoryear{Bae, Zhu  \& Hartmann}{Bae
  et~al.}{2016}]{bae2016planetary}
Bae J.,  Zhu Z.,   Hartmann L.,  2016, The Astrophysical Journal, 819, 134

\bibitem[\protect\citeauthoryear{Bell, Mamajek  \& Naylor}{Bell
  et~al.}{2015}]{bell2015self}
Bell C.~P.,  Mamajek E.~E.,   Naylor T.,  2015, Monthly Notices of the Royal
  Astronomical Society, 454, 593

\bibitem[\protect\citeauthoryear{Benisty et~al.,}{Benisty
  et~al.}{2015}]{benisty2015asymmetric}
Benisty M.,  et~al., 2015, Astronomy \& Astrophysics, 578, L6

\bibitem[\protect\citeauthoryear{Beuzit et~al.,}{Beuzit
  et~al.}{2019}]{beuzit2019sphere}
Beuzit J.-L.,  et~al., 2019, Astronomy \& Astrophysics, 631, A155

\bibitem[\protect\citeauthoryear{Blomgren, Chan, Mulet  \& Wong}{Blomgren
  et~al.}{1997}]{blomgren1997total}
Blomgren P.,  Chan T.~F.,  Mulet P.,   Wong C.-K.,  1997, in Proceedings of
  international conference on image processing. pp 384--387

\bibitem[\protect\citeauthoryear{Boccaletti et~al.,}{Boccaletti
  et~al.}{2020}]{boccaletti2020possible}
Boccaletti A.,  et~al., 2020, Astronomy \& Astrophysics, 637, L5

\bibitem[\protect\citeauthoryear{Boccaletti et~al.,}{Boccaletti
  et~al.}{2021}]{boccaletti2021investigating}
Boccaletti A.,  et~al., 2021, Astronomy \& Astrophysics, 652, L8

\bibitem[\protect\citeauthoryear{Bodrito, Flasseur, Mairal, Ponce, Langlois  \&
  Lagrange}{Bodrito et~al.}{2024}]{bodrito2024modelco}
Bodrito T.,  Flasseur O.,  Mairal J.,  Ponce J.,  Langlois M.,   Lagrange
  A.-M.,  2024, Monthly Notices of the Royal Astronomical Society, p. stae2174

\bibitem[\protect\citeauthoryear{Bowler}{Bowler}{2016}]{bowler2016imaging}
Bowler B.~P.,  2016, Publications of the Astronomical Society of the Pacific,
  128, 102001

\bibitem[\protect\citeauthoryear{Bresson \& Chan}{Bresson \&
  Chan}{2008}]{bresson2008fast}
Bresson X.,  Chan T.~F.,  2008, Inverse Problems \& Imaging, 2, 455

\bibitem[\protect\citeauthoryear{Brown et~al.,}{Brown
  et~al.}{2016}]{brown2016gaia}
Brown A.~G.,  et~al., 2016, Astronomy \& Astrophysics, 595, A2

\bibitem[\protect\citeauthoryear{Brown et~al.,}{Brown
  et~al.}{2021}]{brown2021gaia}
Brown A.~G.,  et~al., 2021, Astronomy \& Astrophysics, 649, A1

\bibitem[\protect\citeauthoryear{Buades, Coll  \& Morel}{Buades
  et~al.}{2005}]{buades2005non}
Buades A.,  Coll B.,   Morel J.-M.,  2005, in IEEE Computer Society Conference
  on Computer Vision and Pattern Recognition. pp 60--65

\bibitem[\protect\citeauthoryear{Carbillet et~al.,}{Carbillet
  et~al.}{2011}]{carbillet2011apodized}
Carbillet M.,  et~al., 2011, Experimental Astronomy, 30, 39

\bibitem[\protect\citeauthoryear{Charbonnier, Blanc-F{\'e}raud, Aubert  \&
  Barlaud}{Charbonnier et~al.}{1997}]{charbonnier1997deterministic}
Charbonnier P.,  Blanc-F{\'e}raud L.,  Aubert G.,   Barlaud M.,  1997, IEEE
  Transactions on image processing, 6, 298

\bibitem[\protect\citeauthoryear{Chen, Wiesel, Eldar  \& Hero}{Chen
  et~al.}{2010}]{chen2010shrinkage}
Chen Y.,  Wiesel A.,  Eldar Y.~C.,   Hero A.~O.,  2010, IEEE Transactions on
  Signal Processing, 58, 5016

\bibitem[\protect\citeauthoryear{Chintarungruangchai, Jiang, Hashimoto, Komatsu
   \& Konishi}{Chintarungruangchai
  et~al.}{2023}]{chintarungruangchai2023possible}
Chintarungruangchai P.,  Jiang G.,  Hashimoto J.,  Komatsu Y.,   Konishi M.,
  2023, New Astronomy, 100, 101997

\bibitem[\protect\citeauthoryear{Christiaens et~al.,}{Christiaens
  et~al.}{2019}]{christiaens2019separating}
Christiaens V.,  et~al., 2019, Monthly Notices of the Royal Astronomical
  Society, 486, 5819

\bibitem[\protect\citeauthoryear{Christiaens et~al.,}{Christiaens
  et~al.}{2023}]{christiaens2023vip}
Christiaens V.,  et~al., 2023, Journal of Open Source Software, 8

\bibitem[\protect\citeauthoryear{Christiaens et~al.,}{Christiaens
  et~al.}{2024}]{christiaens2024minds}
Christiaens V.,  et~al., 2024, Astronomy \& Astrophysics, 685, L1

\bibitem[\protect\citeauthoryear{Conte, Lops  \& Ricci}{Conte
  et~al.}{1995}]{381910}
Conte E.,  Lops M.,   Ricci G.,  1995, IEEE Transactions on Aerospace and
  Electronic Systems, 31, 617

\bibitem[\protect\citeauthoryear{Craven \& Wahba}{Craven \&
  Wahba}{1978}]{craven1978smoothing}
Craven P.,  Wahba G.,  1978, Numerische mathematik, 31, 377

\bibitem[\protect\citeauthoryear{Currie et~al.,}{Currie
  et~al.}{2017}]{currie2017subaru}
Currie T.,  et~al., 2017, The Astrophysical Journal Letters, 836, L15

\bibitem[\protect\citeauthoryear{Currie, Biller, Lagrange, Marois, Guyon,
  Nielsen, Bonnefoy  \& De~Rosa}{Currie et~al.}{2022a}]{currie2022direct}
Currie T.,  Biller B.,  Lagrange A.-M.,  Marois C.,  Guyon O.,  Nielsen E.,
  Bonnefoy M.,   De~Rosa R.,  2022a, arXiv preprint arXiv:2205.05696

\bibitem[\protect\citeauthoryear{Currie et~al.,}{Currie
  et~al.}{2022b}]{currie2022images}
Currie T.,  et~al., 2022b, Nature Astronomy, 6, 751

\bibitem[\protect\citeauthoryear{Dabov, Foi, Katkovnik  \& Egiazarian}{Dabov
  et~al.}{2007}]{dabov2007image}
Dabov K.,  Foi A.,  Katkovnik V.,   Egiazarian K.,  2007, IEEE Transactions on
  Image Processing, 16, 2080

\bibitem[\protect\citeauthoryear{Delorme et~al.,}{Delorme
  et~al.}{2017}]{delorme2017sphere}
Delorme P.,  et~al., 2017, arXiv preprint arXiv:1712.06948

\bibitem[\protect\citeauthoryear{Dohlen et~al.,}{Dohlen
  et~al.}{2008}]{dohlen2008infra}
Dohlen K.,  et~al., 2008, in Ground-based and Airborne Instrumentation for
  Astronomy II. pp 1266--1275

\bibitem[\protect\citeauthoryear{Doucet, Pantin, Lagage  \& Dullemond}{Doucet
  et~al.}{2006}]{doucet2006mid}
Doucet C.,  Pantin E.,  Lagage P.,   Dullemond C.,  2006, Astronomy \&
  Astrophysics, 460, 117

\bibitem[\protect\citeauthoryear{Esposito, Fitzgerald, Graham  \&
  Kalas}{Esposito et~al.}{2013}]{esposito2013modeling}
Esposito T.~M.,  Fitzgerald M.~P.,  Graham J.~R.,   Kalas P.,  2013, The
  Astrophysical Journal, 780, 25

\bibitem[\protect\citeauthoryear{{Esposito} et~al.,}{{Esposito}
  et~al.}{2020}]{Esposito2020}
{Esposito} T.~M.,  et~al., 2020, \mn@doi [The Astronomical Journal]
  {10.3847/1538-3881/ab9199}, \href
  {https://ui.adsabs.harvard.edu/abs/2020AJ....160...24E} {160, 24}

\bibitem[\protect\citeauthoryear{Flasseur, Denis, Thi{\'e}baut  \&
  Langlois}{Flasseur et~al.}{2018}]{flasseur2018exoplanet}
Flasseur O.,  Denis L.,  Thi{\'e}baut {\'E}.,   Langlois M.,  2018, Astronomy
  \& Astrophysics, 618, A138

\bibitem[\protect\citeauthoryear{Flasseur, Denis, Thi{\'e}baut  \&
  Langlois}{Flasseur et~al.}{2020a}]{flasseur2020robustness}
Flasseur O.,  Denis L.,  Thi{\'e}baut {\'E}.,   Langlois M.,  2020a, Astronomy
  \& Astrophysics, 634, A2

\bibitem[\protect\citeauthoryear{Flasseur, Denis, Thi{\'e}baut  \&
  Langlois}{Flasseur et~al.}{2020b}]{flasseur2020paco}
Flasseur O.,  Denis L.,  Thi{\'e}baut {\'E}.,   Langlois M.,  2020b, Astronomy
  \& Astrophysics, 637, A9

\bibitem[\protect\citeauthoryear{Flasseur, Th{\'e}, Denis, Thi{\'e}baut  \&
  Langlois}{Flasseur et~al.}{2021}]{flasseur2021rexpaco}
Flasseur O.,  Th{\'e} S.,  Denis L.,  Thi{\'e}baut {\'E}.,   Langlois M.,
  2021, Astronomy \& Astrophysics, 651, A62

\bibitem[\protect\citeauthoryear{Flasseur, Denis, Thi{\'e}baut, Langlois
  et~al.}{Flasseur et~al.}{2022}]{flasseur2022multispectral}
Flasseur O.,  Denis L.,  Thi{\'e}baut {\'E}.,  Langlois M.,   et~al., 2022, in
  Adaptive Optics Systems VIII. pp 1175--1189

\bibitem[\protect\citeauthoryear{Flasseur, Bodrito, Mairal, Ponce, Langlois  \&
  Lagrange}{Flasseur et~al.}{2023a}]{flasseur2023combining}
Flasseur O.,  Bodrito T.,  Mairal J.,  Ponce J.,  Langlois M.,   Lagrange
  A.-M.,  2023a, in 2023 31st European Signal Processing Conference (EUSIPCO).
  pp 1723--1727

\bibitem[\protect\citeauthoryear{Flasseur, Bodrito, Mairal, Ponce, Langlois  \&
  Lagrange}{Flasseur et~al.}{2023b}]{flasseur2024deep}
Flasseur O.,  Bodrito T.,  Mairal J.,  Ponce J.,  Langlois M.,   Lagrange
  A.-M.,  2023b, Monthly Notices of the Royal Astronomical Society, 527, 1534

\bibitem[\protect\citeauthoryear{Flasseur, Thi{\'e}baut, Denis  \&
  Langlois}{Flasseur et~al.}{2024}]{flasseur2024shrinkage}
Flasseur O.,  Thi{\'e}baut E.,  Denis L.,   Langlois M.,  2024, accepted in
  EUSIPCO, arXiv preprint arXiv:2403.07104

\bibitem[\protect\citeauthoryear{Follette}{Follette}{2023}]{follette2023introduction}
Follette K.~B.,  2023, Publications of the Astronomical Society of the Pacific,
  135, 093001

\bibitem[\protect\citeauthoryear{Gaia et~al.,}{Gaia
  et~al.}{2018}]{gaia2018gaia}
Gaia C.,  et~al., 2018, Astronomy \& Astrophysics, 616

\bibitem[\protect\citeauthoryear{Garufi et~al.,}{Garufi
  et~al.}{2020}]{garufi2020disks}
Garufi A.,  et~al., 2020, Astronomy \& Astrophysics, 633, A82

\bibitem[\protect\citeauthoryear{Genton}{Genton}{2007}]{genton2007separable}
Genton M.~G.,  2007, Environmetrics: The official Journal of the International
  Environmetrics Society, 18, 681

\bibitem[\protect\citeauthoryear{Girard}{Girard}{1989}]{girard1989fmcsure}
Girard D.~A.,  1989, Numerische Mathematik, 56, 1

\bibitem[\protect\citeauthoryear{Gonzalez et~al.,}{Gonzalez
  et~al.}{2017}]{gonzalez2017vip}
Gonzalez C. A.~G.,  et~al., 2017, The Astronomical Journal, 154, 7

\bibitem[\protect\citeauthoryear{Grady et~al.,}{Grady
  et~al.}{2009}]{grady2009revealing}
Grady C.,  et~al., 2009, The Astrophysical Journal, 699, 1822

\bibitem[\protect\citeauthoryear{Haffert, Bohn, de Boer, Snellen, Brinchmann,
  Girard, Keller  \& Bacon}{Haffert et~al.}{2019}]{haffert2019two}
Haffert S.,  Bohn A.,  de Boer J.,  Snellen I.,  Brinchmann J.,  Girard J.,
  Keller C.,   Bacon R.,  2019, Nature Astronomy, 3, 749

\bibitem[\protect\citeauthoryear{Hom et~al.,}{Hom
  et~al.}{2024}]{hom2024uniform}
Hom J.,  et~al., 2024, Monthly Notices of the Royal Astronomical Society, 528,
  6959

\bibitem[\protect\citeauthoryear{Isella, Testi, Natta, Neri, Wilner  \&
  Qi}{Isella et~al.}{2007}]{isella2007millimeter}
Isella A.,  Testi L.,  Natta A.,  Neri R.,  Wilner D.,   Qi C.,  2007,
  Astronomy \& Astrophysics, 469, 213

\bibitem[\protect\citeauthoryear{Isella et~al.,}{Isella
  et~al.}{2018}]{isella2018disk}
Isella A.,  et~al., 2018, The Astrophysical Journal Letters, 869, L49

\bibitem[\protect\citeauthoryear{Isella, Benisty, Teague, Bae, Keppler,
  Facchini  \& P{\'e}rez}{Isella et~al.}{2019}]{Isella2019}
Isella A.,  Benisty M.,  Teague R.,  Bae J.,  Keppler M.,  Facchini S.,
  P{\'e}rez L.,  2019, The Astrophysical Journal Letters, 879, L25

\bibitem[\protect\citeauthoryear{Juillard, Christiaens  \& Absil}{Juillard
  et~al.}{2022}]{juillard2022analysis}
Juillard S.,  Christiaens V.,   Absil O.,  2022, Astronomy \& Astrophysics,
  668, A125

\bibitem[\protect\citeauthoryear{Juillard, Christiaens  \& Absil}{Juillard
  et~al.}{2023}]{juillard2023inverse}
Juillard S.,  Christiaens V.,   Absil O.,  2023, Astronomy \& Astrophysics,
  679, A52

\bibitem[\protect\citeauthoryear{Juillard, Stasevic, Christiaens, Absil  \&
  Milli}{Juillard et~al.}{2024}]{juillard2024combining}
Juillard S.,  Stasevic S.,  Christiaens V.,  Absil O.,   Milli J.,  2024,
  Astronomy \& Astrophysics, 688, A185

\bibitem[\protect\citeauthoryear{Keppler et~al.,}{Keppler
  et~al.}{2018}]{keppler2018discovery}
Keppler M.,  et~al., 2018, Astronomy \& Astrophysics, 617, A44

\bibitem[\protect\citeauthoryear{Kiefer, Bohn, Quanz, Kenworthy  \&
  Stolker}{Kiefer et~al.}{2021}]{kiefer2021spectral}
Kiefer S.,  Bohn A.~J.,  Quanz S.~P.,  Kenworthy M.,   Stolker T.,  2021,
  Astronomy \& Astrophysics, 652, A33

\bibitem[\protect\citeauthoryear{Kingma \& Ba}{Kingma \&
  Ba}{2014}]{kingma2014adam}
Kingma D.~P.,  Ba J.,  2014, arXiv preprint arXiv:1412.6980

\bibitem[\protect\citeauthoryear{Lafreni{\`e}re, Marois, Doyon, Nadeau  \&
  Artigau}{Lafreni{\`e}re et~al.}{2007}]{lafreniere2007new}
Lafreni{\`e}re D.,  Marois C.,  Doyon R.,  Nadeau D.,   Artigau E.,  2007, The
  Astrophysical Journal, 660, 770

\bibitem[\protect\citeauthoryear{Lafreni{\`e}re, Marois, Doyon  \&
  Barman}{Lafreni{\`e}re et~al.}{2009}]{lafreniere2009hst}
Lafreni{\`e}re D.,  Marois C.,  Doyon R.,   Barman T.,  2009, The Astrophysical
  Journal, 694, L148

\bibitem[\protect\citeauthoryear{Lagrange et~al.,}{Lagrange
  et~al.}{2009}]{lagrange2009probable}
Lagrange A.-M.,  et~al., 2009, Astronomy \& Astrophysics, 493, L21

\bibitem[\protect\citeauthoryear{Lagrange et~al.,}{Lagrange
  et~al.}{2010}]{lagrange2010giant}
Lagrange A.-M.,  et~al., 2010, Science, 329, 57

\bibitem[\protect\citeauthoryear{Langlois, Gratton, Lagrange, Delorme,
  Boccaletti, Bonnefoy, Maire  et~al.}{Langlois et~al.}{2020}]{Langlois2020}
Langlois M.,  Gratton R.,  Lagrange A.-M.,  Delorme P.,  Boccaletti A.,
  Bonnefoy M.,  Maire A.-L.,   et~al., 2020, in revision for Astronomy \&
  Astrophysics

\bibitem[\protect\citeauthoryear{Langlois et~al.,}{Langlois
  et~al.}{2021}]{langlois2021sphere}
Langlois M.,  et~al., 2021, Astronomy \& Astrophysics, 651, A71

\bibitem[\protect\citeauthoryear{Lawson et~al.,}{Lawson
  et~al.}{2020}]{lawson2020scexao}
Lawson K.,  et~al., 2020, The Astronomical Journal, 160, 163

\bibitem[\protect\citeauthoryear{Lawson, Currie, Wisniewski, Groff, McElwain
  \& Schlieder}{Lawson et~al.}{2022}]{lawson2022constrained}
Lawson K.,  Currie T.,  Wisniewski J.~P.,  Groff T.~D.,  McElwain M.~W.,
  Schlieder J.~E.,  2022, The Astrophysical Journal Letters, 935, L25

\bibitem[\protect\citeauthoryear{Lebrun, Buades  \& Morel}{Lebrun
  et~al.}{2013}]{lebrun2013nonlocal}
Lebrun M.,  Buades A.,   Morel J.-M.,  2013, SIAM Journal on Imaging Sciences,
  6, 1665

\bibitem[\protect\citeauthoryear{Ledoit \& Wolf}{Ledoit \&
  Wolf}{2004}]{ledoit2004well}
Ledoit O.,  Wolf M.,  2004, Journal of Multivariate Analysis, 88, 365

\bibitem[\protect\citeauthoryear{Lisse, Chen, Wyatt, Morlok, Song, Bryden  \&
  Sheehan}{Lisse et~al.}{2009}]{lisse2009abundant}
Lisse C.~M.,  Chen C.,  Wyatt M.,  Morlok A.,  Song I.,  Bryden G.,   Sheehan
  P.,  2009, The Astrophysical Journal, 701, 2019

\bibitem[\protect\citeauthoryear{Louchet \& Moisan}{Louchet \&
  Moisan}{2008}]{louchet2008total}
Louchet C.,  Moisan L.,  2008, in 2008 16th European Signal Processing
  Conference. pp~1--5

\bibitem[\protect\citeauthoryear{Lu \& Zimmerman}{Lu \&
  Zimmerman}{2005}]{lu2005likelihood}
Lu N.,  Zimmerman D.~L.,  2005, Statistics \& Probability Letters, 73, 449

\bibitem[\protect\citeauthoryear{Mairal, Bach, Ponce, Sapiro  \&
  Zisserman}{Mairal et~al.}{2009}]{mairal2009non}
Mairal J.,  Bach F.,  Ponce J.,  Sapiro G.,   Zisserman A.,  2009, in IEEE
  International Conference on Computer Vision. pp 2272--2279

\bibitem[\protect\citeauthoryear{Maire et~al.,}{Maire
  et~al.}{2017}]{maire2017testing}
Maire A.-L.,  et~al., 2017, Astronomy \& Astrophysics, 601, A134

\bibitem[\protect\citeauthoryear{Marois, Lafreni{\`e}re, Doyon, Macintosh  \&
  Nadeau}{Marois et~al.}{2006}]{marois2006angular}
Marois C.,  Lafreni{\`e}re D.,  Doyon R.,  Macintosh B.,   Nadeau D.,  2006,
  The Astrophysical Journal, 641, 556

\bibitem[\protect\citeauthoryear{Marois, Macintosh, Barman, Zuckerman, Song,
  Patience, Lafreni{\`e}re  \& Doyon}{Marois et~al.}{2008}]{marois2008direct}
Marois C.,  Macintosh B.,  Barman T.,  Zuckerman B.,  Song I.,  Patience J.,
  Lafreni{\`e}re D.,   Doyon R.,  2008, science, 322, 1348

\bibitem[\protect\citeauthoryear{Marois, Zuckerman, Konopacky, Macintosh  \&
  Barman}{Marois et~al.}{2010}]{marois2010images}
Marois C.,  Zuckerman B.,  Konopacky Q.~M.,  Macintosh B.,   Barman T.,  2010,
  Nature, 468, 1080

\bibitem[\protect\citeauthoryear{Marois, Correia, V{\'e}ran  \& Currie}{Marois
  et~al.}{2013}]{marois2013tloci}
Marois C.,  Correia C.,  V{\'e}ran J.-P.,   Currie T.,  2013, International
  Astronomical Union, 8, 48

\bibitem[\protect\citeauthoryear{Marois, Correia, Galicher, Ingraham,
  Macintosh, Currie  \& De~Rosa}{Marois et~al.}{2014}]{marois2014gpi}
Marois C.,  Correia C.,  Galicher R.,  Ingraham P.,  Macintosh B.,  Currie T.,
   De~Rosa R.,  2014, in SPIE Astronomical Intrumentation + Telescopes. p.
  91480U

\bibitem[\protect\citeauthoryear{Mazoyer et~al.,}{Mazoyer
  et~al.}{2020}]{mazoyer2020diskfm}
Mazoyer J.,  et~al., 2020, in Ground-based and Airborne Instrumentation for
  Astronomy VIII. pp 1080--1099

\bibitem[\protect\citeauthoryear{Mesa et~al.,}{Mesa
  et~al.}{2019a}]{mesa2019determining}
Mesa D.,  et~al., 2019a, Monthly Notices of the Royal Astronomical Society,
  488, 37

\bibitem[\protect\citeauthoryear{Mesa et~al.,}{Mesa
  et~al.}{2019b}]{mesa2019vlt}
Mesa D.,  et~al., 2019b, Astronomy \& Astrophysics, 632, A25

\bibitem[\protect\citeauthoryear{Milli, Mouillet, Lagrange, Boccaletti, Mawet,
  Chauvin  \& Bonnefoy}{Milli et~al.}{2012}]{milli2012impact}
Milli J.,  Mouillet D.,  Lagrange A.-M.,  Boccaletti A.,  Mawet D.,  Chauvin
  G.,   Bonnefoy M.,  2012, Astronomy \& Astrophysics, 545, A111

\bibitem[\protect\citeauthoryear{Milli et~al.,}{Milli
  et~al.}{2017}]{milli2017near}
Milli J.,  et~al., 2017, Astronomy \& Astrophysics, 599, A108

\bibitem[\protect\citeauthoryear{Milli et~al.,}{Milli
  et~al.}{2019}]{milli2019optical}
Milli J.,  et~al., 2019, Astronomy \& Astrophysics, 626, A54

\bibitem[\protect\citeauthoryear{M{\"u}ller, van~den Ancker, Launhardt, Pott,
  Fedele  \& Henning}{M{\"u}ller et~al.}{2011}]{muller2011hd}
M{\"u}ller A.,  van~den Ancker M.,  Launhardt R.,  Pott J.-U.,  Fedele D.,
  Henning T.,  2011, Astronomy \& Astrophysics, 530, A85

\bibitem[\protect\citeauthoryear{M{\"u}ller et~al.,}{M{\"u}ller
  et~al.}{2018}]{muller2018orbital}
M{\"u}ller A.,  et~al., 2018, Astronomy \& Astrophysics, 617, L2

\bibitem[\protect\citeauthoryear{Muro-Arena et~al.,}{Muro-Arena
  et~al.}{2018}]{muro2018dust}
Muro-Arena G.,  et~al., 2018, Astronomy \& Astrophysics, 614, A24

\bibitem[\protect\citeauthoryear{Muro-Arena et~al.,}{Muro-Arena
  et~al.}{2020}]{muro2020shadowing}
Muro-Arena G.,  et~al., 2020, Astronomy \& Astrophysics, 635, A121

\bibitem[\protect\citeauthoryear{Nielsen \& Close}{Nielsen \&
  Close}{2010}]{nielsen2010uniform}
Nielsen E.~L.,  Close L.~M.,  2010, The Astrophysical Journal, 717, 878

\bibitem[\protect\citeauthoryear{Nielsen, Close, Biller, Masciadri  \&
  Lenzen}{Nielsen et~al.}{2008}]{nielsen2008constraints}
Nielsen E.~L.,  Close L.~M.,  Biller B.~A.,  Masciadri E.,   Lenzen R.,  2008,
  The Astrophysical Journal, 674, 466

\bibitem[\protect\citeauthoryear{Pairet, Jacques  \& Cantalloube}{Pairet
  et~al.}{2019}]{pairet2019iterative}
Pairet B.,  Jacques L.,   Cantalloube F.,  2019, Signal Processing with
  Adaptive Sparse Structured Representations, 1, 1

\bibitem[\protect\citeauthoryear{Pairet, Cantalloube  \& Jacques}{Pairet
  et~al.}{2021}]{pairet2020mayonnaise}
Pairet B.,  Cantalloube F.,   Jacques L.,  2021, Monthly Notices of the Royal
  Astronomical Society, 503, 3724

\bibitem[\protect\citeauthoryear{Pavlov, M{\"o}ller-Nilsson, Feldt, Henning,
  Beuzit  \& Mouillet}{Pavlov et~al.}{2008}]{pavlov2008sphere}
Pavlov A.,  M{\"o}ller-Nilsson O.,  Feldt M.,  Henning T.,  Beuzit J.-L.,
  Mouillet D.,  2008, Advanced Software and Control for Astronomy II, 7019,
  1093

\bibitem[\protect\citeauthoryear{Pueyo}{Pueyo}{2018}]{pueyo2018direct}
Pueyo L.,  2018, Handbook of Exoplanets, pp 705--765

\bibitem[\protect\citeauthoryear{Ramani, Liu, Rosen, Nielsen  \&
  Fessler}{Ramani et~al.}{2012}]{ramani2012regularization}
Ramani S.,  Liu Z.,  Rosen J.,  Nielsen J.-F.,   Fessler J.~A.,  2012, IEEE
  Transactions on Image Processing, 21, 3659

\bibitem[\protect\citeauthoryear{Reggiani et~al.,}{Reggiani
  et~al.}{2018}]{reggiani2018discovery}
Reggiani M.,  et~al., 2018, Astronomy \& Astrophysics, 611, A74

\bibitem[\protect\citeauthoryear{Ren}{Ren}{2023}]{ren2023karhunen}
Ren B.~B.,  2023, Astronomy \& Astrophysics, 679, A18

\bibitem[\protect\citeauthoryear{Ren, Pueyo, Zhu, Debes  \& Duch{\^e}ne}{Ren
  et~al.}{2018}]{ren2018non}
Ren B.,  Pueyo L.,  Zhu G.~B.,  Debes J.,   Duch{\^e}ne G.,  2018, The
  Astrophysical Journal, 852, 104

\bibitem[\protect\citeauthoryear{Ren, Pueyo, Chen, Choquet, Debes, Duch{\^e}ne,
  M{\'e}nard  \& Perrin}{Ren et~al.}{2020}]{ren2020using}
Ren B.,  Pueyo L.,  Chen C.,  Choquet {\'E}.,  Debes J.~H.,  Duch{\^e}ne G.,
  M{\'e}nard F.,   Perrin M.~D.,  2020, The Astrophysical Journal, 892, 74

\bibitem[\protect\citeauthoryear{Riaud, Mawet, Absil, Boccaletti, Baudoz,
  Herwats  \& Surdej}{Riaud et~al.}{2006}]{riaud2006coronagraphic}
Riaud P.,  Mawet D.,  Absil O.,  Boccaletti A.,  Baudoz P.,  Herwats E.,
  Surdej J.,  2006, Astronomy \& Astrophysics, 458, 317

\bibitem[\protect\citeauthoryear{Ruane et~al.,}{Ruane
  et~al.}{2019}]{ruane2019reference}
Ruane G.,  et~al., 2019, The Astronomical Journal, 157, 118

\bibitem[\protect\citeauthoryear{Schneider et~al.,}{Schneider
  et~al.}{1999}]{schneider1999nicmos}
Schneider G.,  et~al., 1999, The Astrophysical Journal Letters, 513, L127

\bibitem[\protect\citeauthoryear{Sch{\"u}tz, Meeus  \& Sterzik}{Sch{\"u}tz
  et~al.}{2005}]{schutz2005mid}
Sch{\"u}tz O.,  Meeus G.,   Sterzik M.,  2005, Astronomy \& Astrophysics, 431,
  175

\bibitem[\protect\citeauthoryear{Smith \& Terrile}{Smith \&
  Terrile}{1984}]{smith1984circumstellar}
Smith B.~A.,  Terrile R.~J.,  1984, Science, 226, 1421

\bibitem[\protect\citeauthoryear{Soummer, Pueyo  \& Larkin}{Soummer
  et~al.}{2012}]{soummer2012detection}
Soummer R.,  Pueyo L.,   Larkin J.,  2012, The Astrophysical Journal Letters,
  755, L28

\bibitem[\protect\citeauthoryear{Sparks \& Ford}{Sparks \&
  Ford}{2002}]{sparks2002imaging}
Sparks W.~B.,  Ford H.~C.,  2002, The Astrophysical Journal, 578, 543

\bibitem[\protect\citeauthoryear{{Stapper, L. M.} \& {Ginski, C.}}{{Stapper, L.
  M.} \& {Ginski, C.}}{2022}]{stapper2022iterative}
{Stapper, L. M.} {Ginski, C.} 2022, Astronomy \& Astrophysics, 668

\bibitem[\protect\citeauthoryear{Stein}{Stein}{1981}]{stein1981estimation}
Stein C.~M.,  1981, The Annals of Statistics, pp 1135--1151

\bibitem[\protect\citeauthoryear{Teague, Bae, Bergin, Birnstiel  \&
  Foreman-Mackey}{Teague et~al.}{2018}]{teague2018kinematical}
Teague R.,  Bae J.,  Bergin E.~A.,  Birnstiel T.,   Foreman-Mackey D.,  2018,
  The Astrophysical Journal Letters, 860, L12

\bibitem[\protect\citeauthoryear{Thatte, Abuter, Tecza, Nielsen, Clarke  \&
  Close}{Thatte et~al.}{2007}]{thatte2007very}
Thatte N.,  Abuter R.,  Tecza M.,  Nielsen E.~L.,  Clarke F.~J.,   Close L.~M.,
   2007, Monthly Notices of the Royal Astronomical Society, 378, 1229

\bibitem[\protect\citeauthoryear{Thi{\'e}baut}{Thi{\'e}baut}{2002}]{thiebaut2002optimization}
Thi{\'e}baut {\'E}.,  2002, in Astronomical Data Analysis II. pp 174--183

\bibitem[\protect\citeauthoryear{Tilling et~al.,}{Tilling
  et~al.}{2012}]{tilling2012gas}
Tilling I.,  et~al., 2012, Astronomy \& Astrophysics, 538, A20

\bibitem[\protect\citeauthoryear{Traub \& Oppenheimer}{Traub \&
  Oppenheimer}{2010}]{traub2010direct}
Traub W.~A.,  Oppenheimer B.~R.,  2010, Exoplanets, pp 111--156

\bibitem[\protect\citeauthoryear{Van~Leeuwen}{Van~Leeuwen}{2007}]{van2007validation}
Van~Leeuwen F.,  2007, Astronomy \& Astrophysics, 474, 653

\bibitem[\protect\citeauthoryear{Vigan, Moutou, Langlois, Allard, Boccaletti,
  Carbillet, Mouillet  \& Smith}{Vigan et~al.}{2010}]{vigan2010photometric}
Vigan A.,  Moutou C.,  Langlois M.,  Allard F.,  Boccaletti A.,  Carbillet M.,
  Mouillet D.,   Smith I.,  2010, Monthly Notices of the Royal Astronomical
  Society, 407, 71

\bibitem[\protect\citeauthoryear{Vigan et~al.,}{Vigan
  et~al.}{2014}]{vigan2014sphere}
Vigan A.,  et~al., 2014, in Ground-based and Airborne Instrumentation for
  Astronomy V. pp 1568--1577

\bibitem[\protect\citeauthoryear{Wagner, Stone, Spalding, Apai, Dong, Ertel,
  Leisenring  \& Webster}{Wagner et~al.}{2019}]{wagner2019thermal}
Wagner K.,  Stone J.~M.,  Spalding E.,  Apai D.,  Dong R.,  Ertel S.,
  Leisenring J.,   Webster R.,  2019, The Astrophysical Journal, 882, 20

\bibitem[\protect\citeauthoryear{Wagner et~al.,}{Wagner
  et~al.}{2023}]{wagner2023direct}
Wagner K.,  et~al., 2023, Nature Astronomy, 7, 1208

\bibitem[\protect\citeauthoryear{Wahba et~al.}{Wahba
  et~al.}{1985}]{wahba1985comparison}
Wahba G.,  et~al., 1985, The Annals of Statistics, 13, 1378

\bibitem[\protect\citeauthoryear{Wahhaj et~al.,}{Wahhaj
  et~al.}{2015}]{wahhaj2015improving}
Wahhaj Z.,  et~al., 2015, Astronomy \& Astrophysics, 581, A24

\bibitem[\protect\citeauthoryear{Wahhaj et~al.,}{Wahhaj
  et~al.}{2021}]{wahhaj2021search}
Wahhaj Z.,  et~al., 2021, Astronomy \& Astrophysics, 648, A26

\bibitem[\protect\citeauthoryear{Wainwright \& Simoncelli}{Wainwright \&
  Simoncelli}{1999}]{wainwright1999scale}
Wainwright M.~J.,  Simoncelli E.~P.,  1999, in Neural Information Processing
  Systems. pp 855--861

\bibitem[\protect\citeauthoryear{Werner, Jansson  \& Stoica}{Werner
  et~al.}{2008}]{werner2008estimation}
Werner K.,  Jansson M.,   Stoica P.,  2008, IEEE Transactions on Signal
  Processing, 56, 478

\bibitem[\protect\citeauthoryear{Wolf, Jones  \& Bowler}{Wolf
  et~al.}{2024}]{wolf2024direct}
Wolf T.~N.,  Jones B.~A.,   Bowler B.~P.,  2024, The Astronomical Journal, 167,
  92

\bibitem[\protect\citeauthoryear{Xie et~al.,}{Xie
  et~al.}{2022}]{xie2022reference}
Xie C.,  et~al., 2022, arXiv preprint arXiv:2208.07915

\bibitem[\protect\citeauthoryear{Xuan et~al.,}{Xuan
  et~al.}{2018}]{xuan2018characterizing}
Xuan W.~J.,  et~al., 2018, The Astronomical Journal, 156, 156

\bibitem[\protect\citeauthoryear{Yu, Sapiro  \& Mallat}{Yu
  et~al.}{2011}]{yu2011solving}
Yu G.,  Sapiro G.,   Mallat S.,  2011, IEEE Transactions on Image Processing,
  21, 2481

\bibitem[\protect\citeauthoryear{Zhu, Byrd, Lu  \& Nocedal}{Zhu
  et~al.}{1997}]{zhu1997algorithm}
Zhu C.,  Byrd R.~H.,  Lu P.,   Nocedal J.,  1997, ACM Transactions on
  Mathematical Software, 23, 550

\bibitem[\protect\citeauthoryear{Zoran \& Weiss}{Zoran \&
  Weiss}{2011}]{zoran2011learning}
Zoran D.,  Weiss Y.,  2011, in IEEE International Conference on Computer
  Vision. pp 479--486

\makeatother
\end{thebibliography}




\appendix

\section{Derivation of the maximum likelihood estimators for a weighted mixture of multi-variate Gaussian models}
\label{sec:app_proofMLE}

In this appendix, we detail the technical elements yielding to the MLEs
(\ref{eq:MLE_mu})-(\ref{eq:MLE_Cspat}) of the parameters of a weighted mixture of
multi-variate Gaussian, knowing the object of interest $\obj$, see
Sect.~\ref{subsubsec:mles}.

Under the assumptions of Sect.~\ref{sec:covstruct}, the co-log-likelihood of the
4D patch $\data_n$ is given by
  Eq.~\eqref{eq:patchcologlikelihood} and can be rewritten as:
\begin{align}
  \mathscr{L}_n
  &=  \frac{T\,K}{2}\,\log\Abs[\big]{\M C_n^\spec} + \frac{T\,L}{2}\log\Abs[\big]{\M C_n^\spat}
    \notag\\
  &\quad + \sum_{t=1}^T \Paren*{
    \frac{K\,L}{2}\,\log\sigma_{n,t}^2 + \frac{1}{2\,\sigma_{n,t}^2}\,
    \Norm*{\V r_{n,t}}_{{\big( \M C_n^\spec \big)^{-1} \otimes \big( \M C_n^\spat \big)^{-1} }}^2}\,,
  \label{eq:proof_prop1_inter}
\end{align}
with:
\begin{equation}
  \label{eq:patchresiduals}
  \V r_{n,t} = \V v_{n,t} - \V\mu_{n}^\spec - [\M M\,\obj]_{n,t}
\end{equation}
the residuals in the $t$-th frame of the $n$-th patch, and where we used the
following properties of the Kronecker product of any $n\times n$ matrix $\M A$ and
$m\times m$ matrix $\M B$: $|\M A\otimes\M B|=|\M A|^m|\M B|^n$ and
$(\M A\otimes\M B)^{-1}=\M A^{-1}\otimes\M B^{-1}$.

To obtain the MLEs, we differentiate the expression of $\mathscr{L}_{n}$ given in
Eq.~\eqref{eq:proof_prop1_inter}:
\begin{align}
  \dd\mathscr{L}_n
  =&\tfrac{T\,K}{2}\,\Trace\Paren[\Bigg]{\left({\M C_n^\spec}\right)^{-1}\,\dd\M C_n^\spec}
  + \tfrac{TL}{2}\,\Trace\Paren[\Bigg]{\left({\M C_n^\spat}\right)^{-1}\,\dd\M C_n^\spat}
  \notag\\
  + \sum_{t=1}^T
  &\biggl\{\tfrac{KL}{2}\frac{\dd\sigma_{n,t}^2}{\sigma_{n,t}^2}
    -\frac{\dd\sigma_{n,t}^2}{2\sigma_{n,t}^4}\,
    \Norm*{\V r_{n,t}}_{ \big( {\M C_n^\spec}\big)^{-1} \otimes{ \big( \M C_n^\spat \big) }^{-1}}^2
    \notag\\
  & + \frac{1}{\sigma_{n,t}^2}\,\V r_{n,t}\T\,\Paren*{\left({\M C_n^\spec}\right)^{-1}\otimes{\left(\M C_n^\spat\right)}^{-1}}\,
    \dd\V r_{n,t}
  \notag\\
  & - \frac{1}{2\,\sigma_{n,t}^2}\,
    \Trace\Paren*{{\left(\M C_n^\spec\right)}^{-1}\dd\M C_n^\spec\,{\left(\M C_n^\spec\right)}^{-1}\,{\M V}_{n,t}\T\,
    {\left( \M C_n^\spat \right) }^{-1}\,{\M V}_{n,t}}
    \notag\\
  & - \frac{1}{2\,\sigma_{n,t}^2}\,
    \Trace\Paren*{{\left( \M C_n^\spat \right) }^{-1}\,\dd\M C_n^\spat\,{\left( \M C_n^\spat \right)}^{-1}\,{\M V}_{n,t}\,
    {\left( \M C_n^\spec \right)}^{-1}\,{\M V}_{n,t}\T}
    \biggr\}\,,
\end{align}
where we obtained the last two terms by rewriting the squared norm term in
Eq.~\eqref{eq:proof_prop1_inter} as:
\begin{equation}
  \Norm{\V r_{n,t}}_{\big({\M C_n^\spec}\big)^{-1}\otimes{\big( \M C_n^\spat \big)}^{-1}}^2 =
  \Trace\Paren[\Bigg]{{\M V}_{n,t}\T\,\left({\M C_n^\spat}\right)^{-1}\,{\M V}_{n,t}\,\left({\M C_n^\spec}\right)^{-1}}
\end{equation}
with ${\M V}_{n,t}$ the $K\times L$ matrix whose element at row $k$ and column
$\ell$ is $[\V r_{n,t}]_{k,\ell}$. The following set of conditions is sufficient for
the partial derivatives of $\mathscr{L}_{n}$ in respectively $\V\mu^\spec$,
  $\sigma_{n,t}^2$, $\M C_n^\spec$, and $\M C_n^\spat$ to be equal to zero:
\begin{equation}
    \begin{cases}
      \sum_{t=1}^T \sigma_{n,t}^{-2}\,\V r_{n,t} = 0\,,
      \\[2ex]
      \frac{K\,L}{\sigma_{n,t}^2} - \frac{1}{\sigma_{n,t}^4}\,
      \Norm*{\V r_{n,t}}_{ {\big( \M C_n^\spec\big)}^{-1} \otimes {\big( \M C_n^\spat \big)}^{-1} }^2 = 0\,,
      \\[2ex]
      T\,K\,\M I - \sum\limits_{t=1}^T\frac{1}{\sigma_{n,t}^2}\,{\left( \M C_n^\spec \right)}^{-1}\,
      {\M V}_{n,t}\T\,{\left( \M C_n^\spat \right)}^{-1}\,{\M V}_{n,t} = \M 0\,,
      \\[2ex]
      T\,L\,\M I - \sum\limits_{t=1}^T\frac{1}{\sigma_{n,t}^2}\,{\left( \M C_n^\spat \right)}^{-1}\,
      {\M V}_{n,t}\,{\left( \M C_n^\spec \right)}^{-1}\,{\M V}_{n,t}\T = \M 0\,,
    \end{cases}
\end{equation}
with $\M I$ the identity matrix. These conditions hold if:
\begin{align}
	\begin{cases}
      \widehat{\V\mu}_n^{\,\spec} = \frac{\sum_{t=1}^{T} \widehat{\sigma}_{n,t}^{-2} \,
    \Paren*{\V v_{n,t} - [\M M\,\widehat{\obj}]_{n,t}}}%
    {\sum_{t=1}^T \widehat{\sigma}_{n,t}^{-2}}\,,
    \\[2ex]
      \widehat{\sigma}_{n,t}^2 = \tfrac{1}{K\,L} \,
      \Norm*{\V v_{n,t} - \widehat{\V\mu}_{n}^\spec - [\M M\,\obj]_{n,t}}_{{\big(\widehat{\M C}_n^\spec\big)^{-1} \otimes \big( \widehat{\M C}_n^\spat\big)^{-1}}}^2\,,
      \\[2ex]
      \widehat{\M C}_n^\spec = \tfrac{1}{T\,K}\sum\limits_{t=1}^T \widehat{\M V}_{n,t}\T \,
      \Paren*{\widehat{\sigma}_{n,t}^{2} \, \widehat{\M C}_n^\spat}^{-1} \,
      \widehat{\M V}_{n,t}\,,
      \\[2ex]
      \widehat{\M C}_n^\spat = \tfrac{1}{T\,L}\sum\limits_{t=1}^T \widehat{\M V}_{n,t} \,
      \Paren*{\widehat{\sigma}_{n,t}^{2} \, \widehat{\M C}_n^\spec}^{-1} \,
      \widehat{\M V}_{n,t}\T\,.
    \end{cases}
\end{align}
These correspond to the expressions given in Eqs.~\eqref{eq:MLE_mu}--\eqref{eq:MLE_Cspat}.

\section{Additional reconstruction results on simulated synthetic disks}
\label{app:additional_results}

\begin{figure*}
	\centering
	\includegraphics[width=\textwidth]{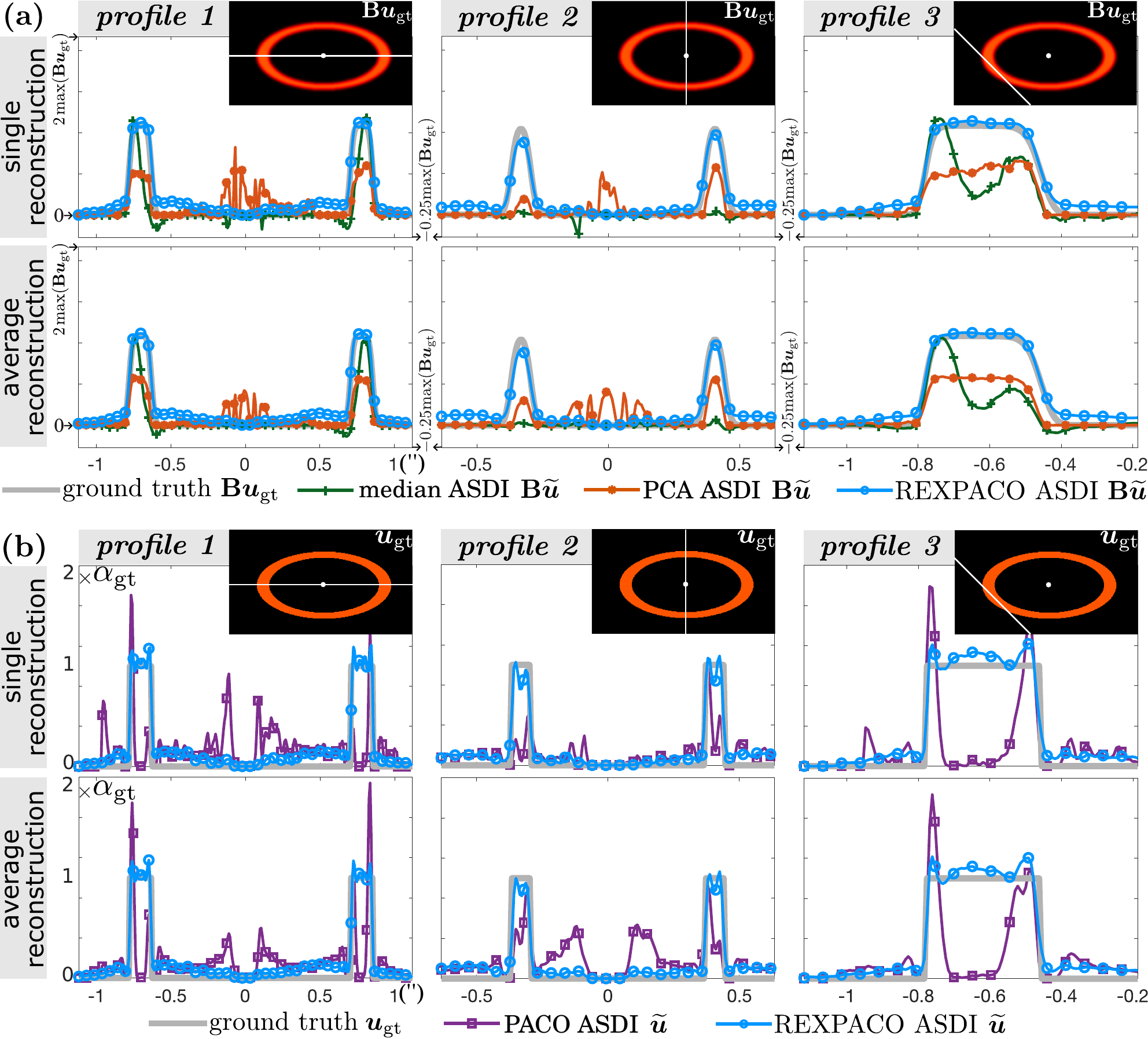}
	\caption{Line cuts along the three profiles defined in Fig. \ref{fig:gt_fullfig} extracted from the reconstructions of synthetic elliptical disks shown in Figs. \ref{fig:ellipse_blurred_fullfig}-\ref{fig:ellipse_deblurred_fullfig} for a contrast $\alpha_{\text{gt}} = 5\times 10^{-6}$. Panel (a) compares median ASDI, PCA ASDI and REXPACO ASDI reconstructions $\M B \, \widetilde{\obj}$ to the ground truth $\M B \, \obj_{\text{gt}}$. These quantities are shown within the range of $[ -0.25\times\text{max}(\M B \, \obj_{\text{gt}}) ; 2 \times \text{max}(\M B \, \obj_{\text{gt}}) ]$ to highlight both over-estimation and under-estimation of the signal of interest. The minimum value (zero) of the ground truth $\M B \, \obj_{\text{gt}}$ is also marked on the right vertical axis. Panel (b) compared PACO ASDI and REXPACO ASDI reconstructions $\widetilde{\obj}$ against the ground truth $\obj_{\text{gt}}$. The ground truth flux distribution $\M B \, {\obj}_{\text{gt}}$ and the associated slice-cut locations are recalled in insets. Dataset: HD 172555 (2015-07-11), see Table \ref{tab:dataset_logs} for the observation parameters.}
	\label{fig:ellipse_cuts_fullfig}
\end{figure*}

\begin{figure*}
	\centering
	\includegraphics[width=\textwidth]{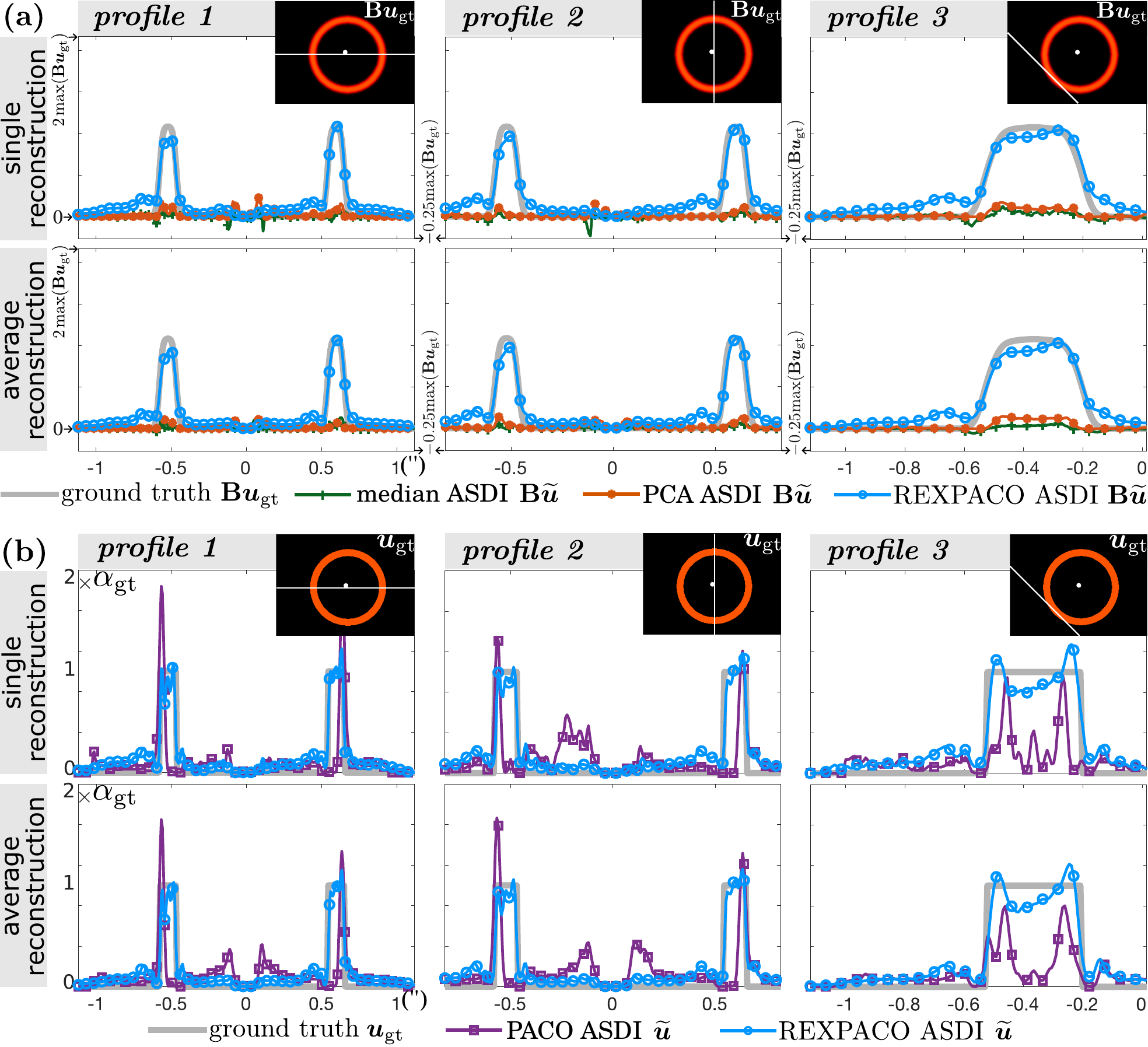}
	\caption{Same than Fig. \ref{fig:ellipse_cuts_fullfig} for synthetic circular disks, see reconstructed flux distributions in Figs. \ref{fig:circle_blurred_fullfig}-\ref{fig:circle_deblurred_fullfig}.}
	\label{fig:circle_cuts_fullfig}
\end{figure*}

\begin{figure*}
	\centering
	\includegraphics[width=\textwidth]{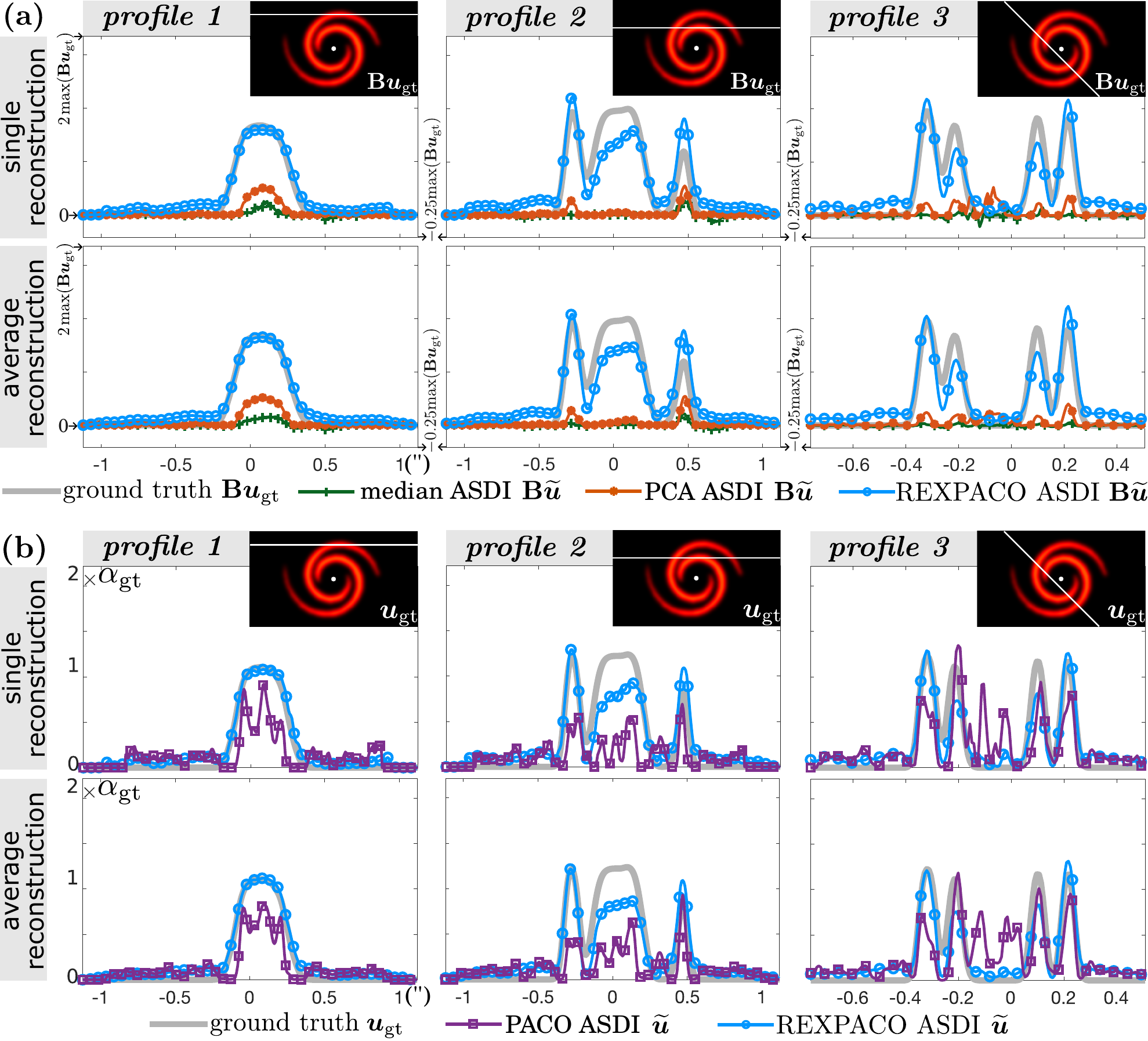}
	\caption{Same than Fig. \ref{fig:ellipse_cuts_fullfig} for synthetic spiral disks, see reconstructed flux distributions in Figs. \ref{fig:spiral_blurred_fullfig}-\ref{fig:spiral_deblurred_fullfig}.}
	\label{fig:spiral_cuts_fullfig}
\end{figure*}

\begin{figure*}
	\centering
	\includegraphics[width=\textwidth]{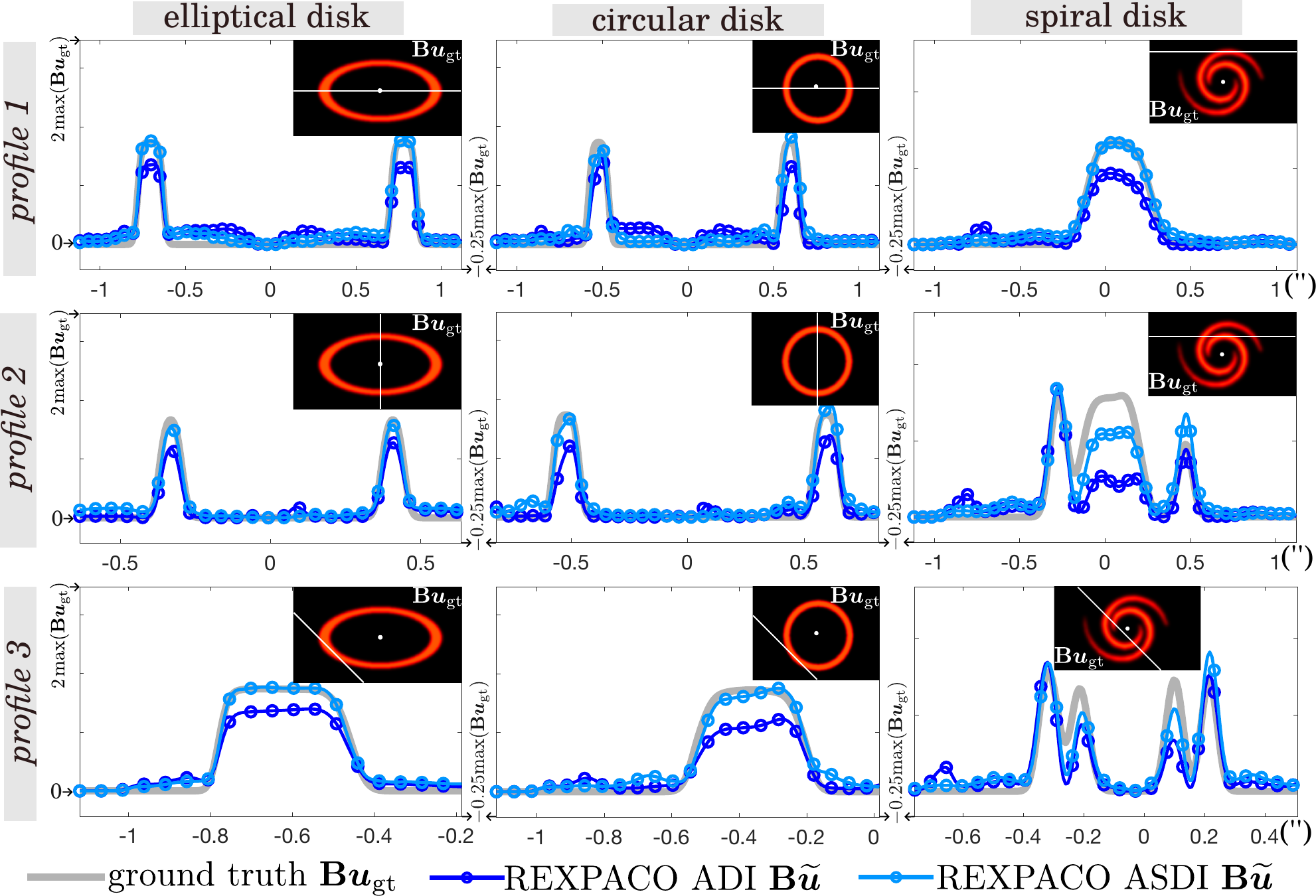}
	\caption{Comparison between ADI and ASDI on the reconstruction of synthetic disks. The considered elliptical, circular, and spiral disks are injected ($\alpha_{\text{gt}} = 1 \times 10^{-5}$, $\Delta_{\text{par}} = 30\degree$, YJ band), within a real SPHERE-IFS dataset and processed with the mono-spectral algorithm REXPACO ADI \citep{flasseur2021rexpaco} and its multi-spectral version REXPACO ASDI proposed in this paper. Line cuts along the three profiles defined in Fig. \ref{fig:gt_fullfig} are extracted from the reconstructions displayed in Fig. \ref{fig:adi_vs_asdi_synthetic_disks_fullfig}.
	 The ground truth flux distribution $\M B \, {\obj}_{\text{gt}}$ and the associated slice-cut locations are recalled in insets. Quantities $\M B \, {\obj}_{\text{gt}}$  and $\M B \, \widetilde{\obj}$ are shown within the range of $[ -0.25\times\text{max}(\M B \, \obj_{\text{gt}}) ; 2 \times \text{max}(\M B \, \obj_{\text{gt}}) ]$ to highlight both over-estimation and under-estimation of the signal of interest. The minimum value (zero) of the ground truth $\M B \, \obj_{\text{gt}}$ is also marked  on the right vertical axis. Dataset: HD 172555 (2015-07-11), see Table \ref{tab:dataset_logs} for the observation parameters.}
	\label{fig:adi_vs_asdi_synthetic_disks_fullfig_cuts_only}
\end{figure*}

This appendix complements the results presented in Sects. \ref{subsec:recons_simulated_disks} and \ref{subsec:importance_spectral_processing} regarding the reconstruction of the flux distributions for synthetic disks. Figures \ref{fig:ellipse_cuts_fullfig}, \ref{fig:circle_cuts_fullfig}, \ref{fig:spiral_cuts_fullfig} report line cuts respectively extracted from Figs. \ref{fig:ellipse_blurred_fullfig}-\ref{fig:ellipse_deblurred_fullfig}, \ref{fig:circle_blurred_fullfig}-\ref{fig:circle_deblurred_fullfig}, and \ref{fig:spiral_blurred_fullfig}-\ref{fig:spiral_deblurred_fullfig} comparing the proposed approach to the median ASDI, PCA ASDI and PACO ASDI baselines. Figure \ref{fig:adi_vs_asdi_synthetic_disks_fullfig_cuts_only} reports line cuts extracted from Fig. \ref{fig:adi_vs_asdi_synthetic_disks_fullfig} comparing the proposed REXPACO ASDI algorithm to its mono-spectral version (REXPACO ADI; \cite{flasseur2021rexpaco}).

\end{document}